\documentclass[aps,prd,groupedaddress]{revtex4-2}
\usepackage{graphicx}
\usepackage{subcaption} 
\usepackage{float}
\usepackage[section]{placeins}

\usepackage{epsfig}
\usepackage{amsfonts}
\usepackage{amsmath}
\usepackage{amssymb}
\usepackage{epsf}
\usepackage{color}
\usepackage{epstopdf}

\usepackage[capitalise]{cleveref}

\newcommand{\insertplot}[5]{\begin{figure}
		\hfill\hbox to 0.05in{\vbox to #5in{\vfill
				\inputplot{#1}{#4}{#5}}\hfill}
		\hfill\vspace{-.1in}
		\caption{#2}\label{#3}
\end{figure}}
\newcommand{\inputplot}[3]{
	\special{ps: plotfile #1}
}

\newcounter{fig}

\newcommand{\rgb}{{{R}_{\mathrm{GB}}^2}}
\newcommand{\coupl}{F}
\newcommand{\diag}{\text{diag}}

\begin{document}

\title{Symmetric wormholes in Einstein-vector-Gauss-Bonnet theory}

\author{Simon Barton}
\email[]{barton@ph1.uni-koeln.de}
\affiliation{Faculty of Mathematics and Natural Sciences, University of Cologne, D-50923 Cologne, Germany}
\author{Claus Kiefer}
\email[]{kiefer@thp.uni-koeln.de}
\affiliation{Faculty of Mathematics and Natural Sciences, University of Cologne, D-50923 Cologne, Germany}
\author{Burkhard Kleihaus}
\email[]{b.kleihaus@uni-oldenburg.de}
\affiliation{Institute of Physics, University of Oldenburg, D-26111 Oldenburg, Germany}
\author{Jutta Kunz}
\email[]{jutta.kunz@uni-oldenburg.de}
\affiliation{Institute of Physics, University of Oldenburg, D-26111 Oldenburg, Germany}

\date{\today}

\begin{abstract}
	We construct wormholes in Einstein-vector-Gauss-Bonnet theory where a real massless vector field is coupled to the higher curvature Gauss-Bonnet invariant.
	We consider three coupling functions which depend on the square of the vector field.
	The respective domains of existence of wormholes possess as their boundaries i) black holes, ii) solutions with a singular throat,
	iii) solutions with a degenerate throat 
	and iv) solutions with cusp singularities.
	Depending on the coupling function wormhole solutions can feature a single throat or an equator surrounded by a double throat.
	The wormhole solutions need a thin shell of matter at the throat, in order to be symmetrically continued into the second asymptotically flat region.
	These wormhole spacetimes allow for bound and unbound particle motion as well as light rings.
\end{abstract}

\maketitle

\section{Introduction}

Like black holes also wormholes have received much attention in recent years, not the least since they may mimic numerous properties of black holes 
\cite{Damour:2007ap,Bambi:2013nla,Azreg-Ainou:2014dwa,Dzhunushaliev:2016ylj,Cardoso:2016rao,Konoplya:2016hmd,Nandi:2016uzg,Bueno:2017hyj,Blazquez-Salcedo:2018ipc}.
The quest for their possible signatures in astrophysical observations has given rise to investigations on gravitational lensing by wormholes
\cite{Cramer:1994qj,Safonova:2001vz,Perlick:2003vg,Nandi:2006ds,Abe:2010ap,Toki:2011zu,Nakajima:2012pu,Tsukamoto:2012xs,Kuhfittig:2013hva,Bambi:2013nla,Takahashi:2013jqa,Tsukamoto:2016zdu},
to studies of their possible shadows
\cite{Bambi:2013nla,Nedkova:2013msa,Ohgami:2015nra,Shaikh:2018kfv,Gyulchev:2018fmd},
to the analysis of accretion disks around wormholes
\cite{Harko:2008vy,Harko:2009xf,Bambi:2013jda,Zhou:2016koy,Lamy:2018zvj,Deligianni:2021ecz,Deligianni:2021hwt}, and more.

The construction of traversable wormholes requires the violation of energy conditions \cite{Morris:1988cz,Visser:1995cc,Alcubierre:2017pqm}.
In General Relativity one therefore should include exotic matter that would provide such a violation. 
A massless real scalar field with a negative kinetic term, i.e., a phantom field, indeed leads to wormhole solutions, as shown by Ellis \cite{Ellis:1973yv,Ellis:1979bh} and Bronnikov \cite{Bronnikov:1973fh}.
However, the need for exotic matter represents a weak point of such wormholes, motivating the search for wormholes that could exist without it.
A direction that has been followed since long and that may provide traversable wormholes without this deficiency is the consideration of wormholes in alternative theories of gravity (see, e.g., \cite{Alcubierre:2017pqm} and references therein).

Alternative theories of gravity have been studied widely in recent years, both in the context of compact objects as well as cosmology \cite{Faraoni:2010pgm,Berti:2015itd,CANTATA:2021ktz}.
Among the plethora of alternative theories, a particular focus has been on scalar-tensor theories that lead to second order equations of motion, i.e., so-called Horndeski theories \cite{Horndeski:1974wa,Kobayashi:2011nu,Charmousis:2011bf,Sotiriou:2015lxa}.
A subset of such theories arises also in the low energy limit of string theory, where higher curvature terms are present in the form of the Gauss-Bonnet (GB) invariant  coupled to a dilatonic scalar field, i.e., Einstein-dilaton-Gauss-Bonnet (EdGB) theories \cite{Zwiebach:1985uq,Gross:1986mw,Metsaev:1987zx}.

While scalar-tensor theories are well-known and have been studied in many contexts, analogous alternative theories that involve vector fields instead of scalar fields, have received much less attention.
Such vector-tensor theories may also lead to second order equations and involve massless or massive vector fields \cite{Horndeski:1976gi,Tasinato:2014eka,Heisenberg:2014rta}.
These provide a largely uncharted area, whose exploration may lead to new options and insights, but these theories should also recover the known phenomenology of the gravitational interaction.

Besides the strong bounds known for the solar system, there are also strong restrictions from pulsars that must be obeyed, while observations of gravitational radiation and black hole shadows impose further constraints (see, e.g., \cite{Will:2018bme,Shao:2014wja,LIGOScientific:2016aoc,LIGOScientific:2017vwq,LIGOScientific:2017ync,EventHorizonTelescope:2019dse}).
While, in particular, for the theoretically well-motivated EdGB theories the observational window has become rather narrow, an attractive related set of theories is less effected by recent observations. 
Here instead of the dilaton some other scalar field is coupled in a specific way, that allows these Einstein-scalar-Gauss-Bonnet (EsGB) theories to retain the solutions of General Relativity as solutions of the new set of field equations, and that leads in addition to scalarized solutions in certain regions of parameter space
\cite{Doneva:2017bvd,Silva:2017uqg,Antoniou:2017acq}.

For wormholes the situation is different, since General Relativity does not allow for wormhole solutions, without the addition of exotic matter.
As shown first in EdGB theories, the effective stress-energy tensor that arises from the GB term, coupled to the scalar field, itself allows for the violation of the energy conditions and therefore gives rise to wormholes \cite{Kanti:2011jz,Kanti:2011yv}.
This remains true when other coupling functions are employed, as demonstrated already for various EsGB theories \cite{Antoniou:2019awm,Ibadov:2020btp,Ibadov:2020ajr}.
The black holes of the corresponding theories represent typically a part of the boundary of the domain of existence (DoE) of the wormhole solutions.

To the best of our knowledge we here explore for the first time the existence of wormhole solutions in vector-tensor theories.
In contrast, black holes and neutron stars have already been addressed and investigated in various vector-tensor theories \cite{Muller:1988,Verbin:2020fzk,Chagoya:2016aar,Babichev:2017rti,Chagoya:2017fyl,Heisenberg:2017xda,Heisenberg:2017hwb,Fan:2016jnz}.
In particular, we here consider Einstein-vector-Gauss-Bonnet (EvGB) theories that retain the vacuum solutions of General Relativity, but allow in addition for vectorized solutions \cite{Ramazanoglu:2017xbl,Ramazanoglu:2018tig,Ramazanoglu:2019gbz,Ramazanoglu:2019jrr,Annulli:2019fzq,Barton:2021wfj,Garcia-Saenz:2021uyv,Silva:2021jya,Demirboga:2021nrc}.
We investigate EvGB theories with a  massless vector field for three coupling functions, for which the associated spontaneously vectorized black holes have been recently obtained \cite{Barton:2021wfj}.
When investigating the DoE of these wormholes we find that the vectorized black holes form part of their boundary of existence, and also the Schwarzschild and Reissner-Nordstr\"om (RN) black holes can be found on parts of the boundary.

In Section II we provide the theoretical setting for the study, and present the action, the field equations, and the Ans\"atze for spherically symmetric wormholes. We discuss the charges, i.e., the mass $M$ and the vector charge $Q$, and the identification of throats and equators, and we recall the equations for the geodesics of particles and light.
Section III contains the presentation of our results. We illustrate the wormhole solutions and discuss the occurrence of singularities, whose presence leads us to consider only symmetrized wormholes here.
To that end we reflect the solutions at the throat or equator to obtain the solution in the second asymptotically flat region on the other side of the throat or equator, respectively.

Subsequently we investigate the DoE and determine its boundary, covering the range of the GB coupling constant $0 \le \lambda/M^2 \le 10^8$. 
We then determine the DoE for the area of the throat, and consider the values of the vector field and the metric at the throat.
We demonstrate the violation of the null energy condition by the wormhole solutions, and we illustrate selected solutions in terms of their embeddings, both for solutions that feature only a single throat and those with an equator and a double throat.
We finally turn to the geodesics in these wormholes spacetimes. 
We show the presence of bound and unbound motion, and we extract their light rings.
We then end with our conclusions.
The Appendix highlights a special region of the DoE, and it discusses the junction conditions that need to be satisfied at the throat or equator, in order to obtain symmetric wormholes.

\section{Theoretical setting}
	
\subsection{Action and equations of motion}
	
We start from the action for EvGB theory
	\begin{equation}
    	S=\frac{1}{16\pi}\int \left[ R-F_{\mu\nu}F^{\mu\nu} + \lambda\coupl(A_\mu A^\mu) \rgb \right]\sqrt{-g}d^4x
    	\label{act}
    \end{equation}
with curvature scalar $R$, field strength tensor $F_{\mu\nu}$ of the massless vector field $A_\mu$,
and	Gauss-Bonnet (GB) term $R^2_{\rm GB}$
	\begin{eqnarray} 
		R^2_{\rm GB} = R_{\mu\nu\rho\sigma} R^{\mu\nu\rho\sigma}
		- 4 R_{\mu\nu} R^{\mu\nu} + R^2 \ . 
	\end{eqnarray} 
In the action the vector field $A_\mu$ is coupled with some coupling function $F(A_\mu A^\mu)$ to the GB invariant in order to obtain non-vanishing contributions to the equations of motion, since the GB invariant $R^2_{\rm GB}$ is topological in four dimensions.
The strength of the coupling is determined by the GB coupling constant $\lambda$, which has dimension length squared.
The coupling function $F(A_\mu A^\mu)$ is chosen to depend only on the square of the vector field, and vanishes for vanishing vector field.
These conditions allow Schwarzschild black holes to remain solutions of the field equations.
RN black holes in contrast are only solutions for vanishing GB coupling constant.
We consider the following choices for $F(A_\mu A^\mu)$ \cite{Barton:2021wfj}
    \begin{align}\label{eq:coupl}
    	(i) ~~\qquad \coupl(A_\mu A^\mu) &= A_\mu A^\mu\\
    	(ii) ~\qquad \coupl(A_\mu A^\mu) &= 1-\exp(-A_\mu A^\mu)\\
    	(iii) \qquad \coupl(A_\mu A^\mu) &= \exp(A_\mu A^\mu)-1 \ .
    \end{align} 

We obtain the field equations from the variational principle.
Varying the action (\ref{act}) with respect to the vector field and to the metric yields the coupled set of EvGB equations
	\begin{equation}
		\nabla_\mu F^{\mu\nu} =  -\frac{1}{2} \frac{dF(A_\mu A^\mu)}{d A_\mu A^\mu} R^2_{\rm GB} A^\nu \ ,
		\label{scleq}
	\end{equation}
	\begin{equation}
		G_{\mu\nu}  = \frac{1}{2}T^{({\rm eff})}_{\mu\nu} \ , 
		\label{Einsteq}
	\end{equation}
where $G_{\mu\nu}$ denotes the Einstein tensor and $T^{({\rm eff})}_{\mu\nu}$ the effective stress-energy tensor	
	\begin{equation}
		T^{({\rm eff})}_{\mu\nu} = T^{(A)}_{\mu\nu} - 2 T^{(GB)}_{\mu\nu} \ ,
		\label{teff}
	\end{equation}
consisting of contributions from the vector field
	\begin{equation}
		T^{(A)}_{\mu\nu} = 4 F_\mu^{\phantom{\mu}\lambda} F_{\nu\lambda} 
		-g_{\mu\nu} F_{\rho\lambda}F^{\rho\lambda} \ , 
		\label{tphi}
	\end{equation}
and the GB invariant 
	\begin{equation}
		T^{(GB)}_{\mu\nu} =
		\frac{1}{2}\left(g_{\rho\mu} g_{\lambda\nu}+g_{\lambda\mu} g_{\rho\nu}\right)
		\eta^{\kappa\lambda\alpha\beta}\tilde{R}^{\rho\gamma}_{\phantom{\rho\gamma}\alpha\beta}
		\nabla_\gamma \nabla_\kappa F(A_\mu A^\mu) + 
		R^2_{\rm GB}\frac{d F(A_\sigma A^\sigma)}{d (A_\sigma A^\sigma)} A_\mu A_\nu\ ,
		\label{teffi}
	\end{equation}
with 
	$\tilde{R}^{\rho\gamma}_{\phantom{\rho\gamma}\alpha\beta}=\eta^{\rho\gamma\sigma\tau}
	R_{\sigma\tau\alpha\beta}$ and $\eta^{\rho\gamma\sigma\tau}= 
	\epsilon^{\rho\gamma\sigma\tau}/\sqrt{-g}$.

	To obtain static, spherically symmetric solutions we consider the line element in isotropic coordinates %
	\begin{equation}\label{eq_met}
		ds^2 = -{f_0}(r) dt^2 +{f_1}(r)\left[dr^2 
		+r^2\left( d\theta^2+\sin^2\theta d\varphi^2\right) \right]\ ,
	\end{equation}
	and we assume for the vector field the form
	\begin{equation}\label{eq_vec}
		A_\mu dx^\mu = A_t(r) dt \ .
	\end{equation}
	When we insert the above ansatz (\ref{eq_met})-(\ref{eq_vec}) for the metric and the vector field into the set of EvGB equations
	we obtain four coupled, nonlinear ordinary differential equations, one of which can be treated as a constraint.
	This leaves us with three independent second order ordinary differential equations (ODEs).

	Inspection of the field equations reveals an invariance under the scaling transformation
	\begin{equation}
		r \to \chi r \ , \ \ \    t \to \chi t \ , \ \ \  F \to \chi^2 F\  , \ \ \  \chi > 0 \ .
	\end{equation}

\subsection{Asymptotic Expansion at Spatial Infinity}

Introducing the constants $M$ and $Q$ we obtain the asymptotic behaviour as
\begin{align}
	g_{tt}&=-f_0(r)=-1+\frac{2M}{r}+\mathcal{O}\left(\frac{1}{r^2}\right),\label{eq:expansion_inf1} \\
	g_{rr}&=\phantom{-}f_1(r)=\phantom{-}1+\frac{2M}{r}+\mathcal{O}\left(\frac{1}{r^2}\right),\label{eq:expansion_inf2} \\
	A_t&=\; \; A_0(r)=\frac{Q}{r}+\mathcal{O}\left(\frac{1}{r^2}\right).\label{eq:expansion_inf3} 
\end{align}

In the static, spherical case the timelike Killing vector is $\xi^\mu=(1,0,0,0)$.
The mass associated with the 3-volume $V_r$ between the throat and the 2-sphere $\mathcal{S}_r$ of constant $r$ is obtained from the boundary $\partial V_r$.
Because the boundary integral over the throat surface does not vanish, the expression for the mass in this volume picks up a contribution from the throat.

Plugging in our choice of the line element and then the expansions at infinity, the Komar mass reads
\begin{align}
	M_{\text{Komar}}(r) &= M_{\text{thr}} + \frac{1}{4\pi}\int_{\partial V_r} d\theta d \varphi \sqrt{\hat{g}_{(2)}}n_t\sigma_r \nabla^t \xi^r
	\label{MKo1}\\
	M_{\text{Komar}}(\infty) &= \frac{1}{8\pi}\lim_{r \to \infty} \int_{\mathcal{S}_r} \frac{f_0'(r)}{\sqrt{f_0(r) f_1(r)}} \sin\theta \, r^2 \, d\theta d\varphi = M
	\label{MKo2}
\end{align}

We therefore identify the constant $M$ with the mass of the solution. 
Here $n_\mu$ is the future-pointing timelike normal vector of $V_r$, $\hat{g}_{(2)}$ is the induced metric on the boundary and $\sigma_\nu$ is its inward pointing spacelike normal vector.
$M_{\text{thr}}$ denotes the Komar integral evaluated at the throat, 
$M_{\text{thr}} = \frac{1}{4\pi}\int_{\text{thr}} d\theta d \varphi \sqrt{\hat{g}_{(2)}}n_t\sigma_r \nabla^t \xi^r$.
Analogously we compute the vector charge from the asymptotic behaviour
\begin{align}
	Q_{\text{Komar}}(r) &= Q_{\text{thr}} + \frac{1}{4\pi}\int_{\partial V_r} d\theta d \varphi \sqrt{\hat{g}_{(2)}}n_\mu\sigma_\nu F^{\mu\nu}
	\label{QKo1} \\
	Q_{\text{Komar}}(\infty) &= \frac{1}{4\pi}\lim_{r \to \infty} \int_{\mathcal{S}_r} \frac{-A_0'(r)}{\sqrt{f_0(r) f_1(r)}} \sin\theta \, r^2 d\theta d\varphi = Q\ ,
	\label{QKo2}
\end{align}
with $Q_{\text{thr}} =  \frac{1}{4\pi}\int_{\text{thr}} d\theta d \varphi \sqrt{\hat{g}_{(2)}}n_\mu\sigma_\nu F^{\mu\nu}$.
Thus, as expected, the constant $Q$ is identified with the vector charge of the wormhole.
In both cases, the contribution from the inner boundary cancels the explicit throat charges.

\FloatBarrier
\subsection{Throats and equators}

To identify and characterize a wormhole one has to consider the circumferential radius $R_c(r)$ as a function of the radial coordinate
\begin{equation}
R_c(r) = \frac{1}{2\pi}\,\int_0^{2 \pi} \sqrt{g_{\varphi \varphi}(r,\theta)}|_{\theta=\pi/2}\, d\varphi\,
\label{circum-radius}
\end{equation}
This yields for the line element (\ref{eq_met})
\begin{equation}
R_c(r) = \sqrt{f_1(r)} r \ .
\label{circum-radius2}
\end{equation}
When $R_c$ develops a minimum at some value $r_0$, this corresponds to the location of a wormhole throat.
However, $R_c$ could also develop a maximum at some value $r_0$, which would then represent an equator.
In mathematical terms this translates into
\begin{equation}
\frac{dR_c}{dr}\Biggr|_{r_0}=0\,, \qquad \frac{d^2R_c}{dr ^2}\Biggr|_{r_0} \gtrless 0\,,
\label{cond-1}
\end{equation}
where the greater sign ($>$) corresponds to the presence of a throat and the smaller sign ($<$) to an equator.
The special case $\left.\frac{d^2R_c}{dr ^2}\right|_{r_0}=0$ corresponds to a saddle point and will be referred to as degenerate throat.
In these coordinates the area of the throat or the equator is given by
    \begin{equation}
        {\cal A}_{t,e} = 4\pi R^2_c(r_0) = 4\pi r_0^2 f_1(r_0) \ .
    \end{equation}

Alternatively, we also consider the line element in isotropic wormhole coordinates
\begin{align}\label{eq_whcoord}
	ds^2 = -F_0(\eta) dt^2 + F_1(\eta)\left( d\eta^2 + (\eta^2+\eta_0^2) (d\theta^2 + \sin^2\theta d\varphi^2)\right).
\end{align}
In contrast to ordinary wormhole coordinates,
this expression is sufficiently general to 
allow for minima (i.e., throats) and maxima
(i.e., equators) of the circumferential radius.
In these coordinates, symmetric wormholes would either possess a single throat at $\eta=0$, or they would possess an equator at $\eta=0$.
In the latter case, the equator would be symmetrically located between two throats.

The coordinate transformation between the two radial coordinates $\eta$ and $r$ is given by
\begin{equation}\label{key}
\eta=r_0\left(\frac{r}{r_0}-\frac{r_0}{r}\right)
\end{equation}
with $\eta_0 = 2r_0$.
Note that $\eta$ scales in the same way as $r$, $\eta \to \chi \eta$.
For a throat or equator that is located at  $\eta=0$, its area is
\begin{equation}
{\cal A}_{t,e} = 4\pi R^2_c(0)=4\pi \eta_0^2 F_1(0) \ ,
\end{equation}
while for a symmetric double-throat wormhole, whose throats would be located at $\pm \eta_{\rm t}$, the area of the throats would be
\begin{equation}
{\cal A}_{t} = 4\pi R^2_c(\eta_{\rm t})=4\pi (\eta_{\rm t}^2 + \eta_0^2 ) F_1(\eta_{\rm t}) \ .
\end{equation}

\subsection{Geodesics} \label{geodesics}
	The Lagrangian for geodesics is $2\mathcal{L}=g_{\mu\nu}\dot{x}^\mu\dot{x}^\nu$, where the dot indicates the derivative with respect to some affine parameter $\tau$.
	In the wormhole coordinates used in \cref{eq_whcoord} it reads 
	\begin{equation}
		2\mathcal{L} = -F_0(\eta) \dot{t}^2 + F_1 \dot{\eta}^2 + F_1 \left(\eta^2+\eta_0^2\right) \dot{\theta}^2 + F_1\sin^2(\theta ) \left(\eta^2+\eta_0^2\right) \dot{\varphi}^2.
	\end{equation}
	The conjugate momenta of the cyclic coordinates $(t, \varphi)$ are conserved and identified with the negative energy $E$ of the orbit and its angular momentum $L$, respectively,
	\begin{align}\label{xx}
		-E &\equiv p_t = \frac{\partial \mathcal{L}}{\partial \dot{t}} = - {F_0} \dot{t}\\	
		&\phantom{{}\equiv{}} p_\eta = \frac{\partial \mathcal{L}}{\partial \dot{\eta}} = F_1 \dot{\eta}\\
		&\phantom{{}\equiv{}} p_\theta = \frac{\partial \mathcal{L}}{\partial \dot{\theta}} = F_1 \left(\eta^2+\eta_0^2\right) \dot{\theta}\\
		L &\equiv p_\varphi = \frac{\partial \mathcal{L}}{\partial \dot{\varphi}} = F_1 \sin^2\theta  \left(\eta^2+\eta_0^2\right) \dot{\varphi}
	\end{align}
    The square $\kappa$ of the tangent vector is conserved along a geodesic.
	We choose ${\theta=\pi/2,\dot{\theta}=0}$ and rewrite
	\begin{align}
		2\mathcal{L} = -\frac{{E}^2}{{F_0}(\eta)} + {F_1}(\eta) \dot{\eta}^2 + \frac{L^2}{F_1(\eta) (\eta^2+\eta_0^2)}	= \kappa
	\end{align}
	The equations of motion then take the simple form
	\begin{align}
		\dot{\eta}^2 &= \frac{{E}^2 - V_{\text{eff}}^2}{{F_0}(\eta) {F_1}(\eta)} \label{eq:geo-eom-1},	\\
		\dot{\varphi} &= \frac{L}{F_1(\eta) (\eta^2+\eta_0^2)}
	\end{align}
	with the effective potential
	\begin{equation}
		V_{\text{eff}}^2 = F_0(\eta)\left(\frac{L^2}{F_1(\eta) (\eta^2+\eta_0^2)}-\kappa\right)
		\label{veff}
	\end{equation}

\subsection{Numerics}
In isotropic coordinates the equations are all of second order. 
Solving the field equations subject to a set of boundary conditions 
is an integration problem.
For the wormholes presented in this work a Runge-Kutta method of fifth order (Bogacki-Shampine \cite{Bogacki1996}) was used.
The black holes were computed using the collocation point solver COLSYS as described in \cite{Barton:2021wfj}.

\newpage

	
	\begin{figure}[t!]
		\begin{subfigure}{.5\textwidth}
			\includegraphics[width=\textwidth]{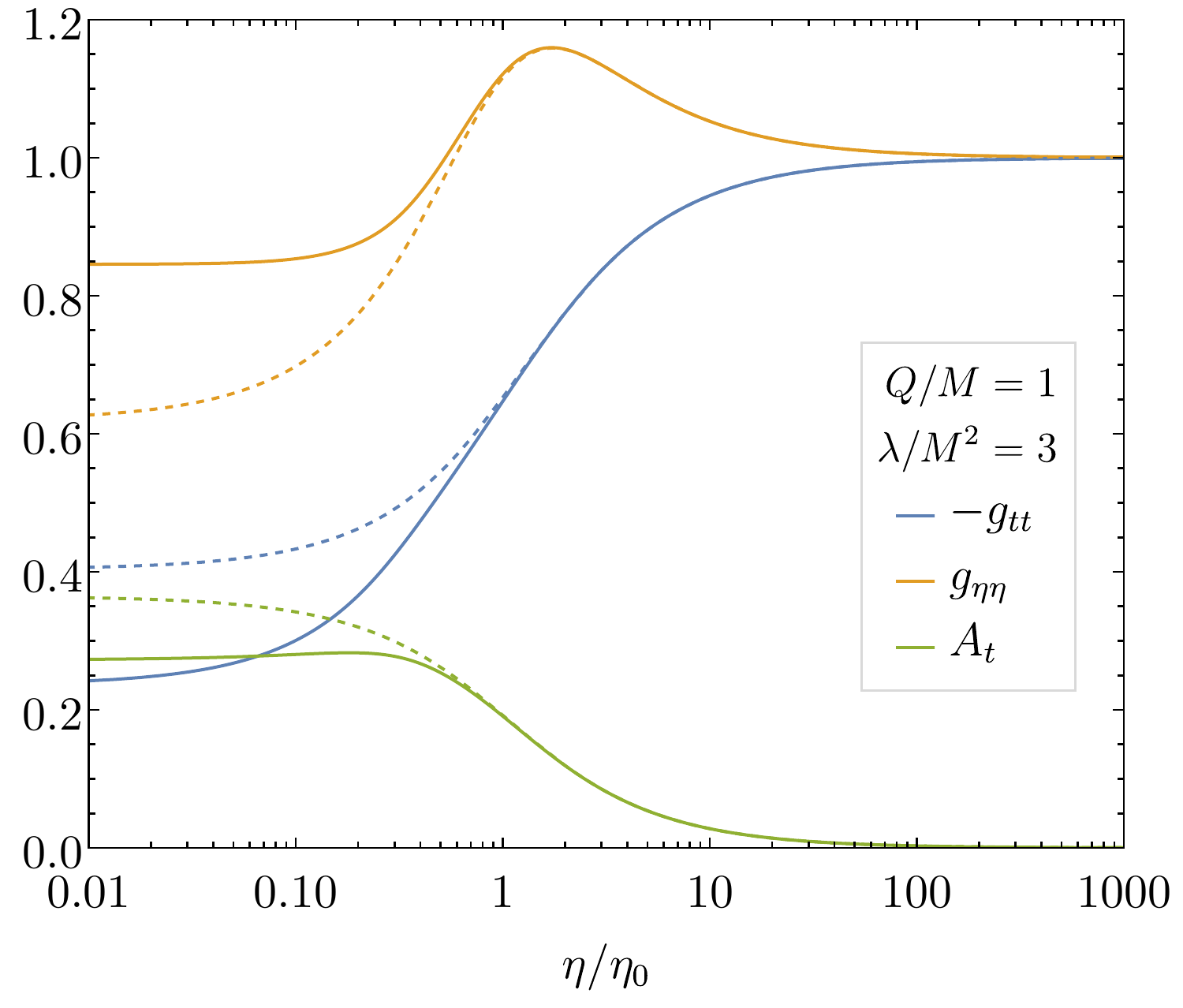}
			\caption{}
		\end{subfigure}%
		\begin{subfigure}{.5\textwidth}
			\includegraphics[width=\textwidth]{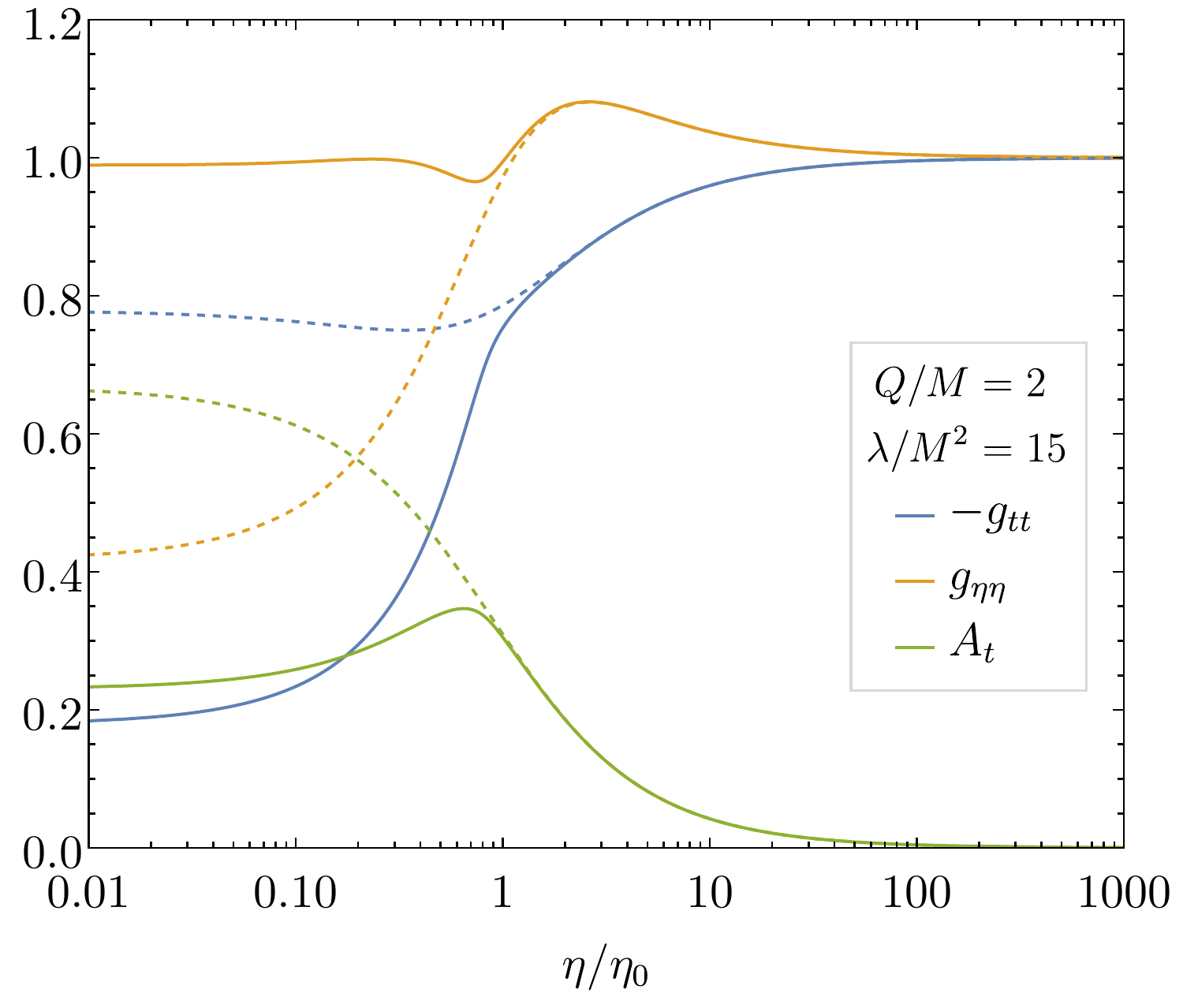}
			\caption{}
		\end{subfigure}\\
		\begin{subfigure}{.5\textwidth}
			\includegraphics[width=\textwidth]{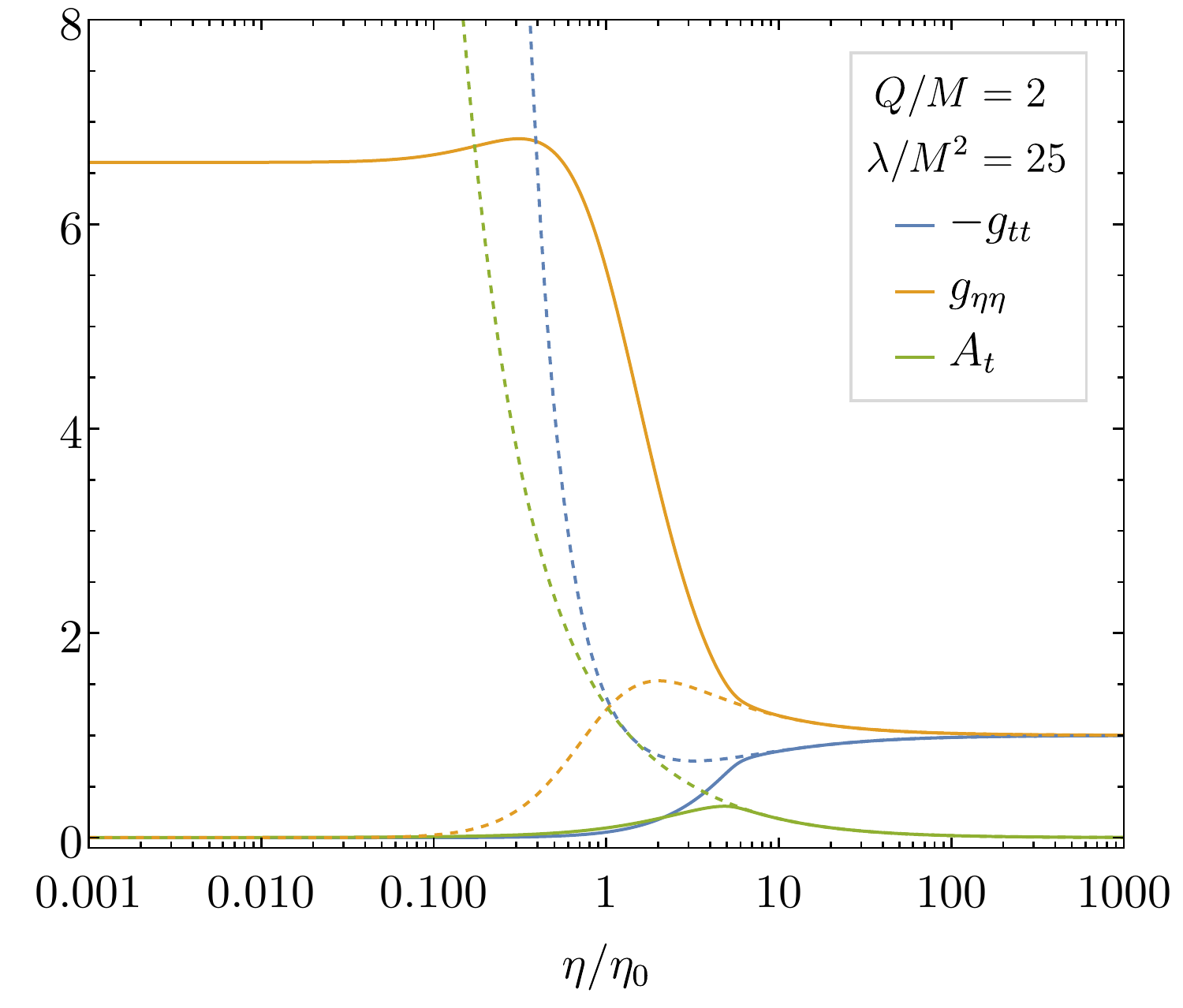}
			\caption{}
		\end{subfigure}%
		\begin{subfigure}{.5\textwidth}
			\includegraphics[width=\textwidth]{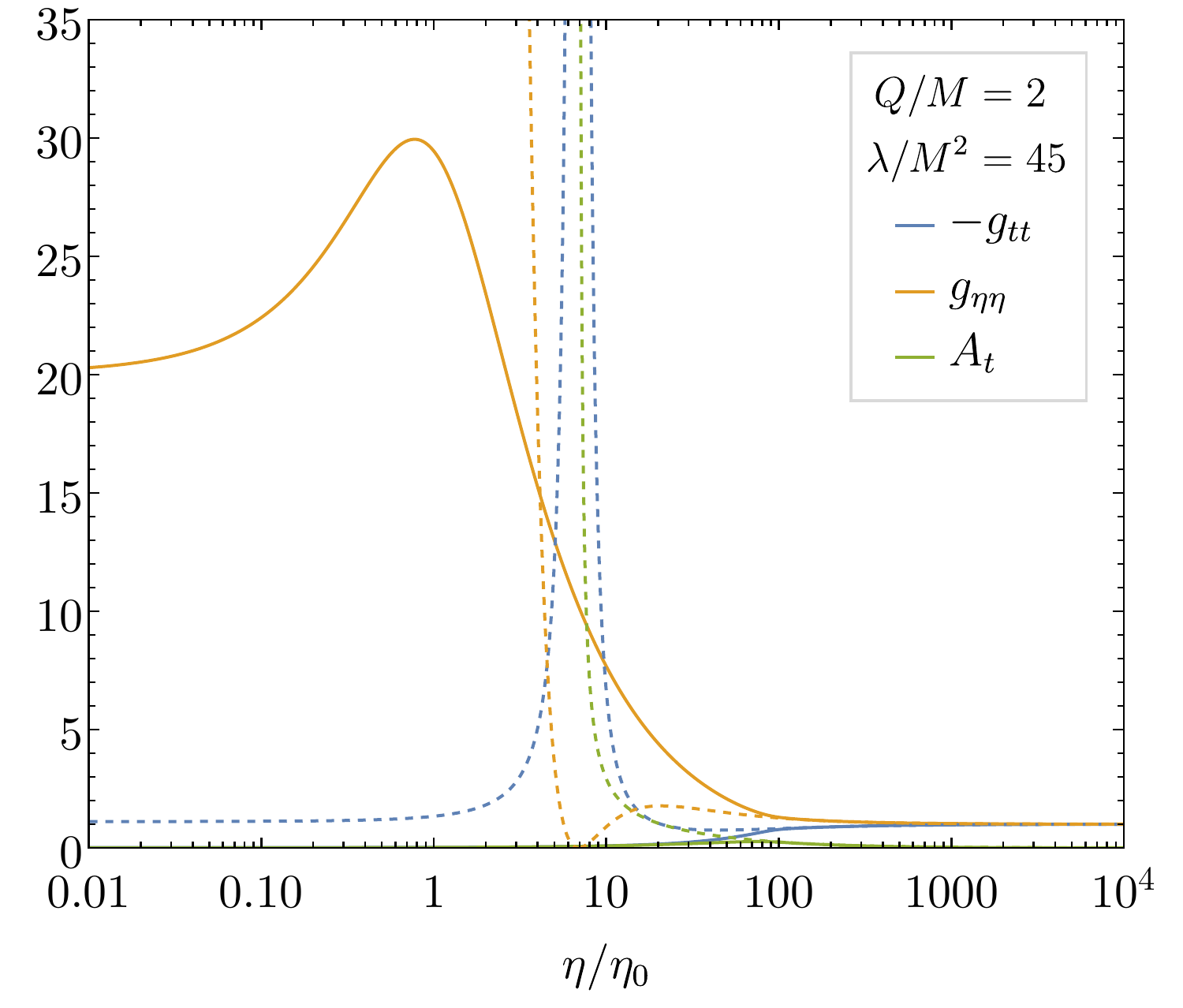}
			\caption{}
		\end{subfigure}\\
		\begin{subfigure}{.5\textwidth}
			\includegraphics[width=\textwidth]{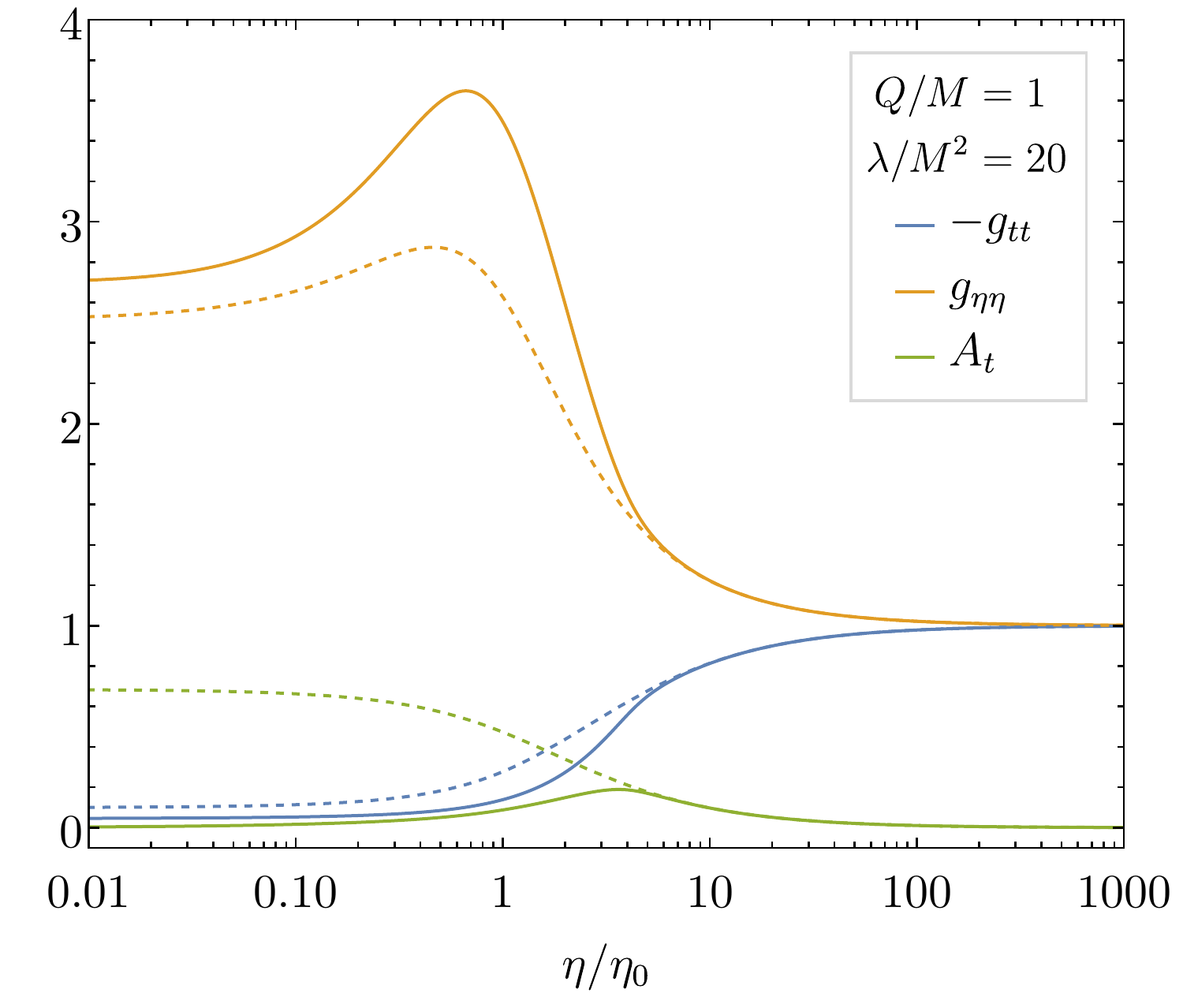}
			\caption{}
		\end{subfigure}%
		\begin{subfigure}{.5\textwidth}
			\includegraphics[width=\textwidth]{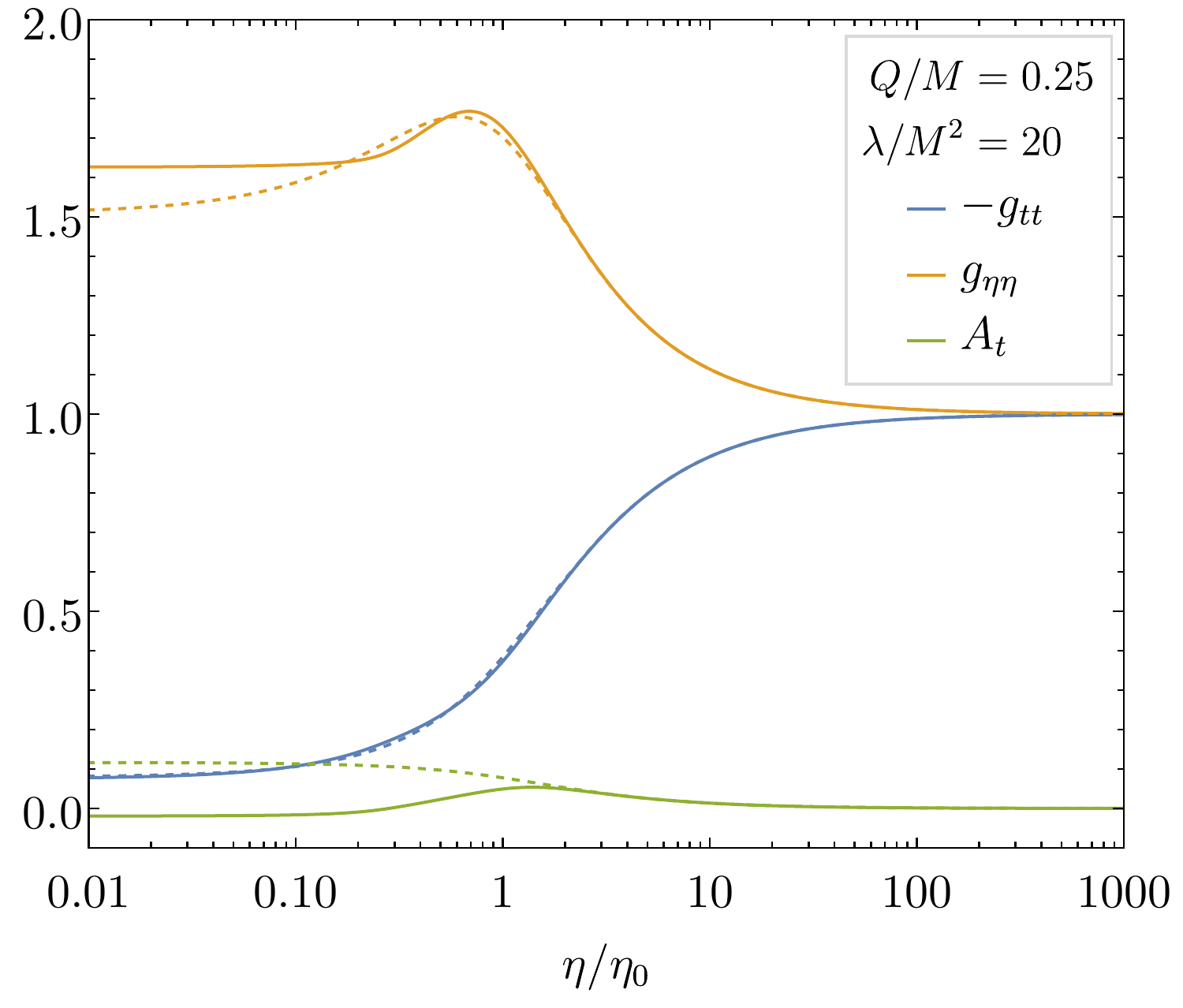}
			\caption{}
		\end{subfigure}\\
		\caption{
        Several examples for symmetric wormhole solutions are shown for coupling function (i):
        The metric coefficients $g_{tt}$, $g_{\eta\eta}$ and the vector field component $A_t$ vs the wormhole radial coordinate $\eta$.
        The RN solution of the same mass and vector charge is shown in dashed.}
			\label{fig:fun}
	\end{figure}
	
	\begin{figure}[t!]
		\begin{subfigure}{.5\textwidth}
			\includegraphics[width=\textwidth]{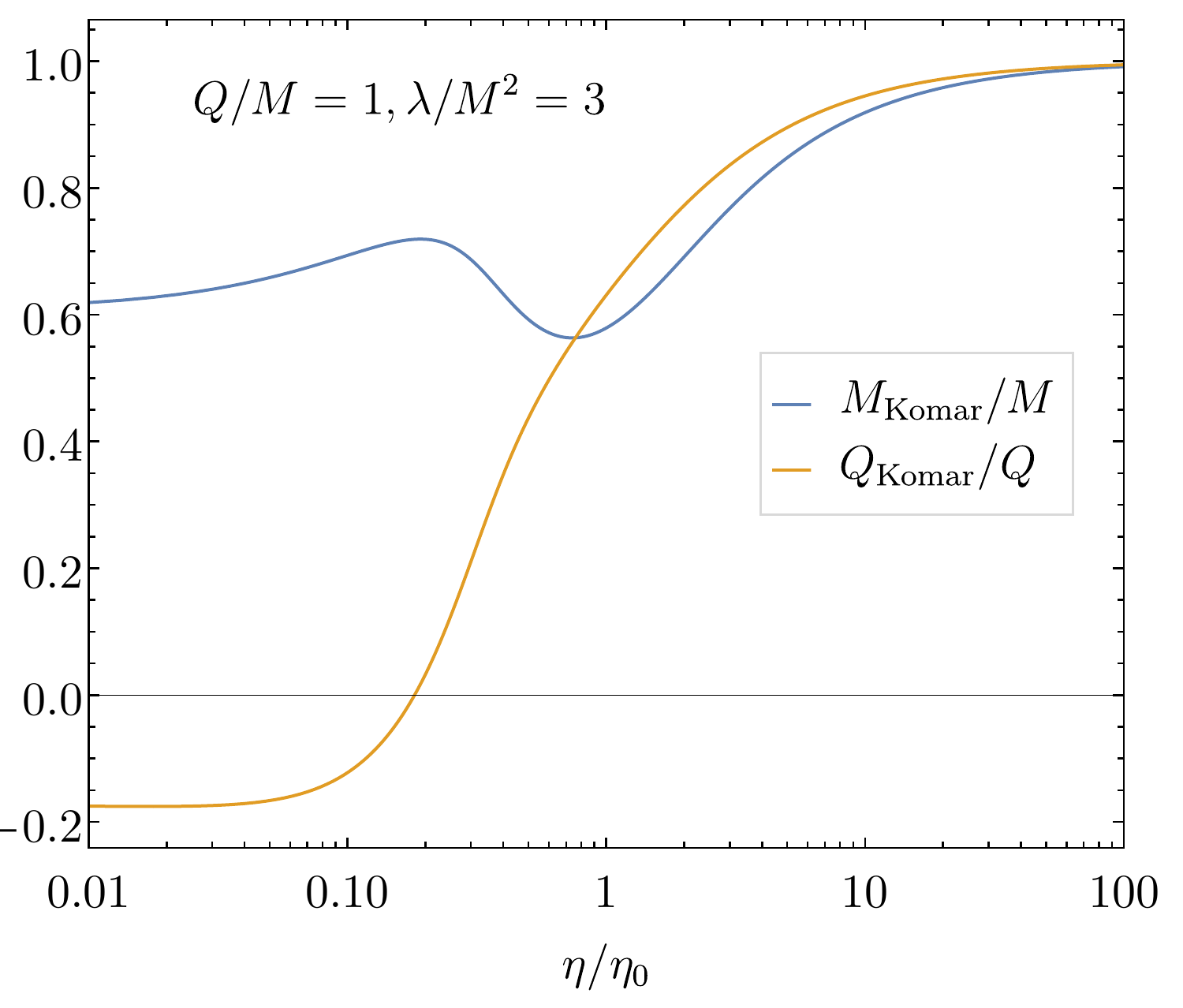}
			\caption{}
		\end{subfigure}%
		\begin{subfigure}{.5\textwidth}
			\includegraphics[width=\textwidth]{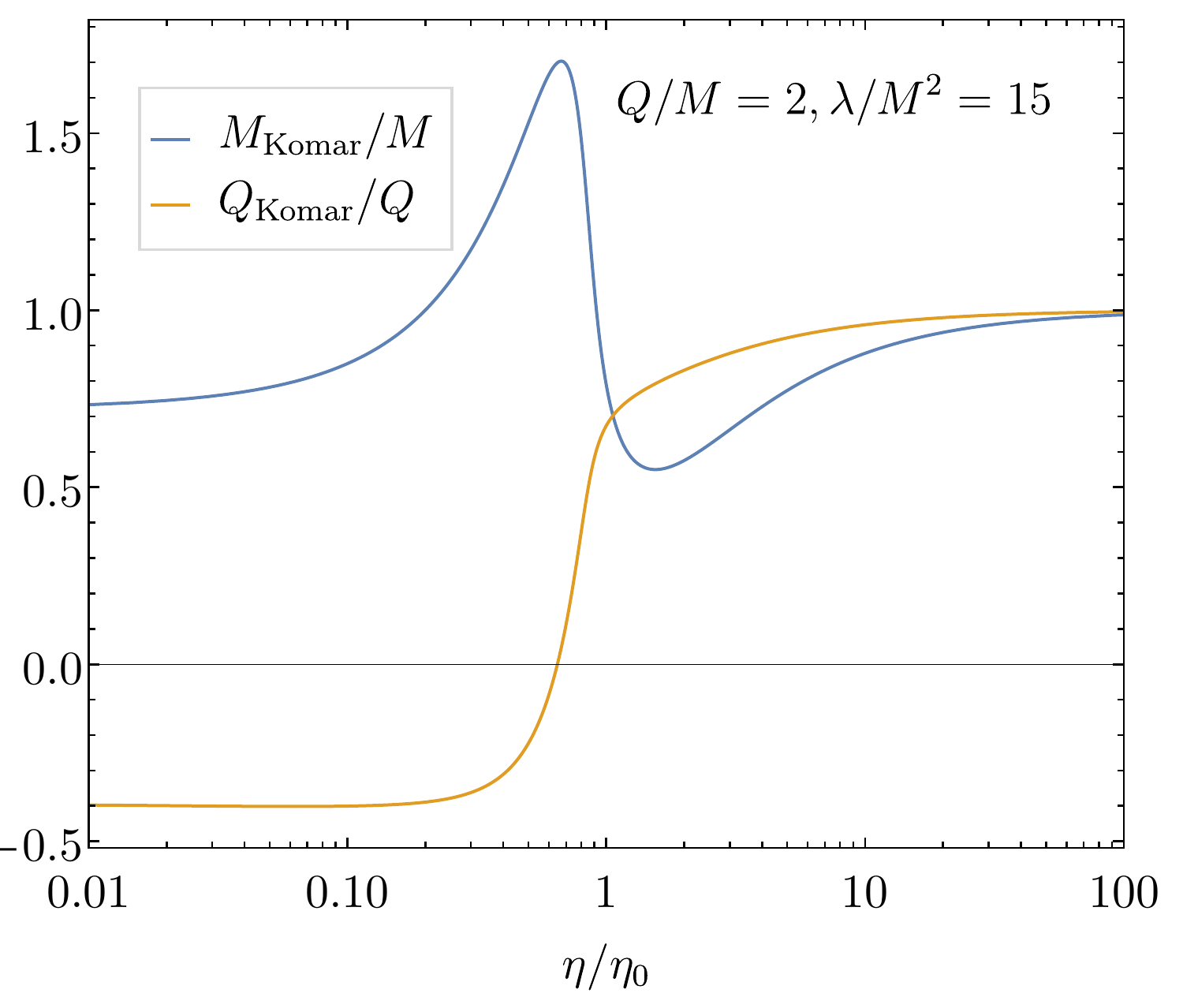}
			\caption{}
		\end{subfigure}\\
		\begin{subfigure}{.5\textwidth}
			\includegraphics[width=\textwidth]{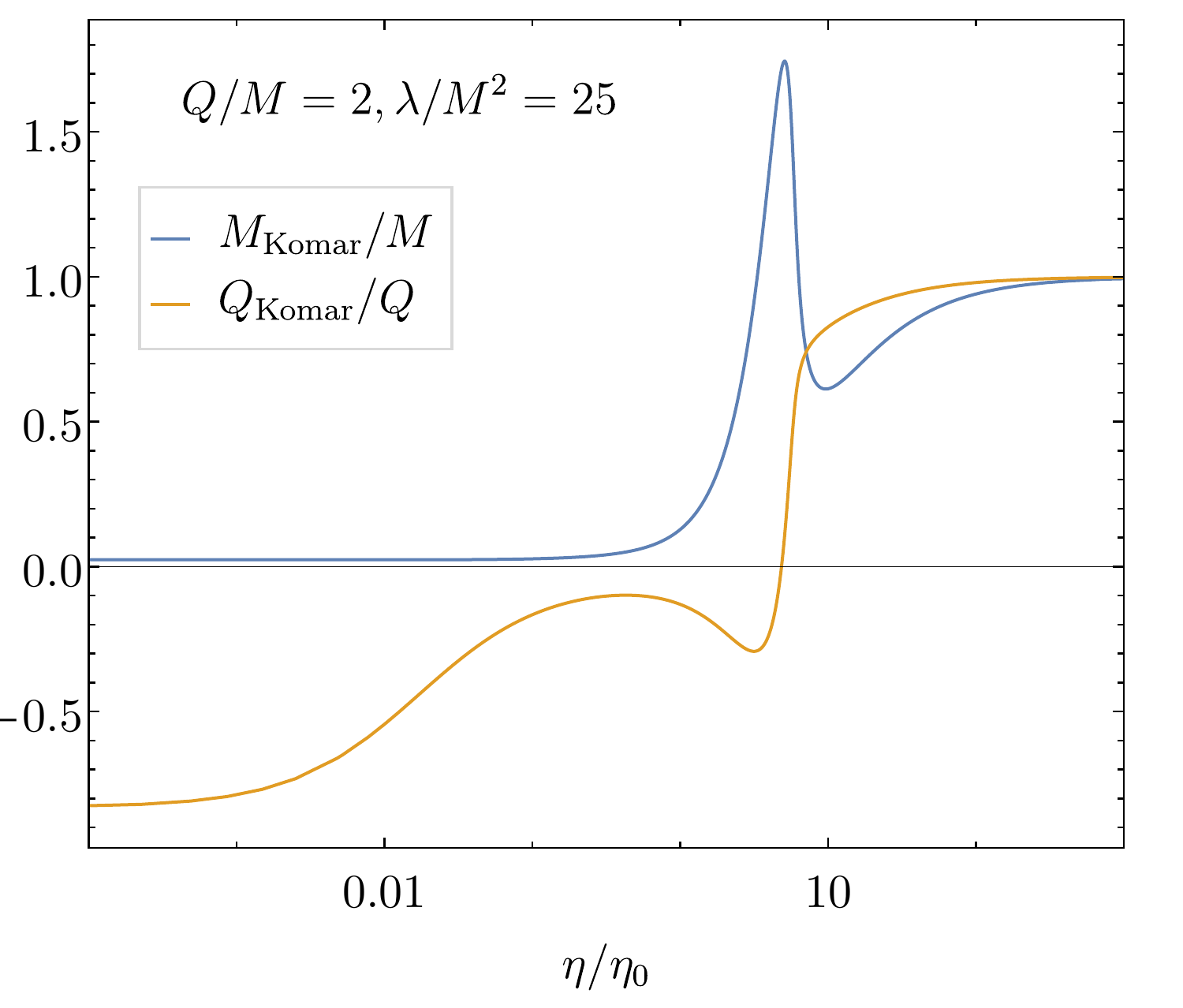}
			\caption{}
		\end{subfigure}%
		\begin{subfigure}{.5\textwidth}
			\includegraphics[width=\textwidth]{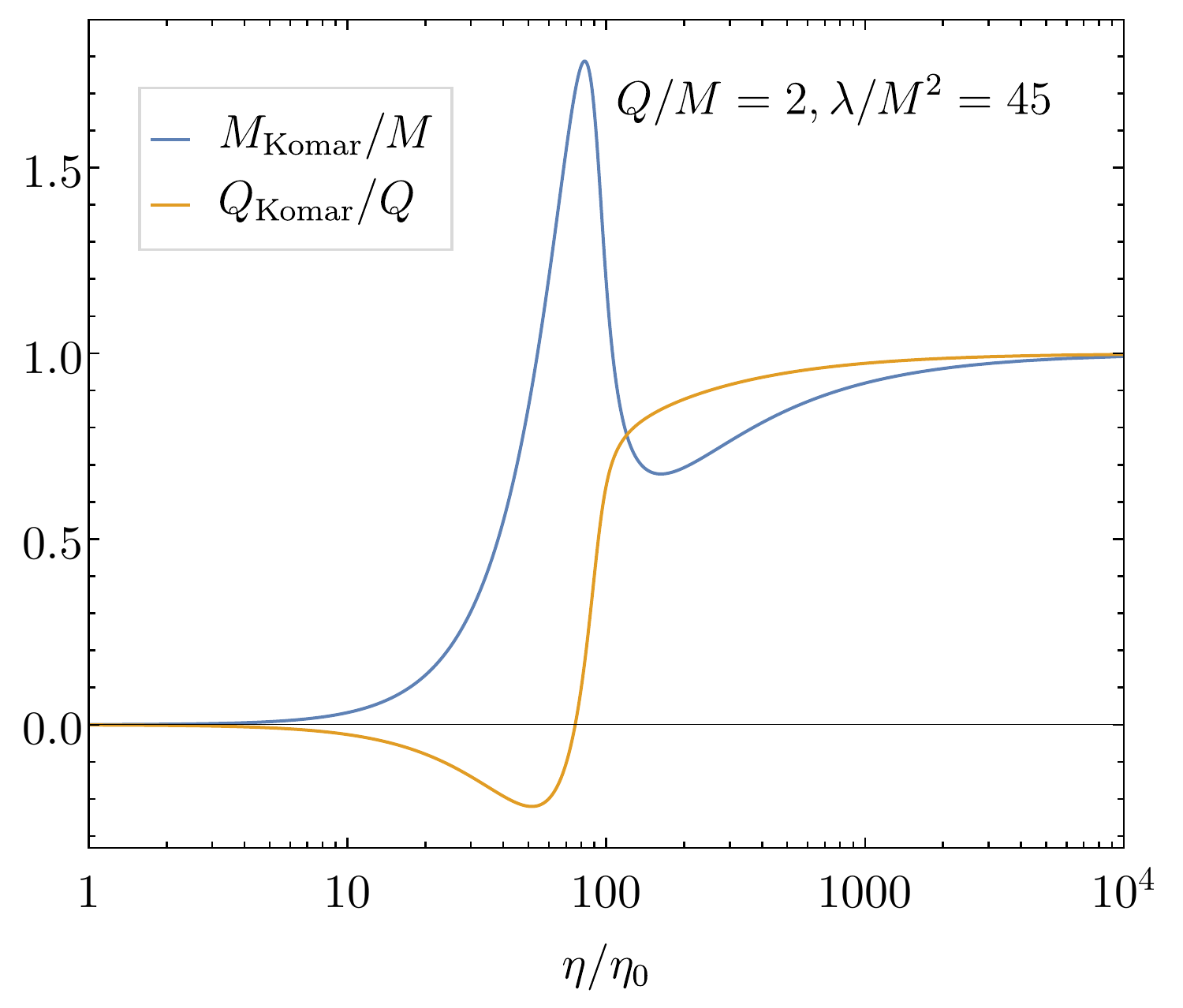}
			\caption{}
		\end{subfigure}\\
		\begin{subfigure}{.5\textwidth}
			\includegraphics[width=\textwidth]{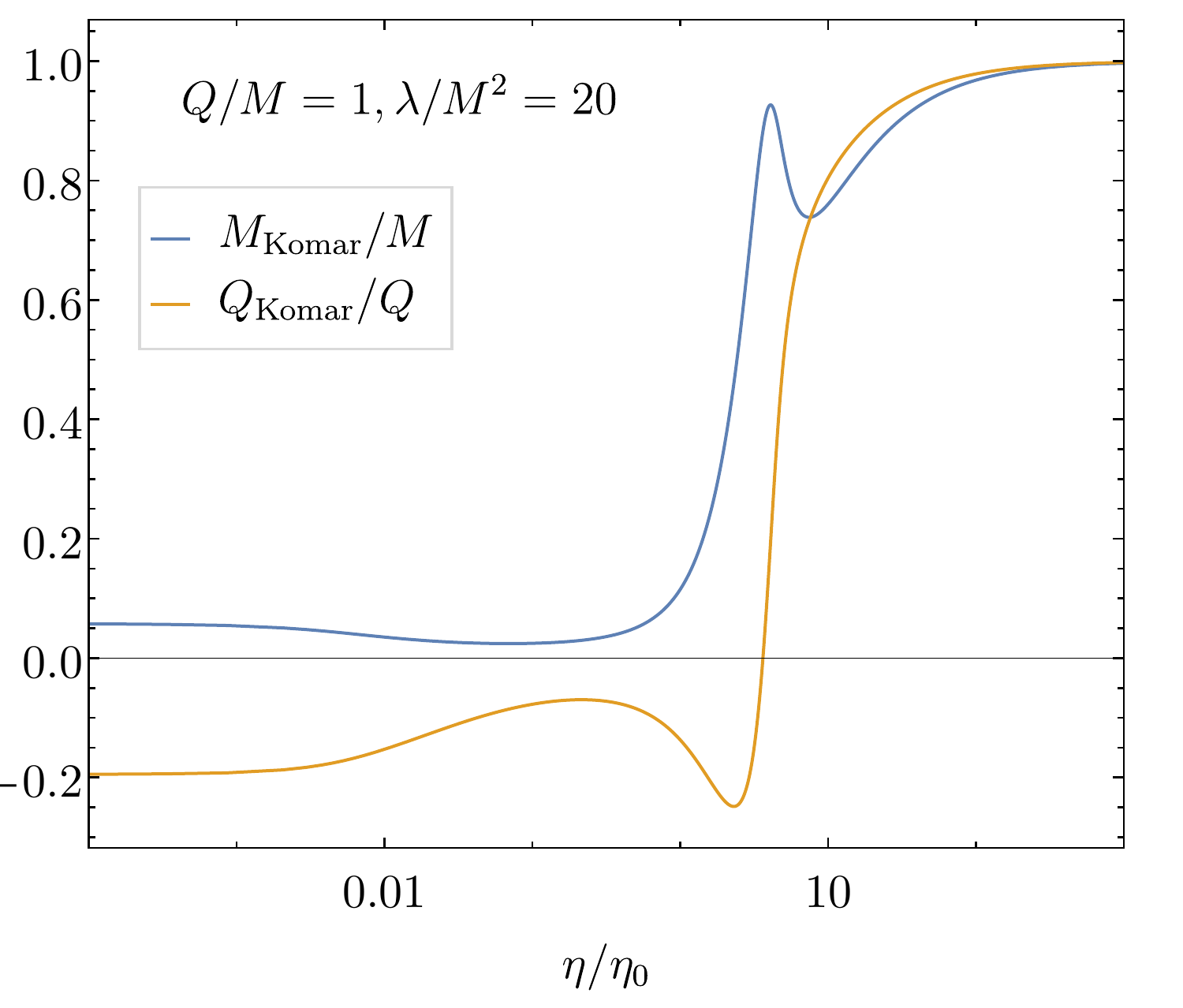}
			\caption{}
		\end{subfigure}%
		\begin{subfigure}{.5\textwidth}
		    \includegraphics[width=\textwidth]{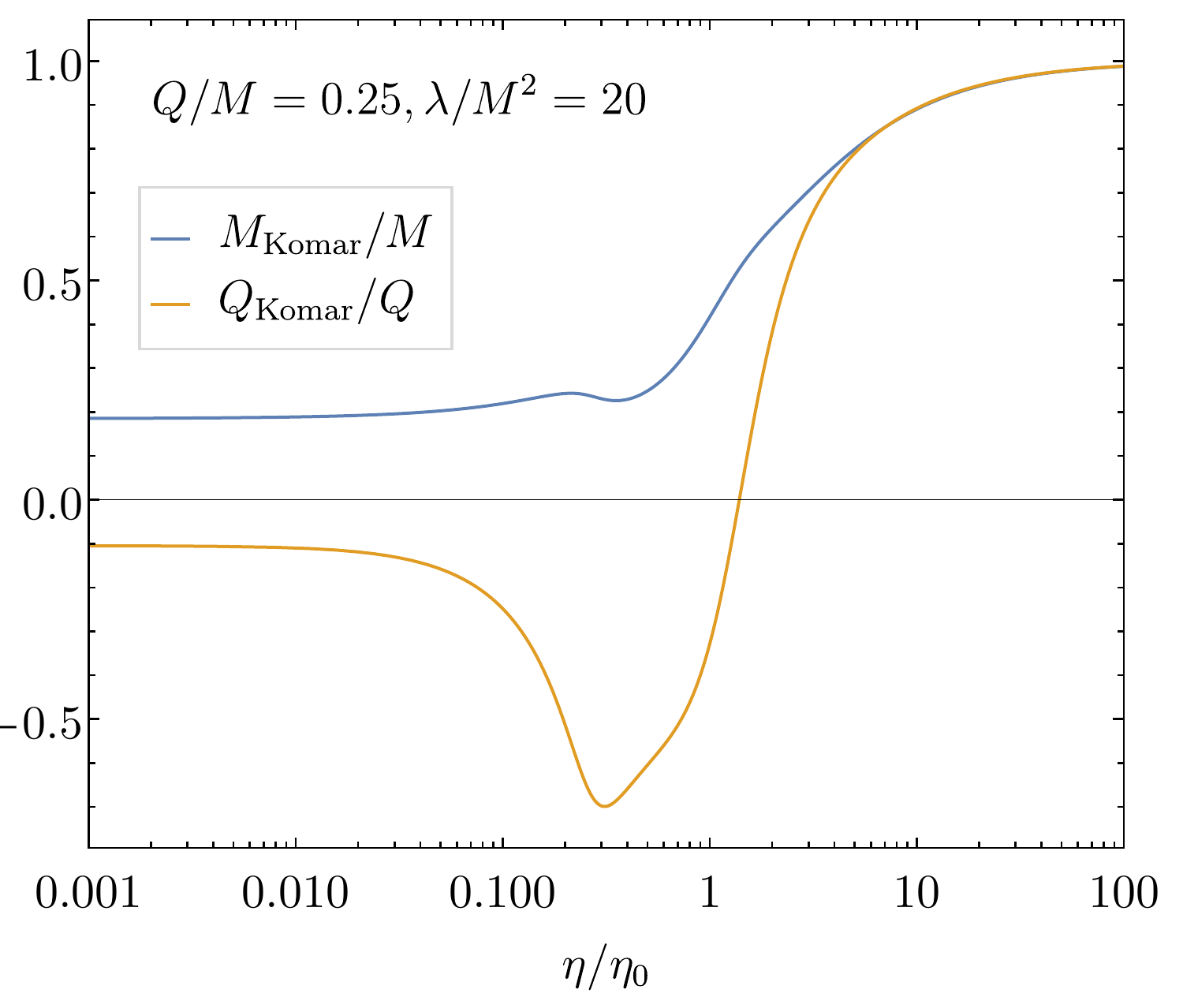}
			\caption{}
		\end{subfigure}%
		\caption{Komar mass and charge evaluated at the dimensionless isotropic wormhole radius $\eta/\eta_0$ for symmetric wormholes.}
		\label{fig:MQ}
	\end{figure}

\section{Results}

In the following we present our results for the EvGB wormholes obtained for the coupling functions (i), (ii) and (iii), 
given by Eqs.~(\ref{eq:coupl}), focusing mostly on (i).
We start our discussion by exhibiting a selection of typical solutions. 
Then we provide an overview of the solutions by presenting their DoE.
Subsequently we give further details of these types of solutions and discuss the emergence of cusps. 
Then we turn to the energy conditions and provide embeddings for some of the solutions.
Finally, we analyze their geodesics.

\subsection{Sets of Solutions}
	
For illustration we display in Fig.~\ref{fig:fun} a selection of the symmetric vectorized wormhole solutions for coupling function (i), 
associated with various points in the DoE.
The wormhole solutions are uniquely determined by their values of the dimensionless charge $Q/M$ and coupling constant $\lambda/M^2$, noted in the legend.
Shown are the metric coefficients $g_{tt}$ and $g_{\eta\eta}$ as well as the time component of the vector field $A_t$ as functions of the radial coordinate $\eta$.
The throat is located at $\eta=0$, and it splits the total Universe into two parts, that are symmetric with respect to reflection at the throat, $\eta \to -\eta$.
Therefore only one side of the solutions is depicted.

The solutions are asymptotically flat, finite, and continuously differentiable for $0 < \eta < \infty$.
Reflection at the throat leads to symmetric wormholes with two asymptotically flat infinities, which are everywhere continuous.
However, due to the reflection the solutions are not differentiable at the throat, and therefore a thin shell of matter at the throat will be necessary to amend this.
Without reflection at the throat in contrast, a singularity would be encountered somewhere beyond the throat, where $\eta<0$.

Since the wormholes carry mass and charge, the figure also provides a comparison with the RN solutions that possess the corresponding same values of the mass and charge in each case.
For better comparison we have chosen the same radial coordinate $\eta/\eta_0$ for the RN solutions as the one employed for the respective wormholes.
Depending on the value of $Q/M$ the RN solutions represent black holes ($Q/M \le 1$) or naked singularities ($Q/M > 1$).
For the black holes the metric coefficients are monotonic functions, whereas for the naked singularities they exhibit extrema.
In the next subsection we will see that the RN black holes form a part of the boundary of the DoE of the wormhole solutions.
In fact, they are reached 
when the limit $\lambda/M^2 \to 0$ is taken and $Q/M \le 1$.

	We end this subsection with a brief discussion on the mass and the charge of the wormholes.
	The expansion at spatial infinity, Eqs.~(\ref{eq:expansion_inf1})-(\ref{eq:expansion_inf3}), ($r>0$), 
	defines the mass $M$ and the vector charge $Q$ of the wormholes.
	On the other hand, for these spherically symmetric systems, we can also consider a mass function $M(r)$ and a charge function $Q(r)$, that we define via the respective Komar integrals evaluated at the radial coordinate $r$, Eqs.~(\ref{MKo1})-(\ref{QKo2}).
	For $r \to \infty$ these functions $M(r)$ and $Q(r)$ then converge towards the respective charges $M$ and $Q$. This is demonstrated in Fig.~\ref{fig:MQ} for a set of 4 wormhole solutions, where we employed the isotropic wormhole coordinate $\eta$ instead of $r$.
	
\subsection{Domain of Existence}

\begin{figure}[t]
	\begin{subfigure}{0.5\textwidth}
		\includegraphics[width=\textwidth]{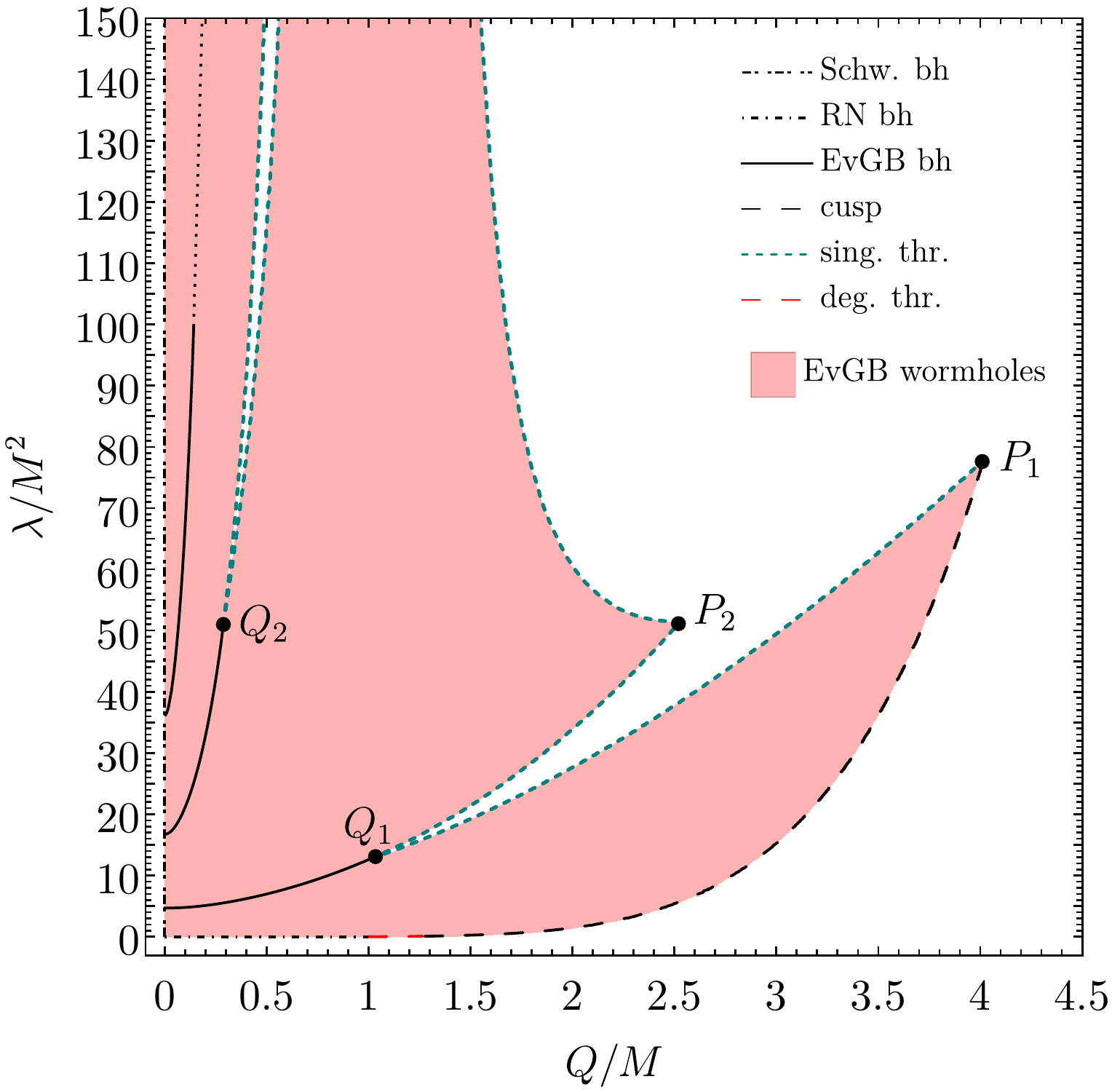}
		\caption{Coupling (i)}
	\end{subfigure}%
	\begin{subfigure}{0.5\textwidth}
		\includegraphics[width=\textwidth]{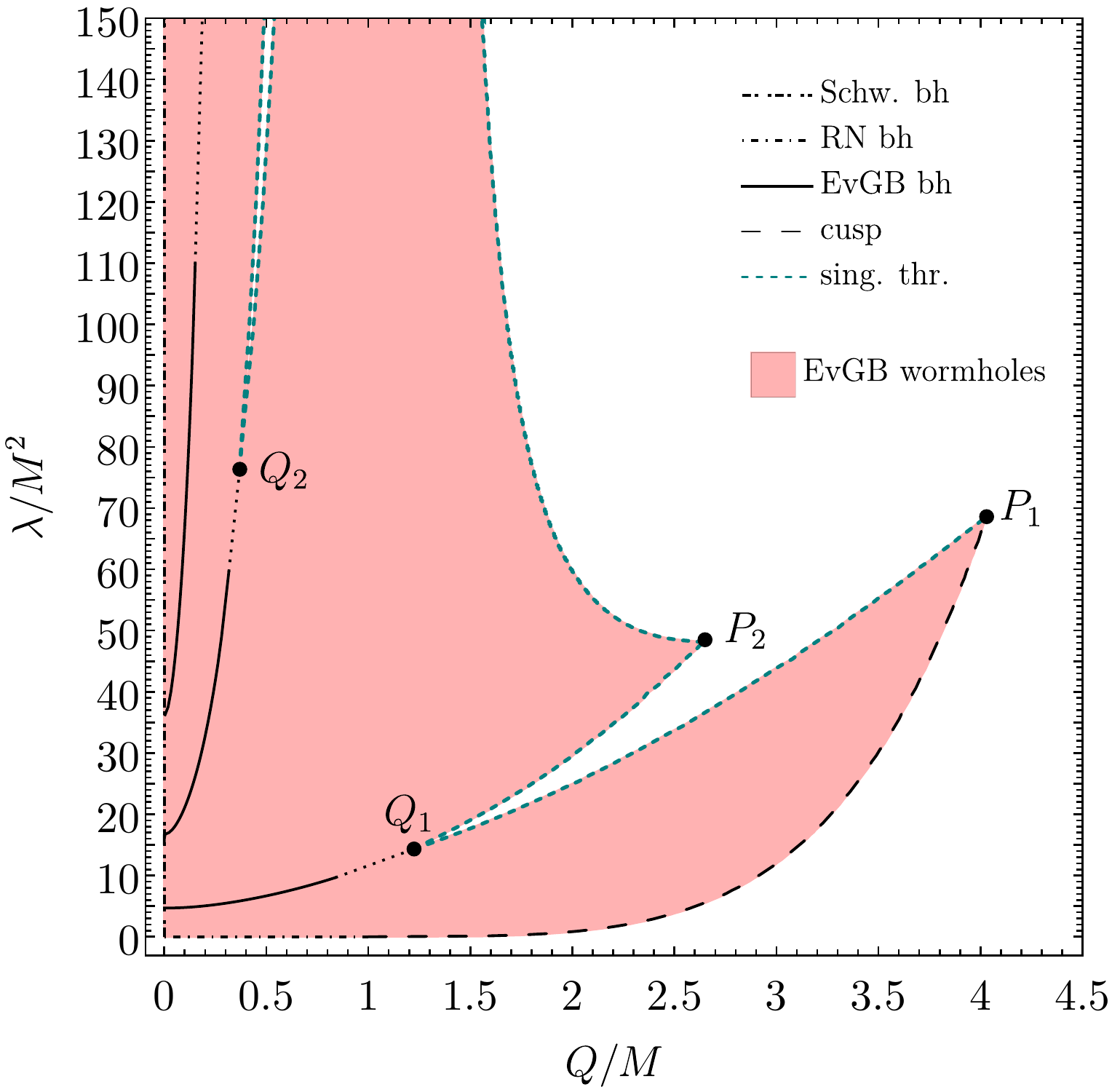}
		\caption{Coupling (ii)}
	\end{subfigure}	
	\caption{Domain of existence (DoE): $\lambda/M^2$ vs $Q/M$ for symmetric wormhole solutions for coupling function (i) in (a) and (ii) in (b).
	The DoE are bounded by black holes, singular throats, degenerate throats and the emergence of cusps.}
	\label{fig:phaseswh}
\end{figure}

\begin{figure}[t!]
	\begin{subfigure}{0.50\textwidth}
		\includegraphics[width=\textwidth]{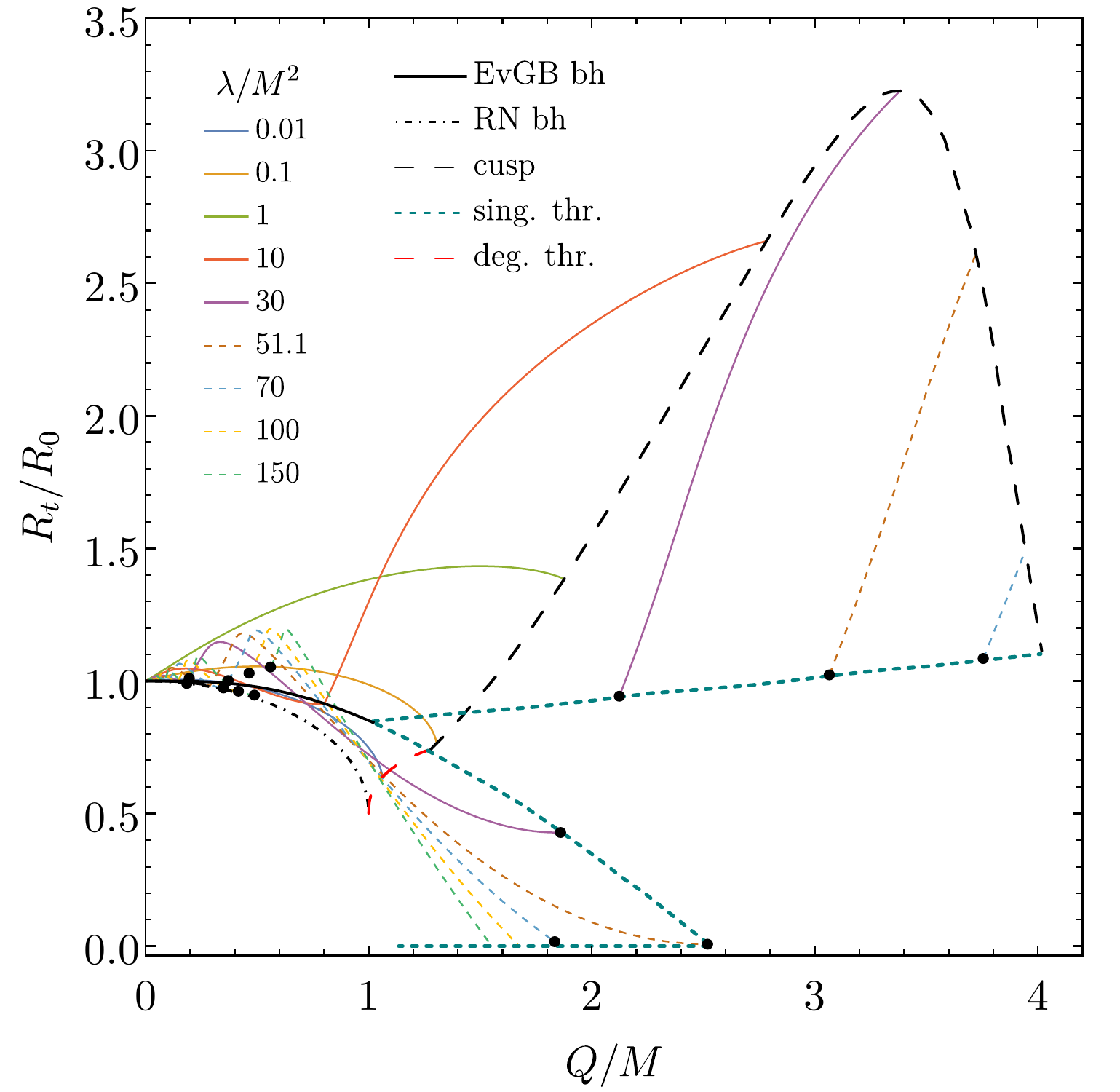}
		\caption{}
	\end{subfigure}%
	\begin{subfigure}{0.50\textwidth}
		\includegraphics[width=\textwidth]{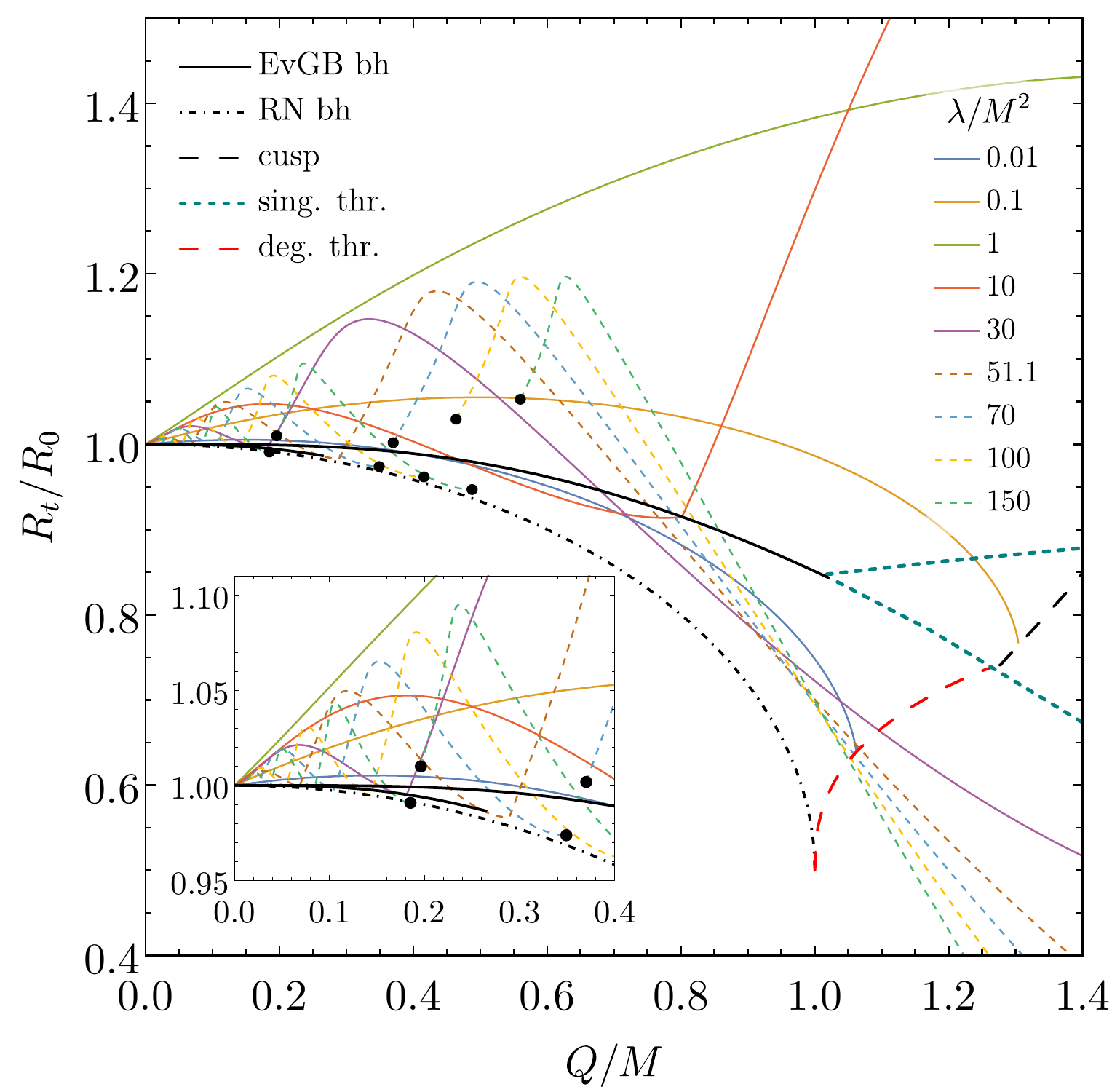}
		\caption{}
	\end{subfigure}
	\caption{Throat radius vs. vector charge.
	    Shown are wormhole solutions for coupling function (i) in the upper subfigures and (ii) in the lower ones, 
	    {(b) represent a zoom of (a).}
	    Curves of constant coupling $\lambda/M^2$ are also indicated.
	    {The dots correspond to the endpoints at the boundary of the DoE, where the wormhole throats become singular.}
		}
	\label{fig:qa}
\end{figure}

\begin{figure}[t]
	\begin{subfigure}{0.5\textwidth}
		\includegraphics[width=\textwidth]{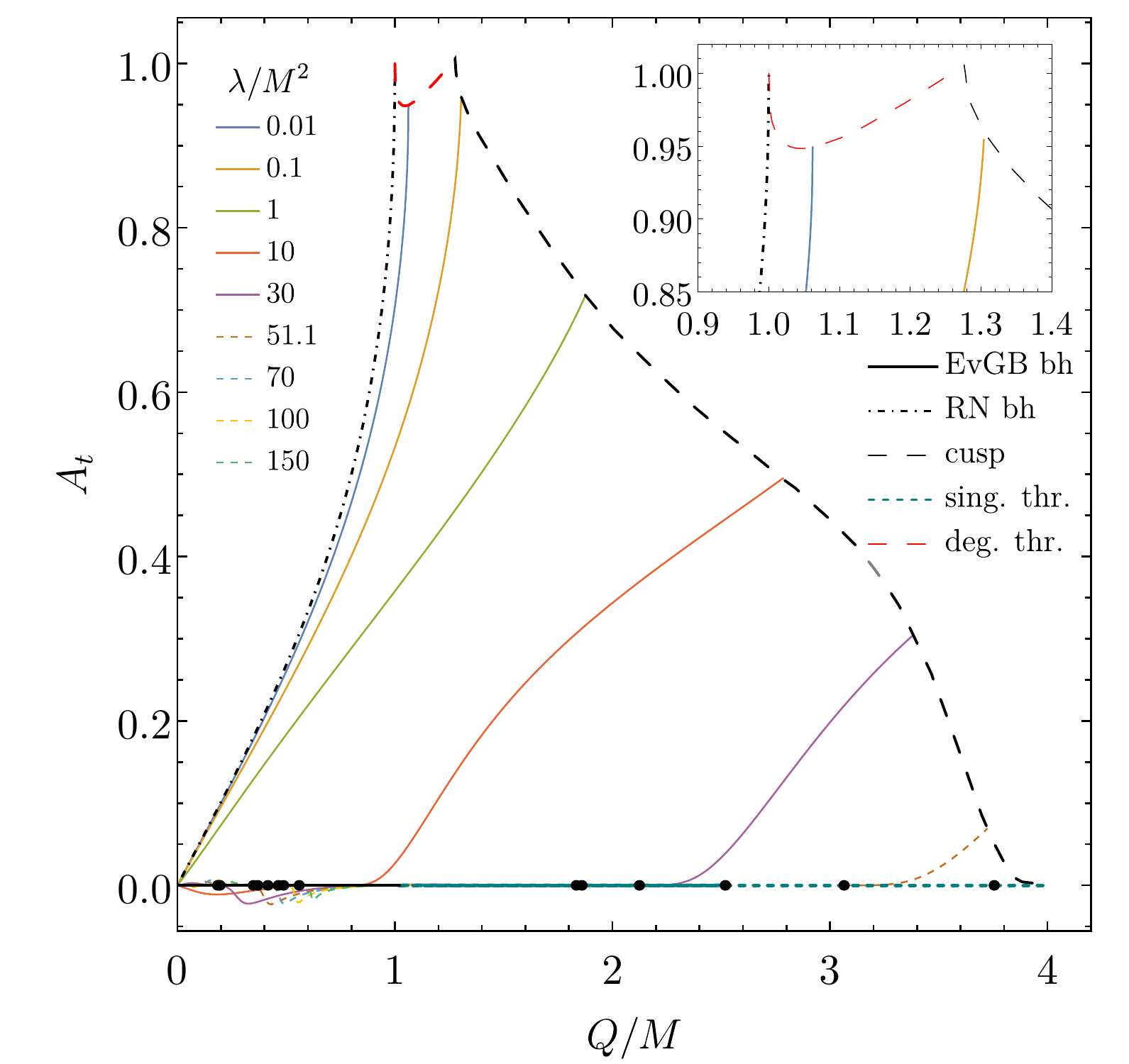}
		\caption{}
		\label{fig:qv}
	\end{subfigure}%
	\begin{subfigure}{0.5\textwidth}
		\includegraphics[width=\textwidth]{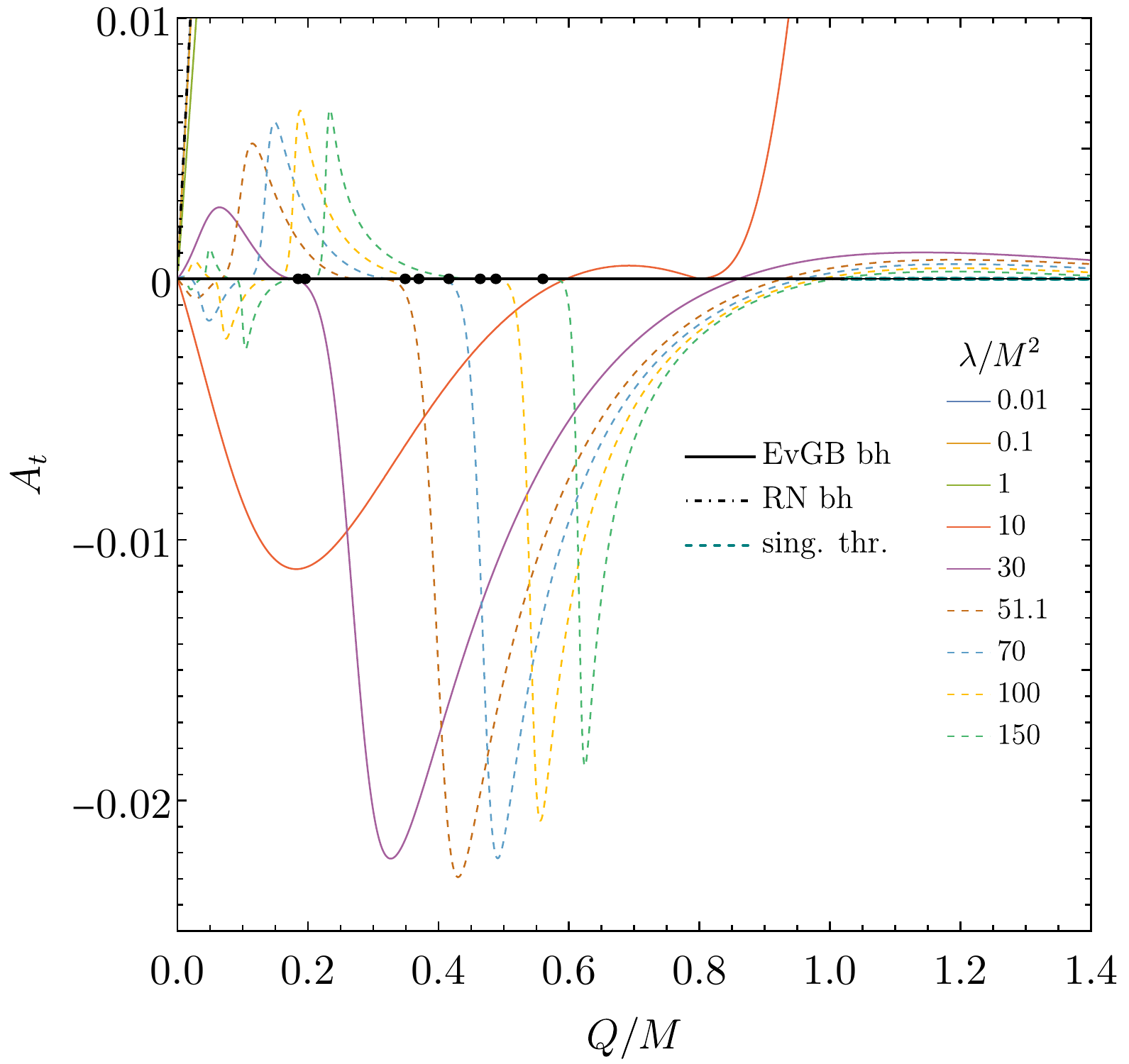}
		\caption{}
	    \label{fig:qv2}
	\end{subfigure}	
	\caption{Vector field at the throat vs. scaled vector charge for coupling function (i). Analogous to Fig.~\ref{fig:qa}(a) and (b).}
	\label{fig:A_t}
\end{figure}

\begin{figure}[t]
	\begin{subfigure}{0.5\textwidth}
		\includegraphics[width=\textwidth]{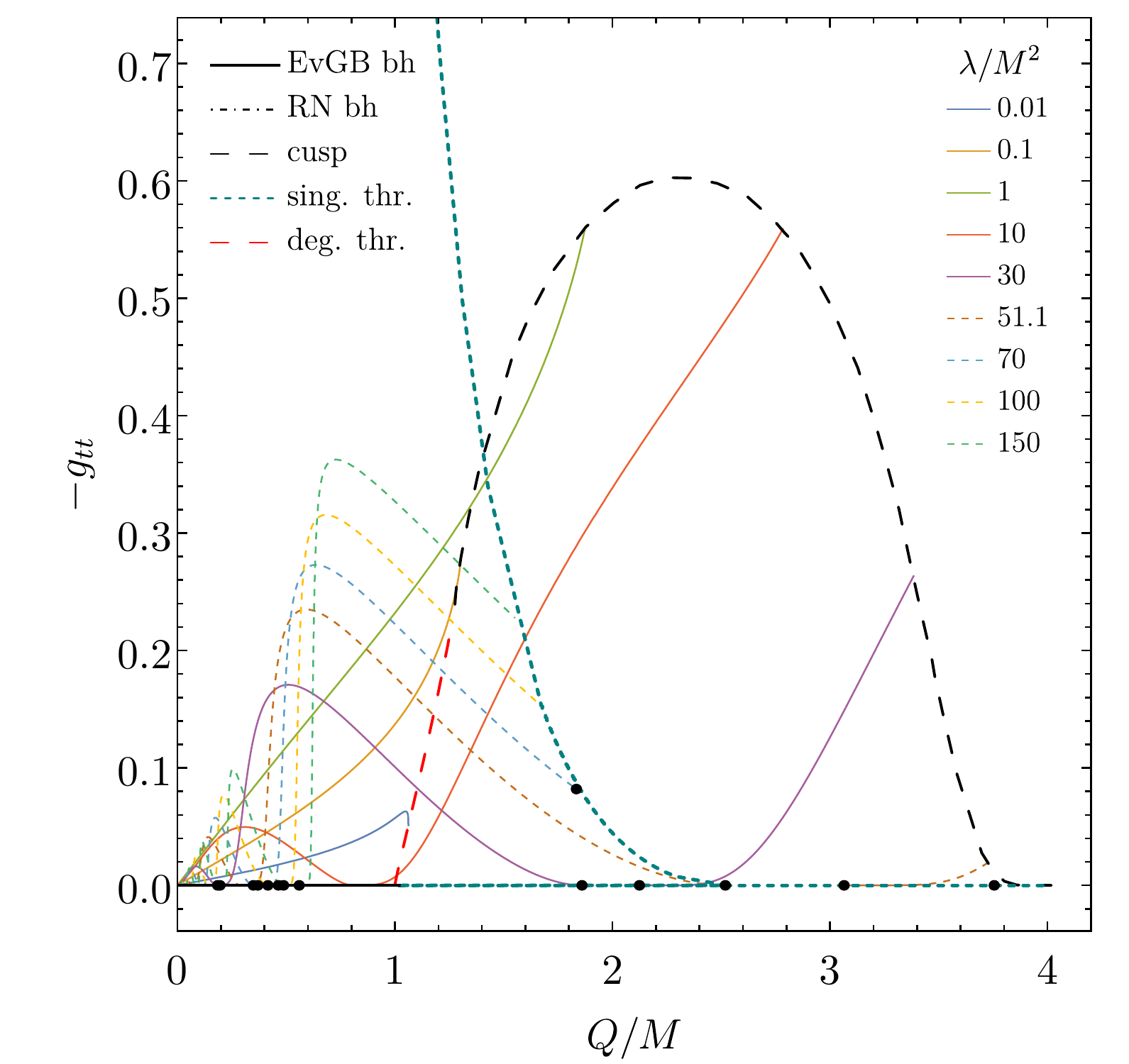}
		\caption{}
		\label{fig:qf0}
	\end{subfigure}%
	\begin{subfigure}{0.5\textwidth}
		\includegraphics[width=\textwidth]{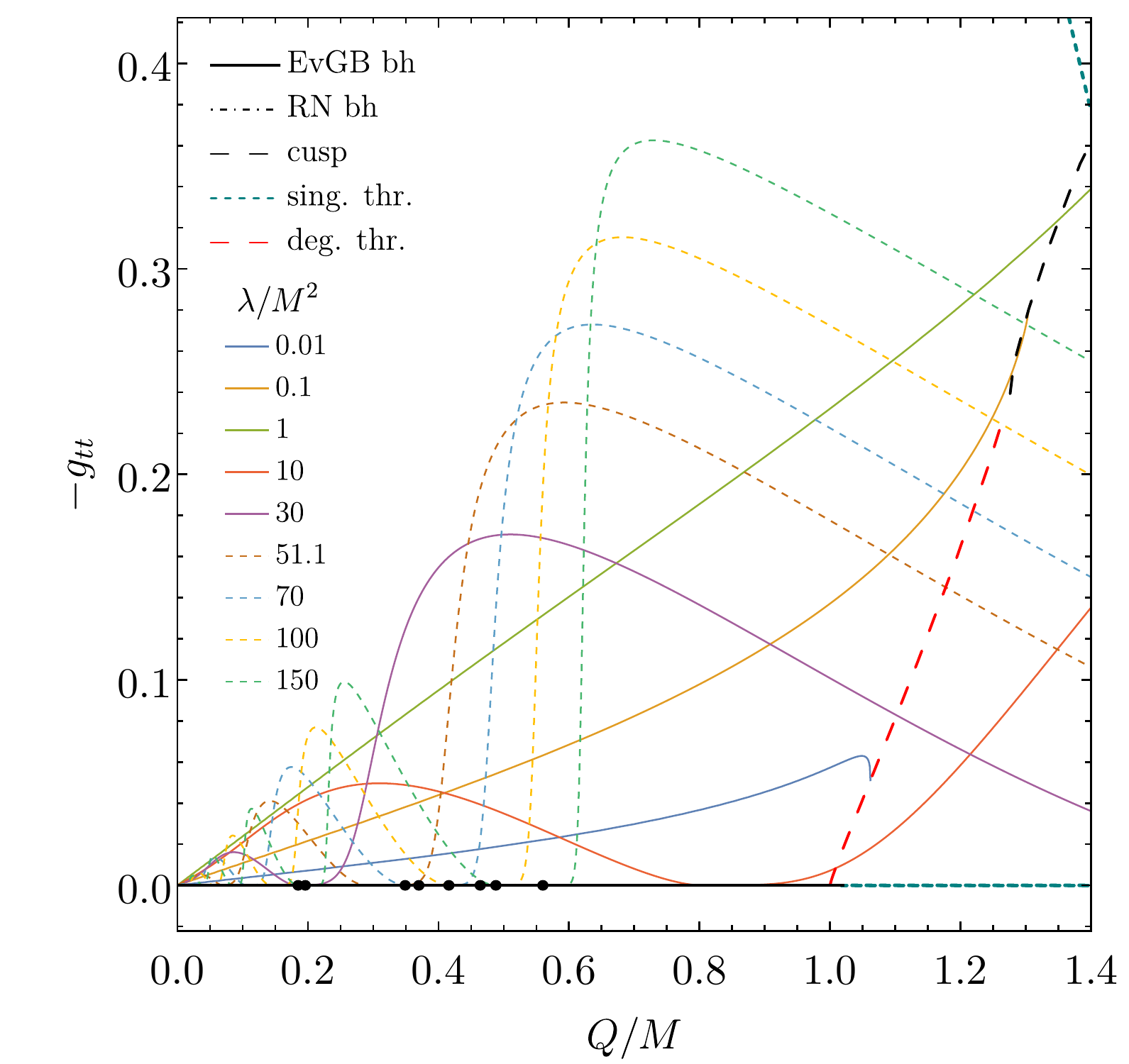}
		\caption{}
	    \label{fig:qf02}
	\end{subfigure}	
	\caption{$-g_{tt}$ at the throat vs. scaled vector charge for coupling function (i). Analogous to Fig.~\ref{fig:qa}(a) and (b).}
	\label{fig:gtt}
\end{figure}

As discussed above, EvGB wormholes depend on two independent parameters. 
This is in contrast to EvGB black holes which are characterized 
by a single parameter.  Taking advantage of the invariance under scaling 
transformations, we demonstrate the DoE of symmetric 
EvGB wormholes in terms of the dimensionless coupling constant 
$\lambda/M^2$ and the dimensionless vector charge 
$Q/M$ in Fig.~\ref{fig:phaseswh} for the coupling functions (i) and (ii).
Here we restrict to parameter values $\lambda/M^2 \leq 150$ for convenience.
Symmetric wormholes exist in the colored regions.
Since the theory is symmetric with respect to $A_\mu \to - A_\mu$, the DoE can be reflected through the axis $Q=0$.
Therefore only the $Q/M \ge 0$ part of the DoE is shown.
Also shown are the families of EvGB black holes (thick-solid black),
the Schwarzschild black holes (dash-dotted black) and 
the RN black holes (dash-dotted black). 
The boundaries of the DoE consist of  singular wormholes 
(short-dashed), wormholes with a cusp-singularity outside the throat 
(long-dashed black) and wormholes with a degenerate throat (long-dashed red).

We now consider these domains and their boundaries in more detail, 
starting with the coupling function (i) and addressing first the 
boundaries of the domain.
For vanishing coupling constant, no wormholes arise.
Relevant solutions of the field equations are only the Schwarzschild and the 
RN black holes. The Schwarzschild black holes form the left boundary of the 
DoE of wormholes, while the RN black holes form a part of 
the lower boundary of the DoE. 

Beyond the extremal RN black hole the lower boundary is formed by solutions 
with a regular throat.
For small values of the coupling parameter, $0\leq \lambda/M^2 \leq 0.162$,
the boundary consists of solutions whose
throats degenerate to saddle points. These wormholes possess no cusps nor 
singularities. They are shown by the long-dashed red curve in Fig.~\ref{fig:phaseswh}. 

For larger values of the coupling parameter the boundary is formed by solutions, 
that develop a cusp-singularity, i.e., a certain curvature singularity.
Since this cusp arises before the throat ($\eta>0$), these solutions cannot be 
smoothly continued to infinity and must be discarded. 
The onset of the cusp-singularity is marked by a long-dashed black curve.
The mechanism leading to a cusp-singularity will be discussed below.

When following the cusp-singularity along the boundary while 
increasing $\lambda/M^2$, the cusp moves closer to the throat until it 
hits the throat at the maximal value of the vector charge $Q/M$, 
indicated by $P_1$ in Fig.~\ref{fig:phaseswh}.
The throat itself is then singular. No symmetric wormholes exist beyond 
this maximal value of the vector charge.
Solutions with a singular throat (short-dashed) also form most of the 
remaining parts of the boundary of the DoE.

Continuing along the boundary to smaller values of the coupling and the charge, 
a local minimal value of the charge $Q/M$ of the singular boundary solutions 
is reached. Let us denote it as $Q_1$ for later reference.
At this point we observe a bifurcation with a second branch of singular
wormhole solutions, which extends up to a local maximum of the charge $Q/M$
at some point $P_2$. Here a third branch of singular wormholes emerges
which extends to arbitrarily large values of $\lambda/M^2$.
We note that for small $Q/M$ and large $\lambda/M^2$ more void regions exist
in the DoE, which are bounded by singular wormholes.
The emergence of these regions are indicated by the point $Q_2$ 
in Fig.~\ref{fig:phaseswh}. 

Interestingly, the families of EvGB black holes seem to extend exactly up
to the bifurcation point $Q_1$ and  $Q_2$ of the singular wormholes.
EvGB black holes exist between vanishing charge and a maximal value of the 
charge, where a singularity arises \cite{Barton:2021wfj}.
Since the boundary curve consists also of singular solutions it cannot be 
precisely determined, where EvGB black holes end and singular boundary 
solutions start.
Possible transition points could be the points $Q_n$.
But the transition could also arise later.

We expect that to
each  point $Q_n$ a branch of black holes is associated, such that
the vector field component $A_t$ possesses $n$ nodes (including the zero
at the horizon).

The DoE of coupling function (ii) is shown 
in Fig.~\ref{fig:phaseswh}b. It is very similar in structure to the domain of 
existence of coupling function (i). 
{However, there are no wormholes with
degenerate throat at the boundary of the DoE.}

In Fig.~\ref{fig:qa} we give an alternative presentation of the DoE.
Here we show the dimensionless throat radius  $R_t/R_0 = \sqrt{{\cal A}_t/(16 \pi M^2)}$ 
versus the dimensionless charge $Q/M$ for several values of the 
dimensionless coupling parameter  $\lambda/M^2$. Also shown is $R_t/R_0$
along (part of) the boundary of the DoE in  Fig.~\ref{fig:phaseswh}.
In addition the dimensionless Schwarzschild radius of the RN black holes and 
the EvGB black holes is shown.

We note that for the black holes $R_t/R_0$ is a decreasing function of $Q/M$ 
with values $R_t/R_0 = 1$ at $Q/M = 0$.
For the RN black holes $R_t/R_0= \left(1+\sqrt{1-Q^2/M^2}\right)/2$,
which implies $R_t/R_0= 1/2$ at $Q/M = 1$.
For small values of $\lambda/M^2$ the branches of wormholes follow closely 
the curve of RN black holes. However, they end at some wormhole with 
degenerate throat or cusp-singularity with $Q/M > 1$. 
For larger values of $\lambda/M^2$ 
the quantity $R_t/R_0$ is no longer a decreasing function of $Q/M$, 
but possesses maxima and minima.
If $\lambda/M^2$ is larger than some value given by the point $Q_1$, 
the curves $R_t/R_0$  vs $Q/M$ possess gaps corresponding
to the voids in the DoE. 
For example for $\lambda/M^2 = 30$ one part of the curves $R_t/R_0$  extends
from the Schwarzschild black hole up to some point on the lower boundary 
of singular wormholes. The second part to the curve starts at some point 
on the upper boundary of singular wormholes and ends at a point 
of the boundary of wormholes with cusp singularities. 
If $\lambda/M^2 > 51.1$ (corresponding to $Q_2$ and $P_2$ ) the first
part of the curve again ends at some singular wormhole solution, 
but with vanishing $R_t/R_0$. The second part
again connects singular wormholes and wormholes with cusp singularities, 
provided  $\lambda/M^2$ is smaller than
the value of $P_1$.
In Fig.~\ref{fig:qa}b we give a closer look at the curves $R_t/R_0$ vs $Q/M$, 
restricting to small $Q/M$ and  $R_t/R_0$. The round
dots indicate the gaps {corresponding to the voids.} 
We note that the functions $R_t/R_0$ are oscillating around one, 
and that the number of oscillations increases with increasing $\lambda/M^2$. 
Interestingly, the minima of the oscillations are close to the values of the 
RN black holes, but never below, as can be seen in the inset.

As can be seen in  Fig.~\ref{fig:qa} the dimensionless throat radius  $R_t/R_0$
equals zero for singular wormholes along the curve in Fig.~\ref{fig:phaseswh}
extending from the point $P_2$ up to arbitrarily large values of the 
coupling parameter $\lambda/M^2$. Although the curvature invariants at the
throat diverge for these solutions, the metric functions are finite at the
throat. As a consequence, the throat degenerates to a single point,
which corresponds to the origin of the coordinate system. Thus the 
total spacetime may be considered as two copies of a topologically trivial
spacetime glued together at their singular points.  

{We note that for the coupling function (ii) the DoE is very similar
to the case (i), except that there are no wormhole solutions with 
degenerate throat.}

In Fig.~\ref{fig:A_t} we show the vector field $A_t$ at the throat 
versus the dimensionless charge $Q/M$ for several values of the 
dimensionless coupling parameter $\lambda/M^2$. 
Also shown is $A_t$ for the RN black holes, given by the simple expression 
$A_t = (Q/M)/\left(1+\sqrt{1-Q^2/M^2}\right)$. 
Note that $A_t=0$ for the EvGB black holes and the 
singular wormholes along the boundary of the DoE.
For the RN black holes the maximum value of $A_t$ is one.
We observe that for the wormhole solutions $A_t$ exceeds this value only slightly at the point where the 
boundaries with degenerate wormholes and wormholes with cusp singularities
meet. 

In  Fig.~\ref{fig:A_t}b we show $A_t$  for large coupling parameter $\lambda/M^2$
and small $Q/M$. The dots indicate the voids in the DoE.
We note that  $A_t$  oscillates around zero. The number of nodes of $A_t$ at
the throat indicates the number of nodes of $A_t$, when considered as function 
of $r$: with decreasing values of $Q/M$ the number of nodes of $A_t$ increases
by one exactly when  $A_t(r_{\rm th})$ passes through zero.
  
We end our discussion of the DoE by considering the metric component $g_{tt}$ at the throat, shown for coupling function (i) in Fig.~\ref{fig:gtt}.
Here we show the metric component  $g_{tt}$ at the throat 
versus the dimensionless charge $Q/M$ for several values of the 
dimensionless coupling parameter $\lambda/M^2$. We note that 
 $g_{tt}$ vanishes for the black holes and the singular wormholes along the 
boundary of the DoE, but it is finite for wormholes with degenerate
throats and cusp singularities.

{Finally, we note that for the coupling function (ii) the 
graphs $A_t$ vs $Q/M$ and $g_{tt}$ vs $Q/M$ are very similar to the case of coupling function (i).}

\FloatBarrier	

	\begin{figure}[t!]
		\begin{subfigure}{.5\textwidth}
			\includegraphics[width=\textwidth]{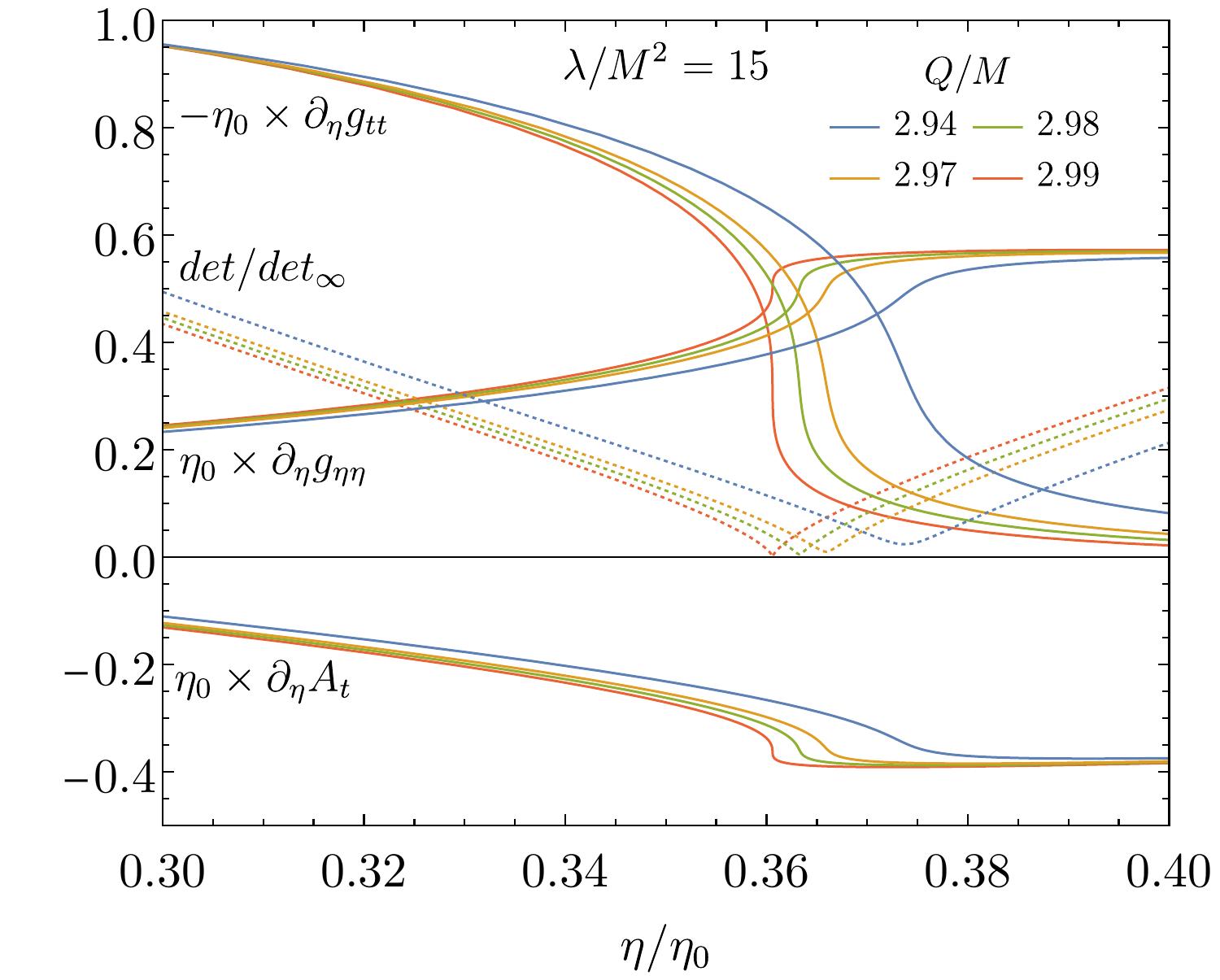}
			\caption{}
			\label{fig:cusp1_16}
		\end{subfigure}%
		\begin{subfigure}{.5\textwidth}
			\includegraphics[width=\textwidth]{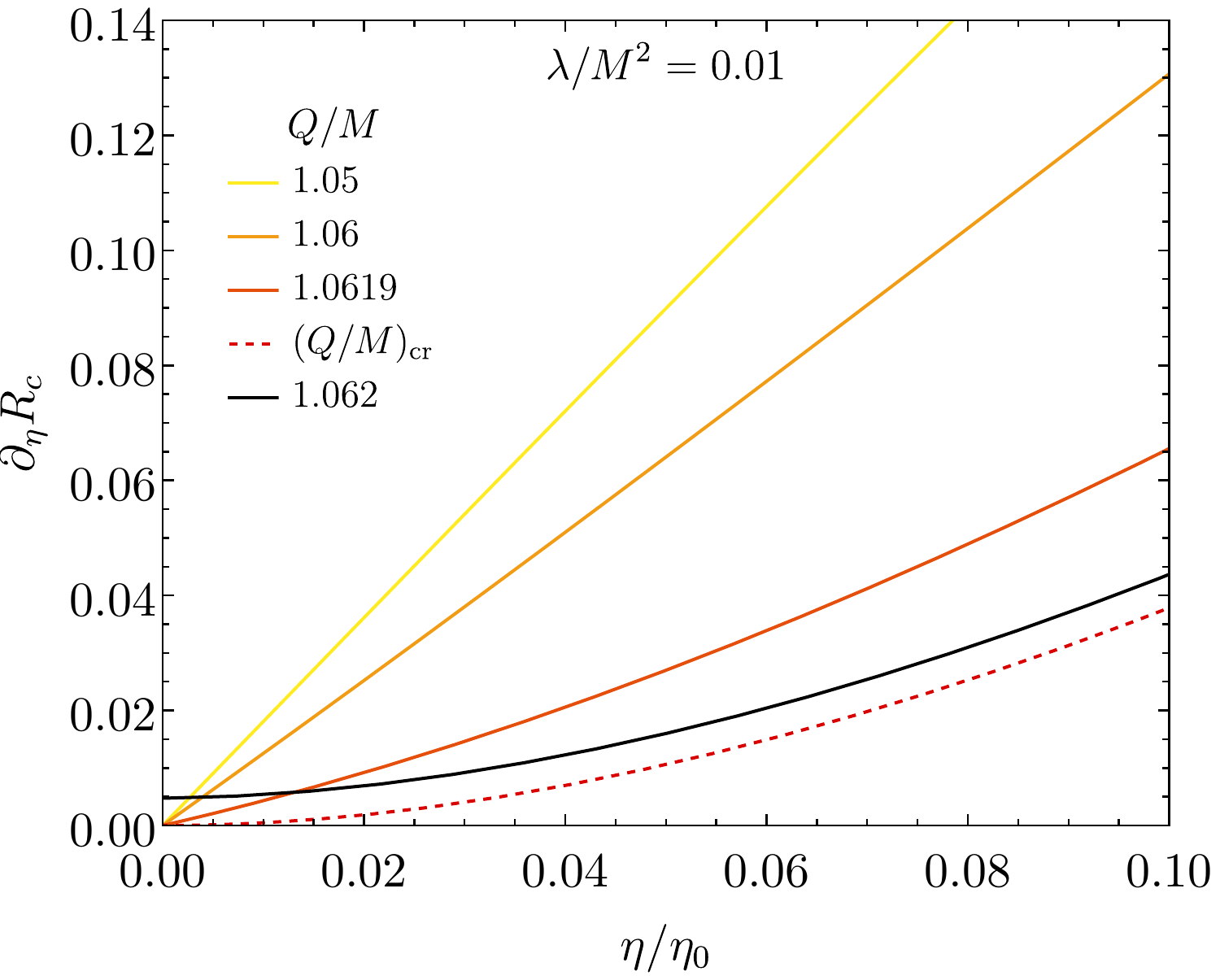}
			\caption{}
			\label{fig:deg1_01}
		\end{subfigure}
		\caption{Emergence of cusp singularities and degenerate throats:
			(a) Formation of a cusp for $\lambda/M^2=15$. Shown are several functions and derivatives
			as function of $\eta/\eta_0$ for several values of $Q/M$, approaching the critical
			value $Q/M=3.01$.
			(b) Formation of a degenerate throat $\lambda/M^2=0.01$. Shown is the derivative 
			{$\partial_\eta R_{c}$ as function of $\eta/\eta_0$} for several values of $Q/M$, approaching the critical
			value $Q/M\approx 1.062$
		}
		\label{fig:cusps}
	\end{figure}
\subsection{Cusps and Degenerate Throats}
	
We now turn to the cusps that arise when the wormhole solutions are constructed. 
This is similar to the appearence of cusps for particle-like solution in EsGB theories \cite{Kleihaus:2019rbg,Kleihaus:2020qwo}.
The wormhole solutions are obtained by solving the coupled set of ODEs in the radial variable $r$, integrating numerically from asymptotic infinity towards zero.
The second order equations are not diagonal with respect to the second derivatives of the functions.
Diagonalization of these equations then implies that a determinant arises, containing the respective coefficients.
Since these coefficients are functions of the radial variable, it may happen that the determinant possesses a node at some value of the radial variable $r_{\star}$.
However, the diagonalization procedure involves division by this determinant and, consequently, the respective solution will possess a cusp singularity at $r_{\star}$.
The emergence of cusps is demonstrated in Fig.~\ref{fig:cusps}a , where we show the derivatives
of some functions together with the scaled determinant
$det/det_{\infty}$, with 
$det_\infty = - r^5/(16\lambda M Q^2)$.

Next we turn to the discussion of degenerate throats which appear for small values of the coupling parameter $\lambda/M^2$, when the 
dimensionless charge $Q/M$ approaches a critical value. In this limit the derivative of the circumferential radius $R_c$ with respect to the 
radial coordinate $r$, i.~e.~$R_{c,r}(r)$, does not possess a zero any more. This is demonstrated in  Fig.~\ref{fig:cusps}b for $\lambda/M^2=0.01$.

\FloatBarrier	
\subsection{Energy Conditions}

	\begin{figure}[t]
		\begin{subfigure}{.5\textwidth}
			\includegraphics[width=\textwidth]{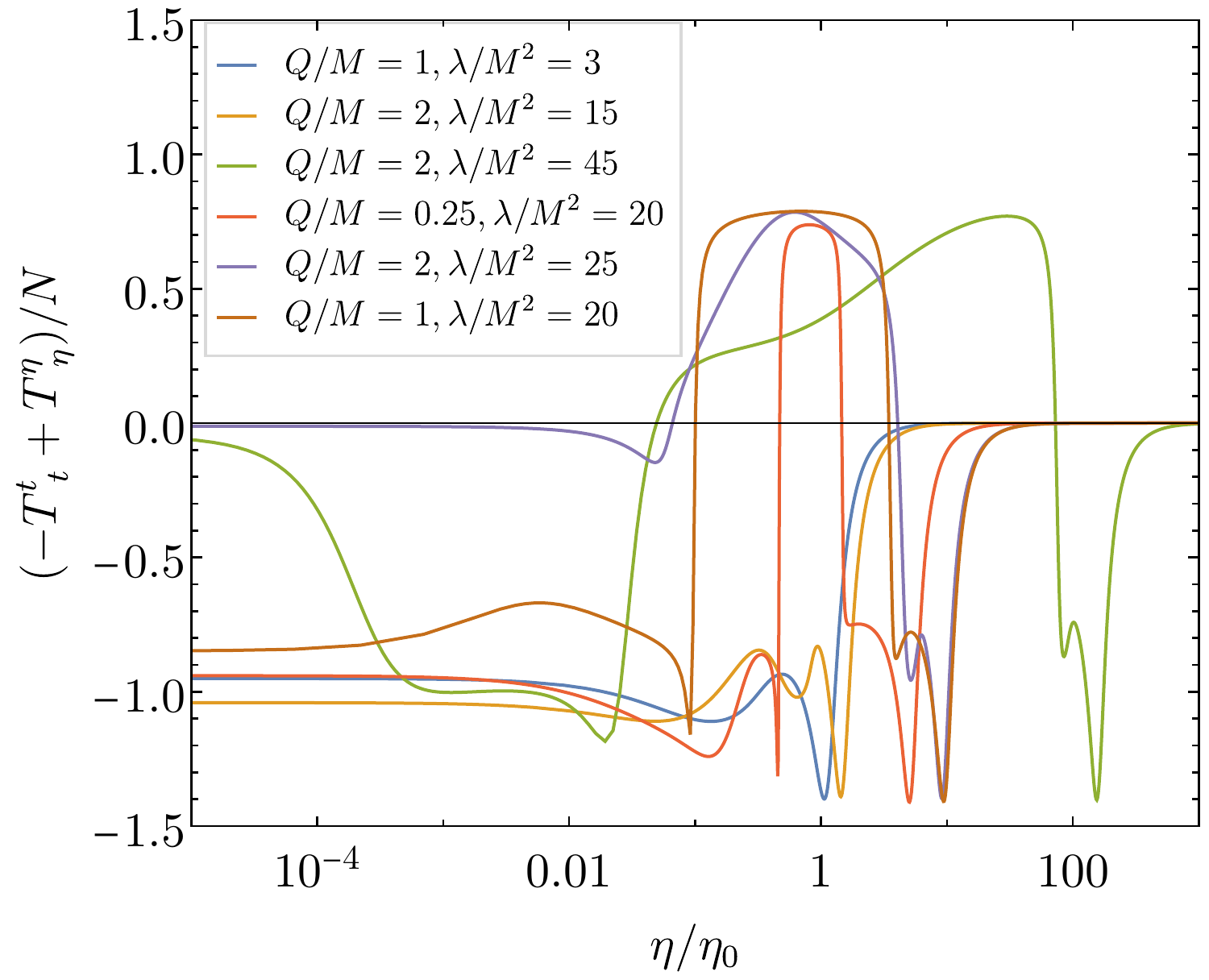}
			\caption{}
		\end{subfigure}%
		\begin{subfigure}{.5\textwidth}
			\includegraphics[width=\textwidth]{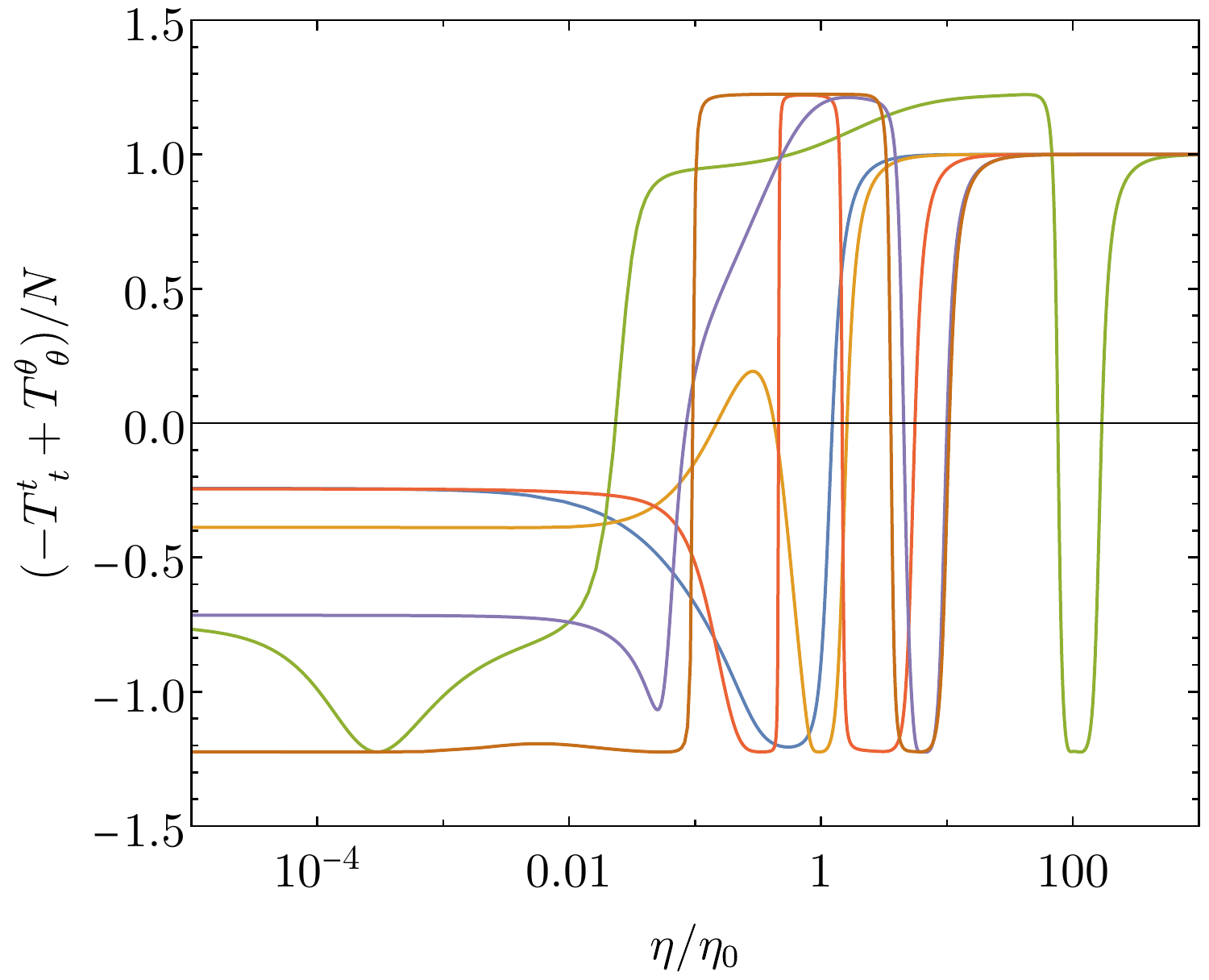}
			\caption{}
		\end{subfigure}\\
			\begin{subfigure}{.5\textwidth}
			\includegraphics[width=\textwidth]{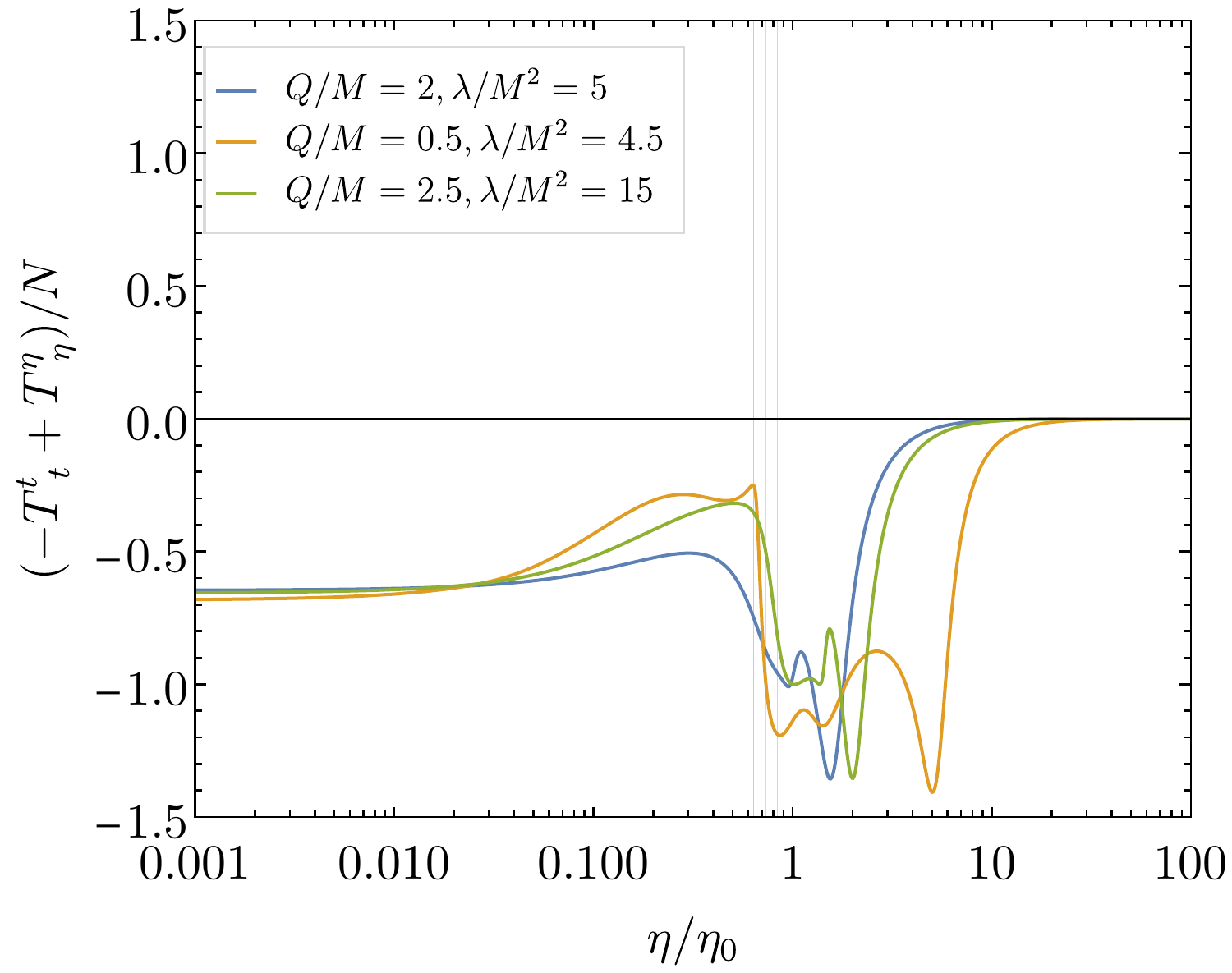}
			\caption{}
		\end{subfigure}%
		\begin{subfigure}{.5\textwidth}
			\includegraphics[width=\textwidth]{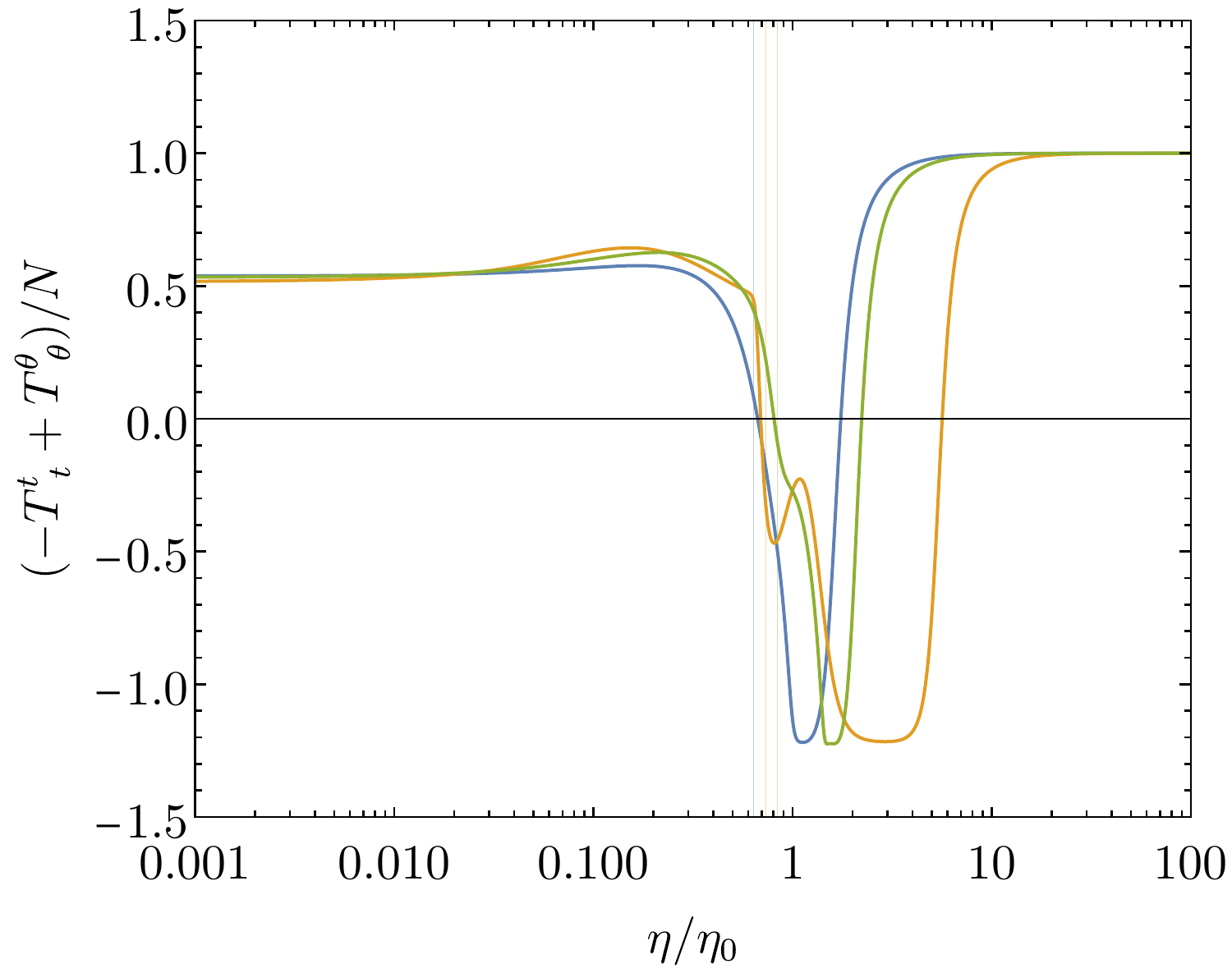}
			\caption{}
		\end{subfigure}
		\caption{Violation of the null energy condition at the throat and equator. Top: coupling (i). Here $\eta=0$ refers to the throat.
		Bottom: coupling (iii). Here $\eta=0$ refers to the  equator. The vertical lines indicate the respective throat. 
		}
		\label{fig:ec}
	\end{figure}

We next turn to the violation of the energy conditions for the symmetric wormholes solutions.
In particular, we demonstrate the violation of the null energy condition (NEC)
\begin{equation}
    T_{\mu\nu} n^\mu n^\nu \geq 0 \ .
\end{equation}
Here $n^\mu$ is an arbitrary null vector, $n^\mu n_\mu=0$. Choosing the null vectors
\begin{eqnarray}
    n^\mu &=&\left(1,\sqrt{-g_{tt}/g_{\eta\eta}},0,0\right) \ , \\
    n^\mu &=&\left(1,0,\sqrt{-g_{tt}/g_{\theta \theta}},0\right) \ ,
\end{eqnarray}
and inserting  these into the NEC,
e.g., $T_{\mu\nu}n^\mu n^\nu=T^t_t n^t n_t + T^\eta_\eta n^\eta n_\eta =-g_{tt}\,(-T^t_t +T^\eta_\eta)$
we find for the NEC to hold the respective conditions
\begin{eqnarray}
 -T_t^t + T_\eta^\eta &\geq& 0 \ , \\
 -T_t^t + T_\theta^\theta &\geq& 0 \ .
\end{eqnarray}

These conditions for the NEC are always violated for the symmetric wormhole solutions.
We demonstrate this violation with some examples for coupling function (i) in Fig.~\ref{fig:ec}, where we show the combinations of the stress-energy tensor $(- T^t_t+T^\eta_{~\eta})/N$ (a) and $(- T^t_t+T^\theta_{~\theta})/N$ (b) with normalization factor $N=\sqrt{( T^t_{~t})^2+(T^\eta_{~\eta})^2+2(T^\theta_{~\theta})^2}$ versus the radial wormhole coordinate $\eta$.
Clearly, both conditions are always violated at the throat, but violation occurs also in other regions.
We note that for wormhole solutions in the presence of an equator, as obtained for coupling function (iii), the angular condition is satisfied at the equator.
But since the radial condition is violated at the equator, the NEC is violated there, as well.

\FloatBarrier

	\begin{figure}[h!]
		\begin{subfigure}{.5\textwidth}
			\includegraphics[width=\textwidth]{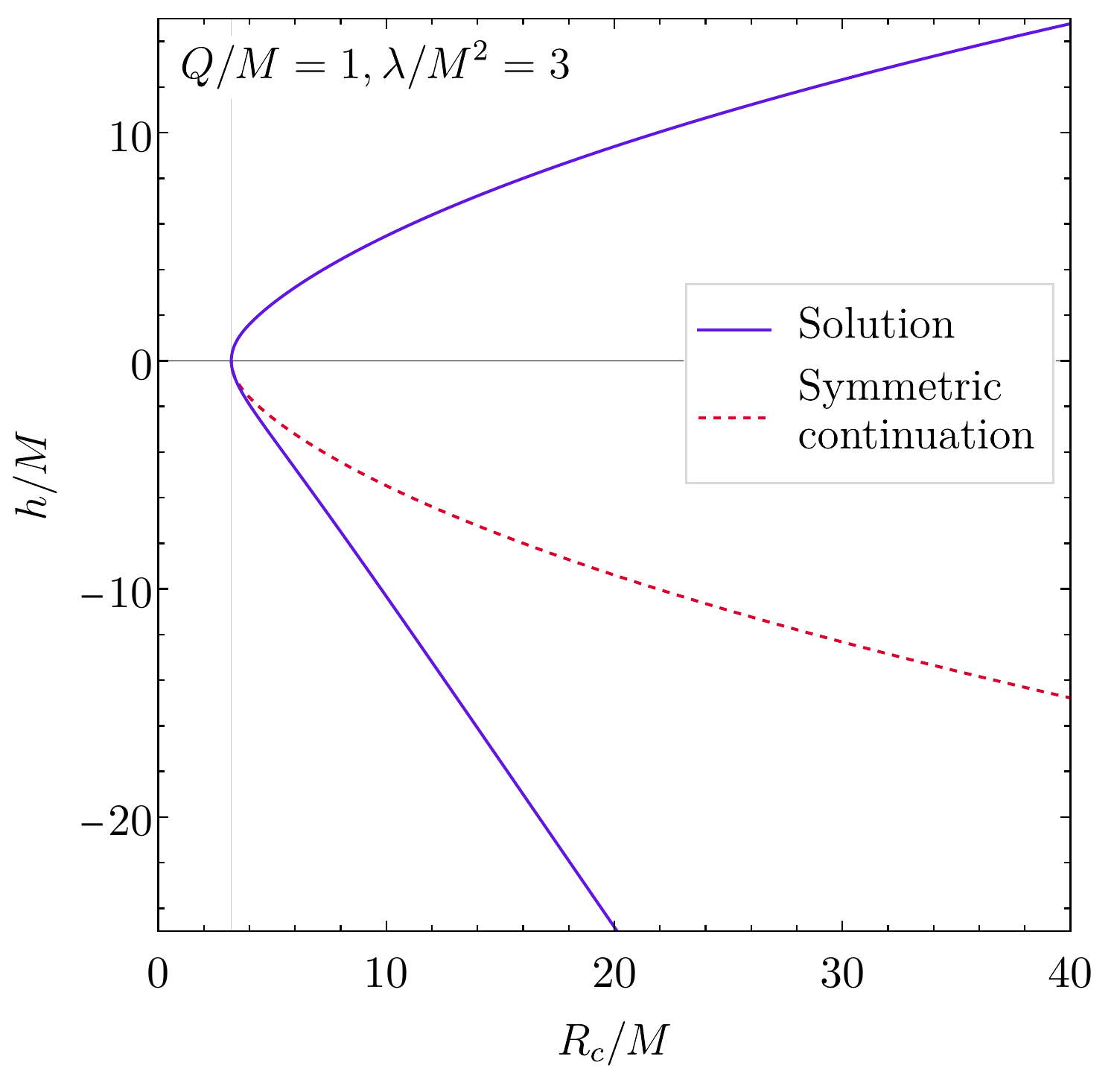}
			\caption{}
		\end{subfigure}%
		\begin{subfigure}{.5\textwidth}
			\includegraphics[width=\textwidth]{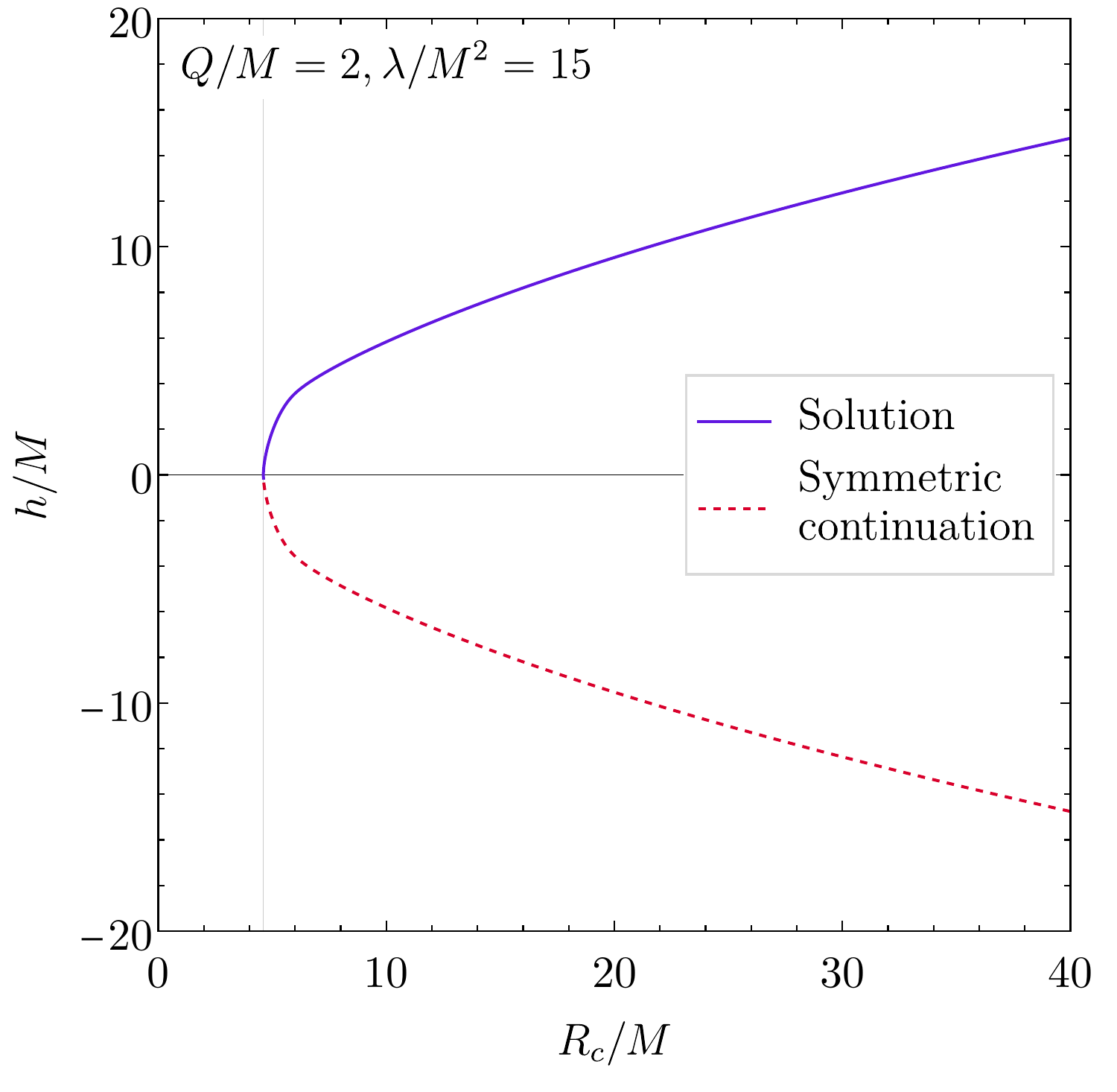}
			\caption{}
		\end{subfigure}\\
		\begin{subfigure}{.5\textwidth}
			\includegraphics[width=\textwidth]{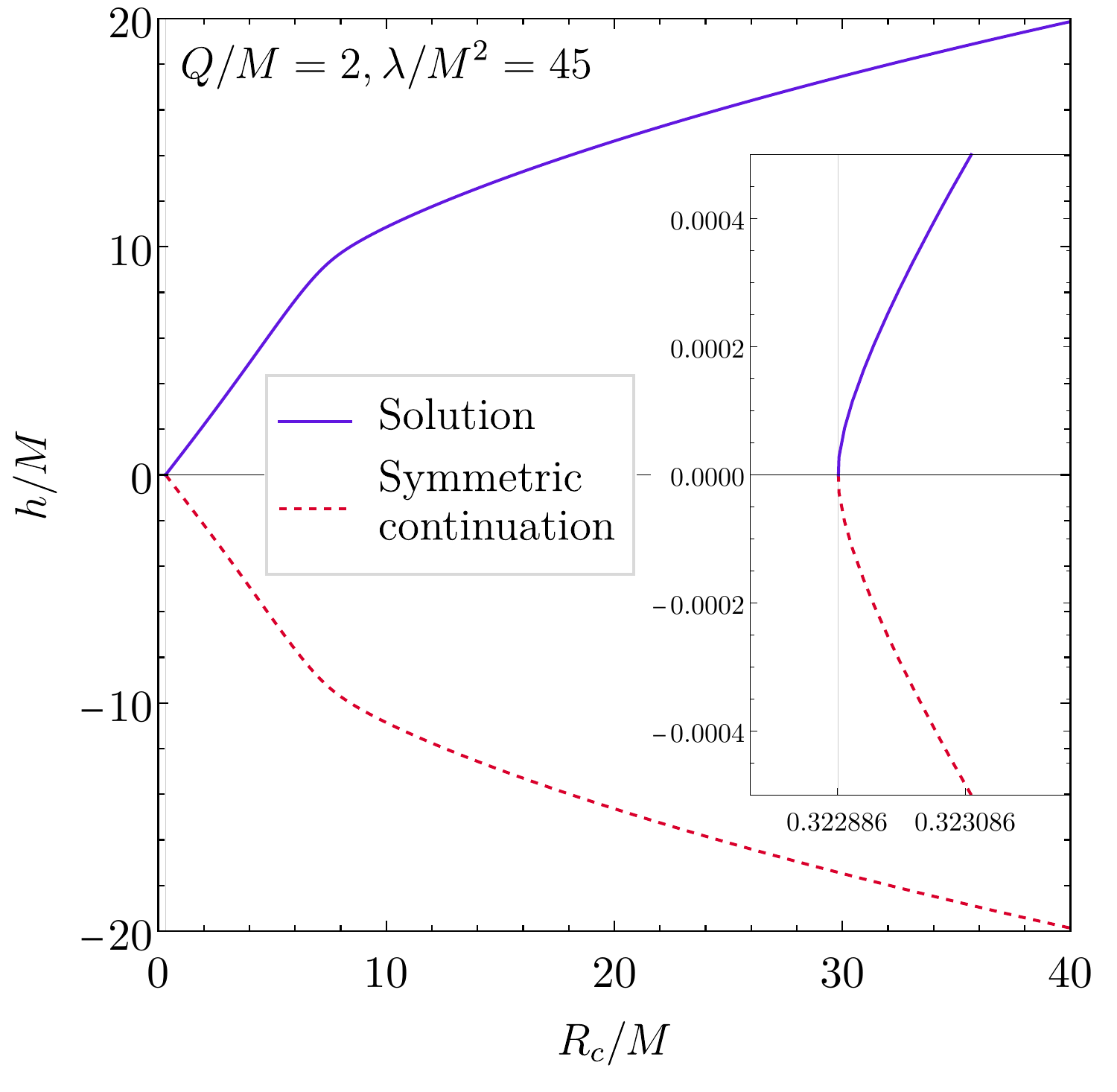}
			\caption{}
		\end{subfigure}%
		\begin{subfigure}{.5\textwidth}
			\includegraphics[width=\textwidth]{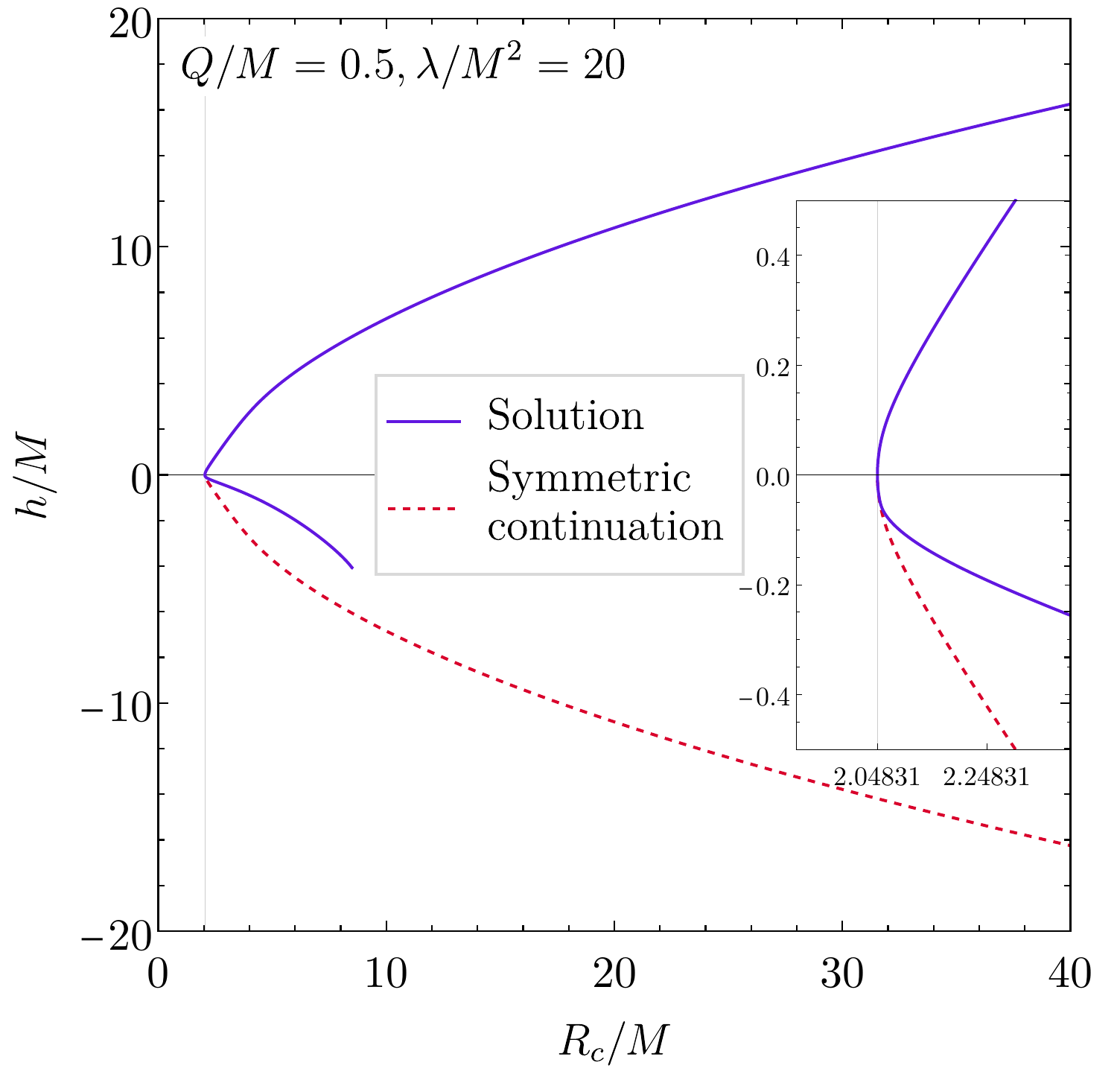}
			\caption{}
		\end{subfigure}
		\caption{Embedding diagrams for several examples of wormhole solutions.
}
			\label{fig:emb2dim}
	\end{figure}
	
\subsection{Embeddings}

\begin{figure}[t]
		\begin{subfigure}{.5\textwidth}
			\includegraphics[width=\textwidth]{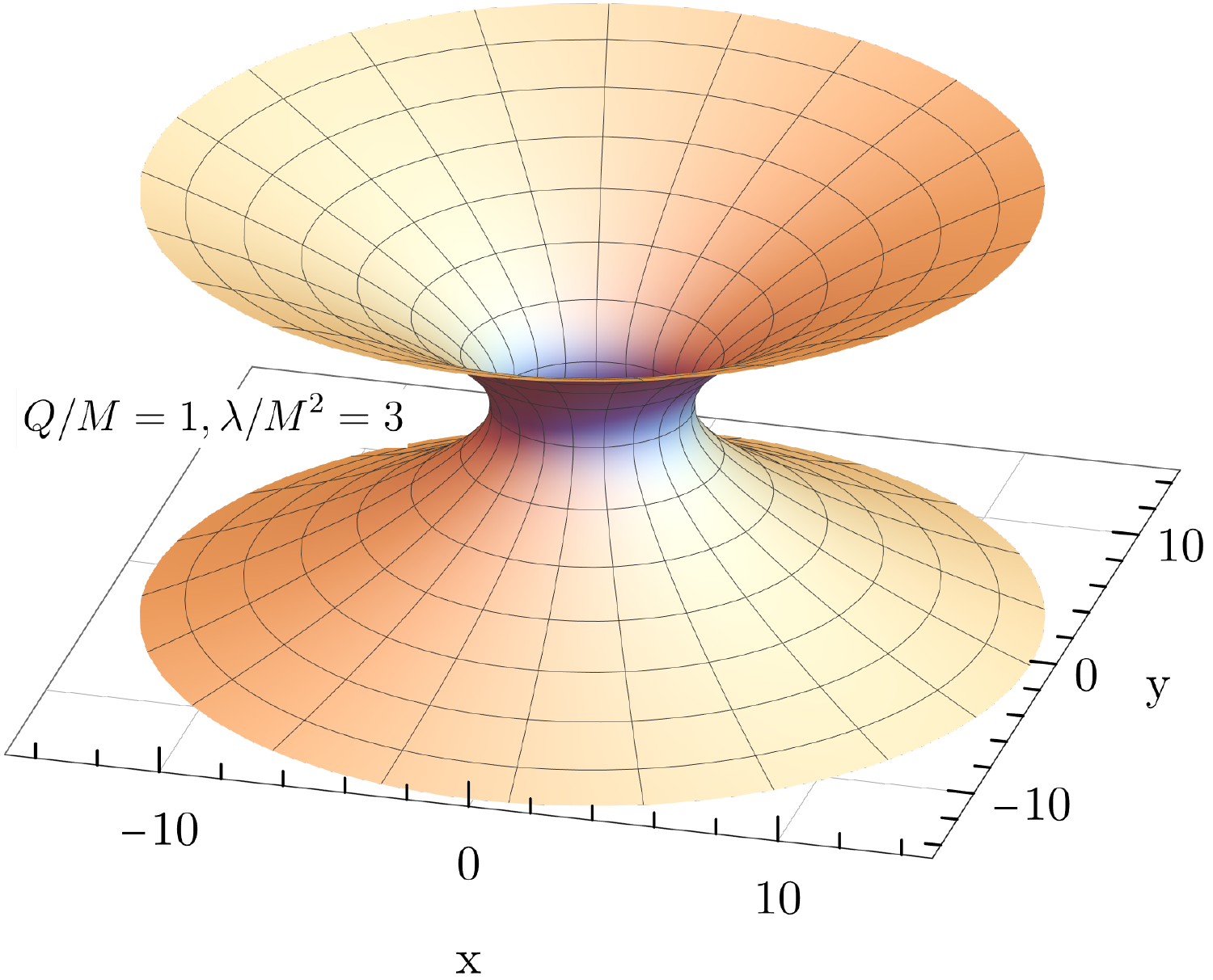}
			\caption{}
		\end{subfigure}%
		\begin{subfigure}{.5\textwidth}
			\includegraphics[width=\textwidth]{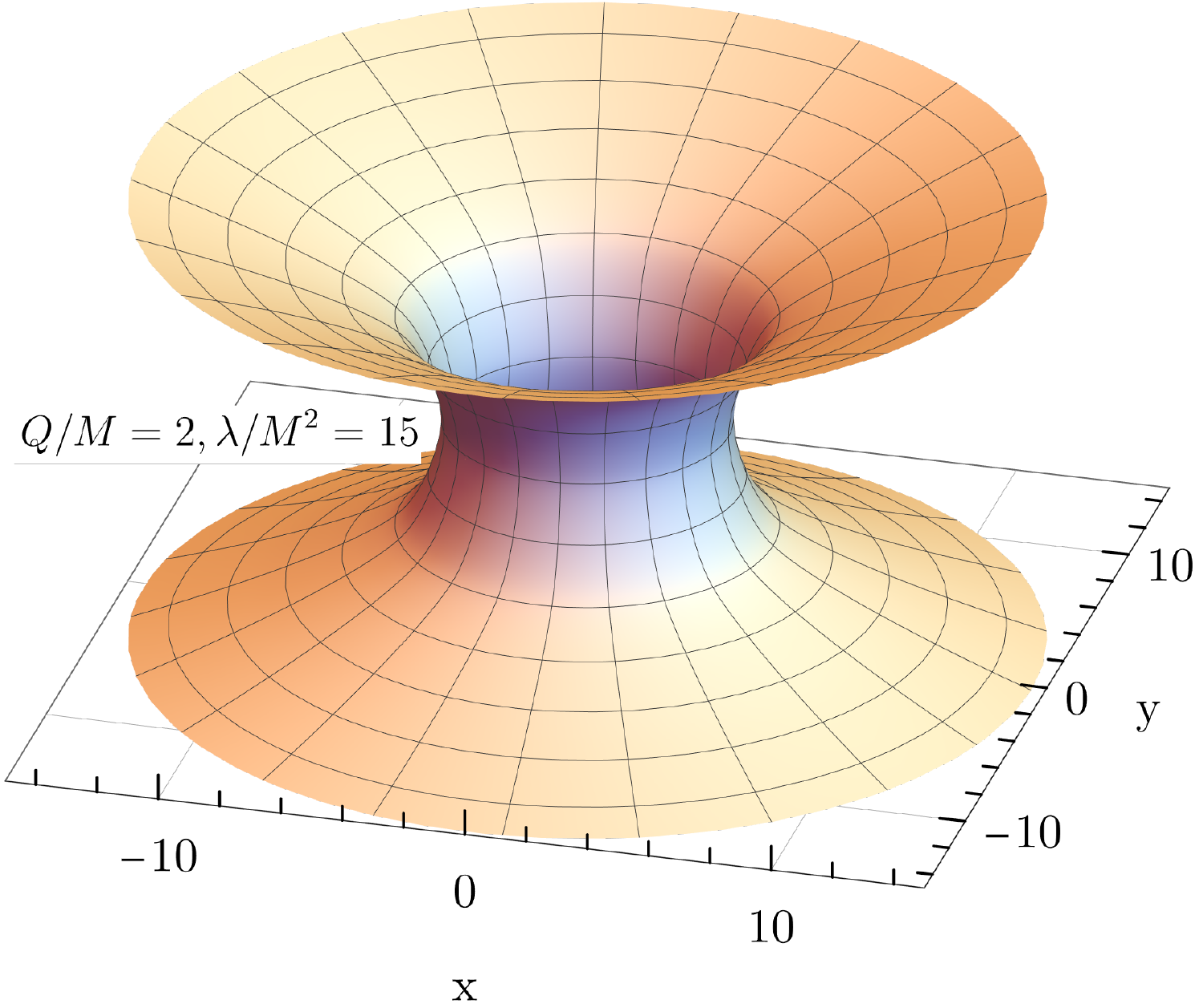}
			\caption{}
		\end{subfigure}\\
		\begin{subfigure}{.5\textwidth}
			\includegraphics[width=\textwidth]{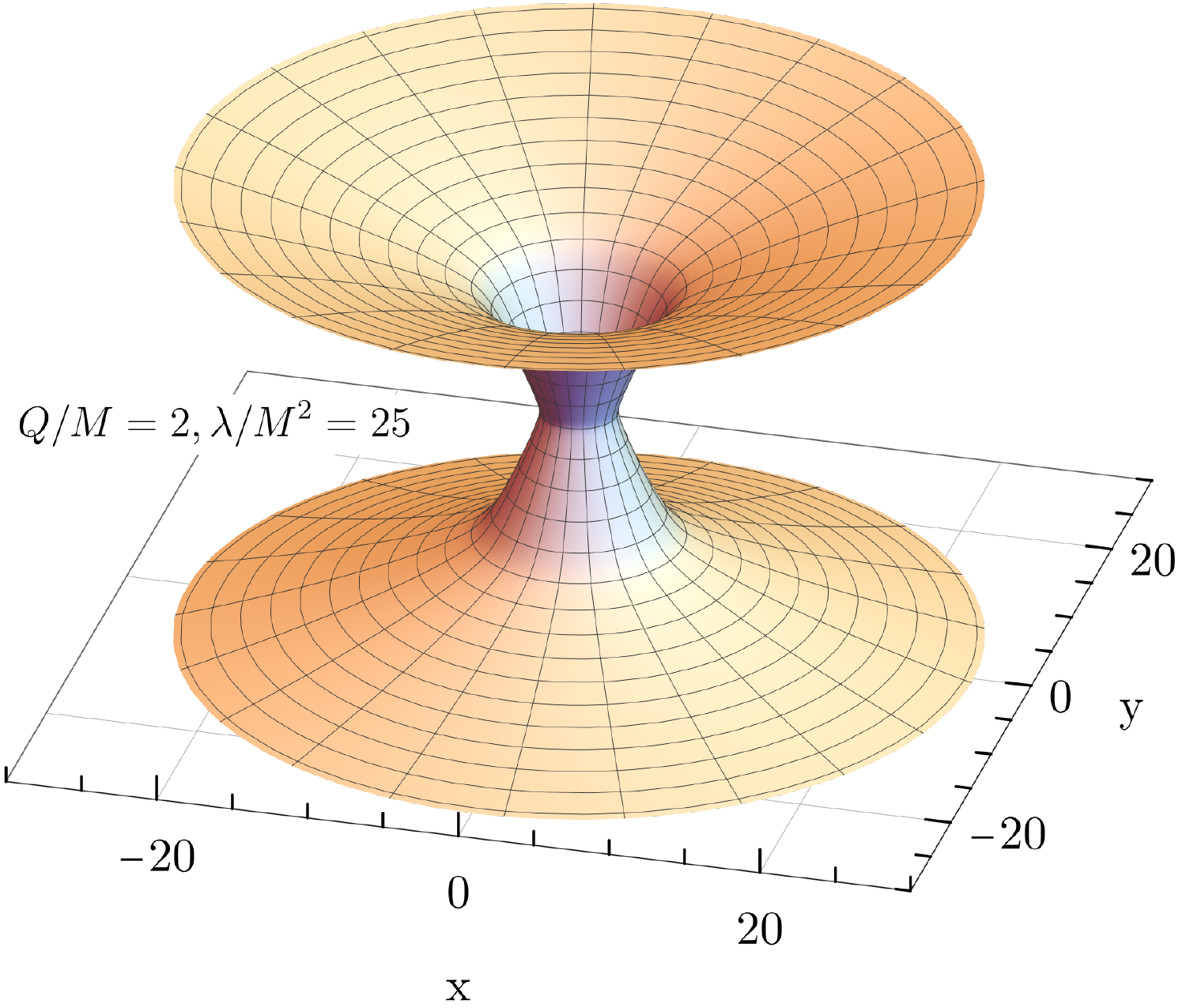}
			\caption{}
		\end{subfigure}%
		\begin{subfigure}{.5\textwidth}
			\includegraphics[width=\textwidth]{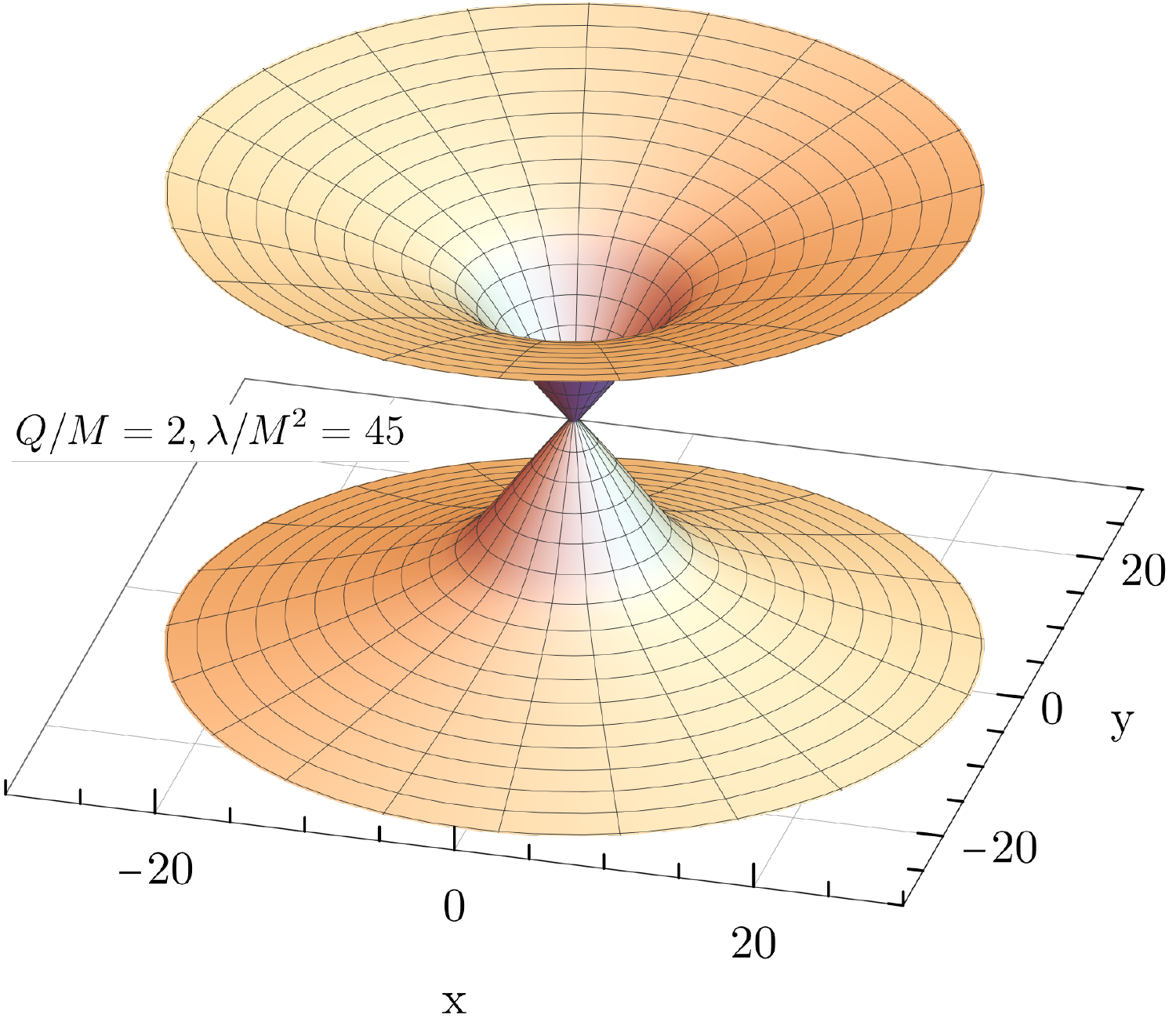}
			\caption{}
		\end{subfigure}\\
		\begin{subfigure}{.5\textwidth}
			\includegraphics[width=\textwidth]{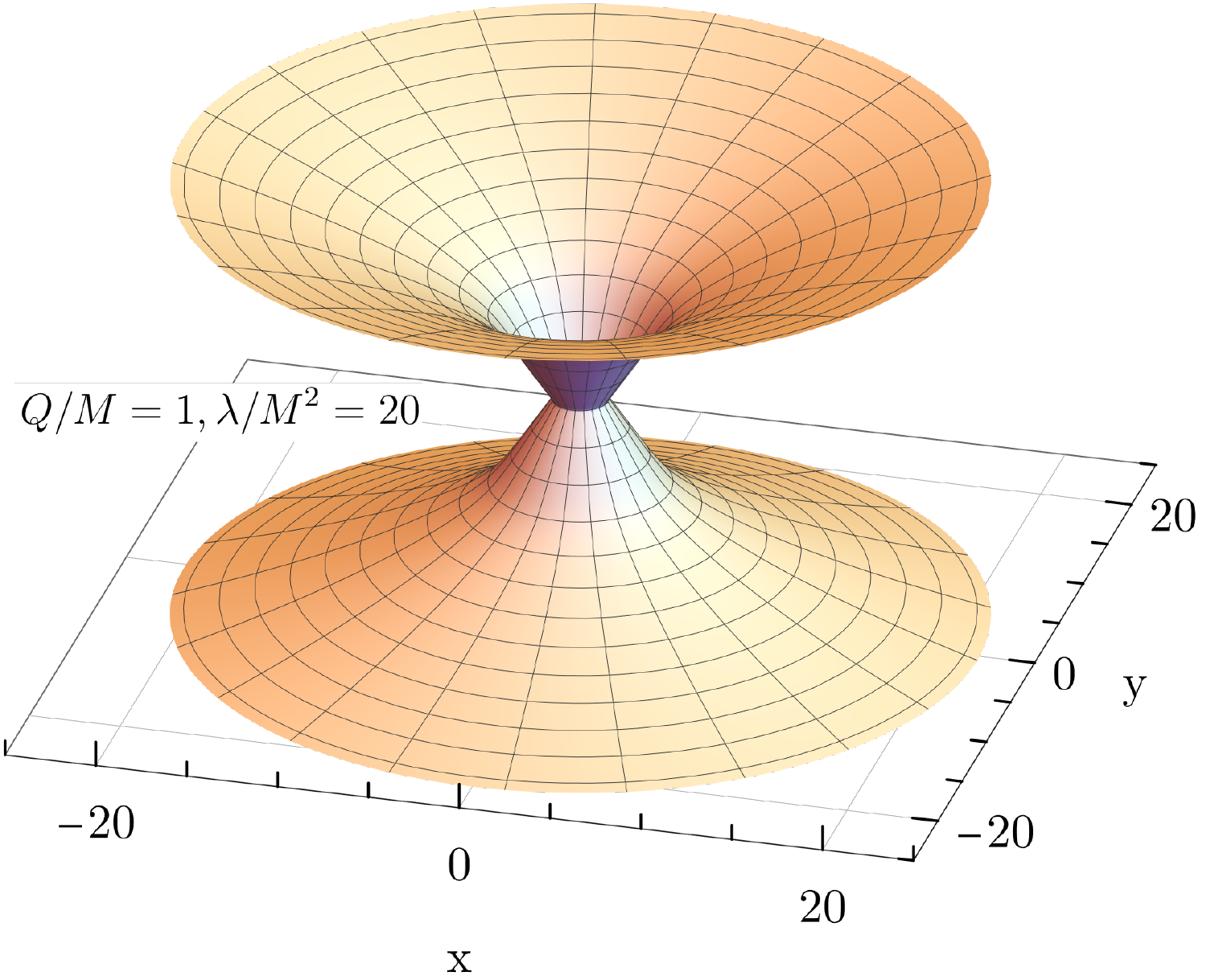}
			\caption{}
		\end{subfigure}%
		\begin{subfigure}{.5\textwidth}
			\includegraphics[width=\textwidth]{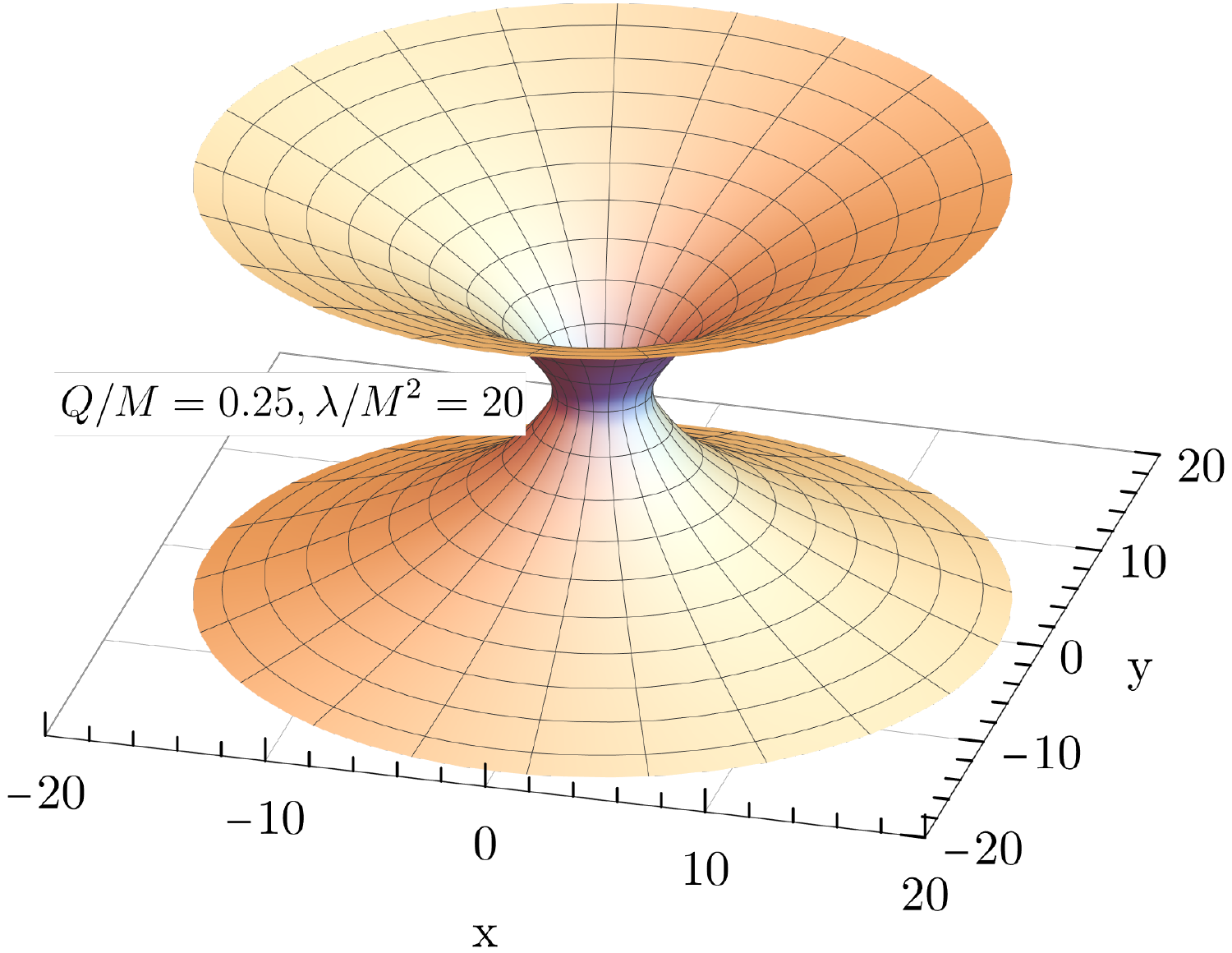}
			\caption{}
		\end{subfigure}\\
		\caption{3D embeddings for several examples of wormhole solutions .
		The proper distance between adjacent horizontal grid lines is $2$.}
		\label{fig:emb3dimi}
\end{figure}

	\begin{figure}[t]
		\begin{subfigure}{.5\textwidth}
			\includegraphics[width=\textwidth]{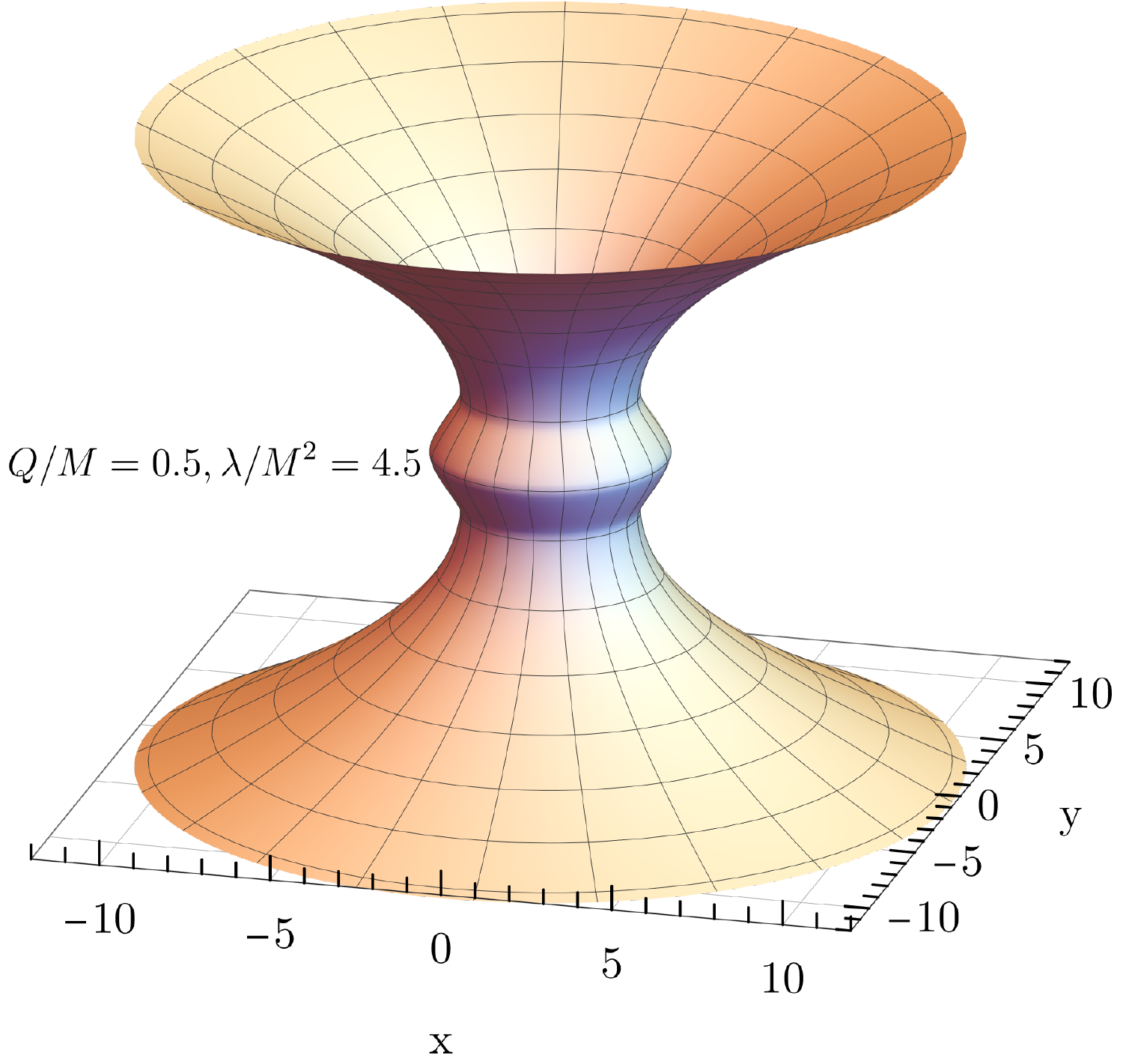}
			\caption{}
		\end{subfigure}%
		\begin{subfigure}{.5\textwidth}
			\includegraphics[width=\textwidth]{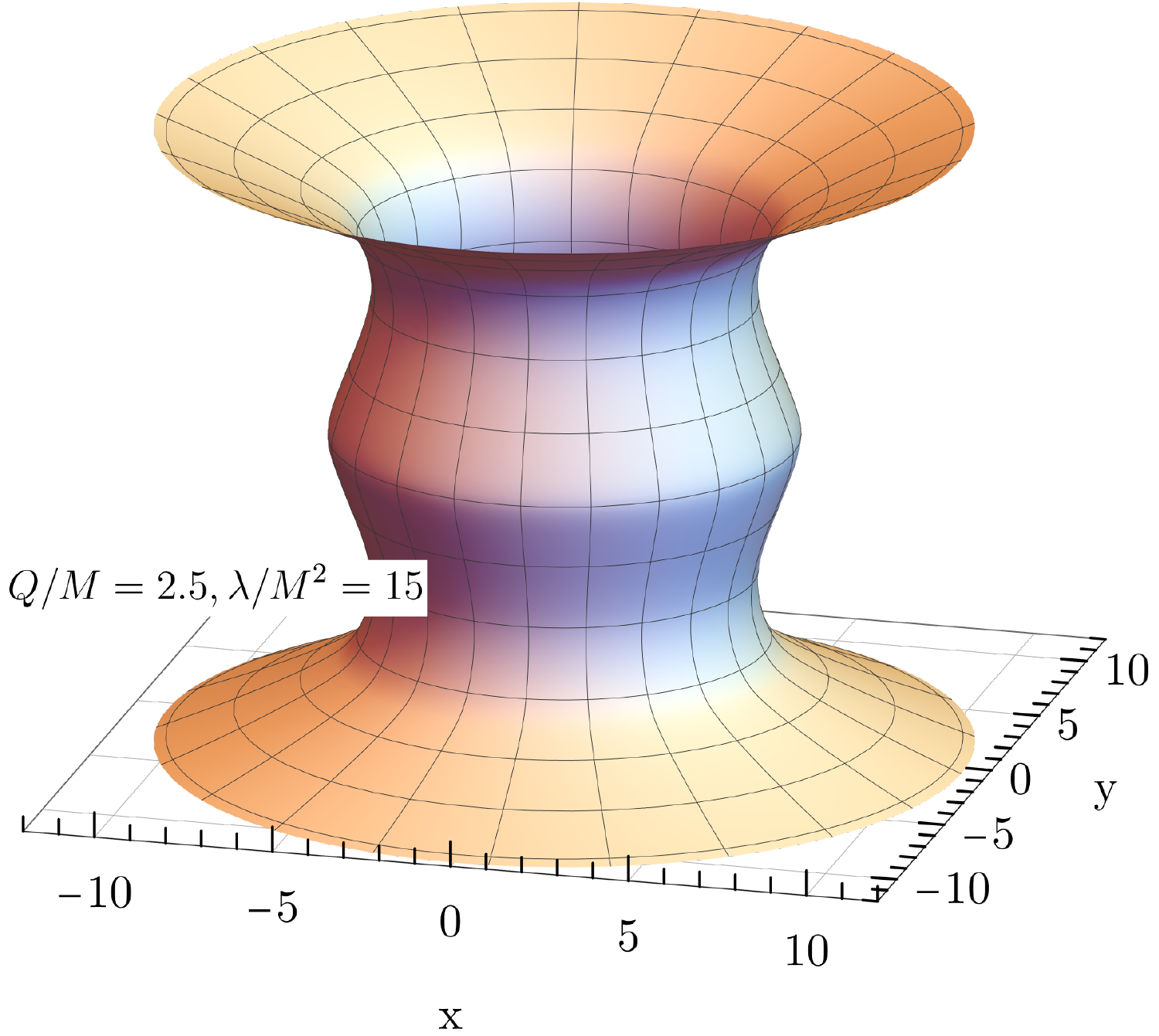}
			\caption{}
		\end{subfigure}\\
		\caption{3D embeddings for solutions with equator (coupling (iii)).}
		\label{fig:emb3dimiii}
\end{figure}

Embeddings of wormholes with throats and equators are helpful means to visualize the corresponding geometries.
To obtain the isometric embedding of the equatorial plane of the solutions, we start from the line element (\ref{eq_whcoord}) with $t$ constant, and $\theta=\pi/2$. 
We then equate this line element with a hypersurface of the
3-dimensional Euclidean space with cylindrical coordinates ($\rho$, $\varphi$, $h$).
This yields
\begin{equation}
F_1(\eta) \,[d\eta^2 +(\eta^2+\eta_0^2)\,d\varphi^2]=
d\rho^2 +\rho^2 d\varphi^2 + dh^2 \ . \label{emb}
\end{equation}
We now consider the coordinates $\rho$ and $h$ to be functions of the wormhole coordinate $\eta$.
Comparing coefficients we obtain
\begin{eqnarray}
    \rho(\eta) &=&\sqrt{F_1(\eta) \left(\eta^2+\eta_0^2 \right)} \ ,
\label{rhodef} \\ 
    \left(\frac{d\rho }{d\eta}\right)^2+\left(\frac{dh}{d\eta}\right)^2  &=&F_1(\eta) \ . 
\label{zzde}
\end{eqnarray}
Finally, we solve for the function $h(\eta)$
\begin{equation}
   h(\eta)=\pm \int_0^\eta \sqrt{ F_1(\tilde\eta)
    -\left( \frac{d}{d\tilde \eta}
   \left[  \sqrt{F_1(\tilde\eta) \left( \tilde \eta^2+\eta_0^2 \right) } \right]  \right)^2}d\tilde \eta \ .
 \label{zeq}
\end{equation} 
The functions $\rho(\eta)$ and $h(\eta)$ then provide a parametric representation of the equatorial plane.

We show such 2-dimensional embeddings for a fixed angle $\varphi$ in Fig.~\ref{fig:emb2dim}.
The coordinate $\rho$ corresponds to the circumferential radius $R_c$, shown on the abscissa, $h$ corresponds to the ordinate of the figures.
In particular, we compare the symmetrized wormhole solutions with their respective `parent' solutions, which still possess singularities in the $\eta<0$ region.
The figure contains wormholes with a single throat from coupling function (i) as well as wormholes with an equator and a double throat obtained with coupling function (iii).
Figs.~\ref{fig:emb3dimi} and \ref{fig:emb3dimiii} show sets of 3-dimensional embeddings, where we have included the azimuthal coordinate $\varphi$.

\FloatBarrier
\subsection{Geodesics}

\begin{figure}[t]
	\begin{subfigure}{0.54731\textwidth}
		\includegraphics[width=\textwidth]{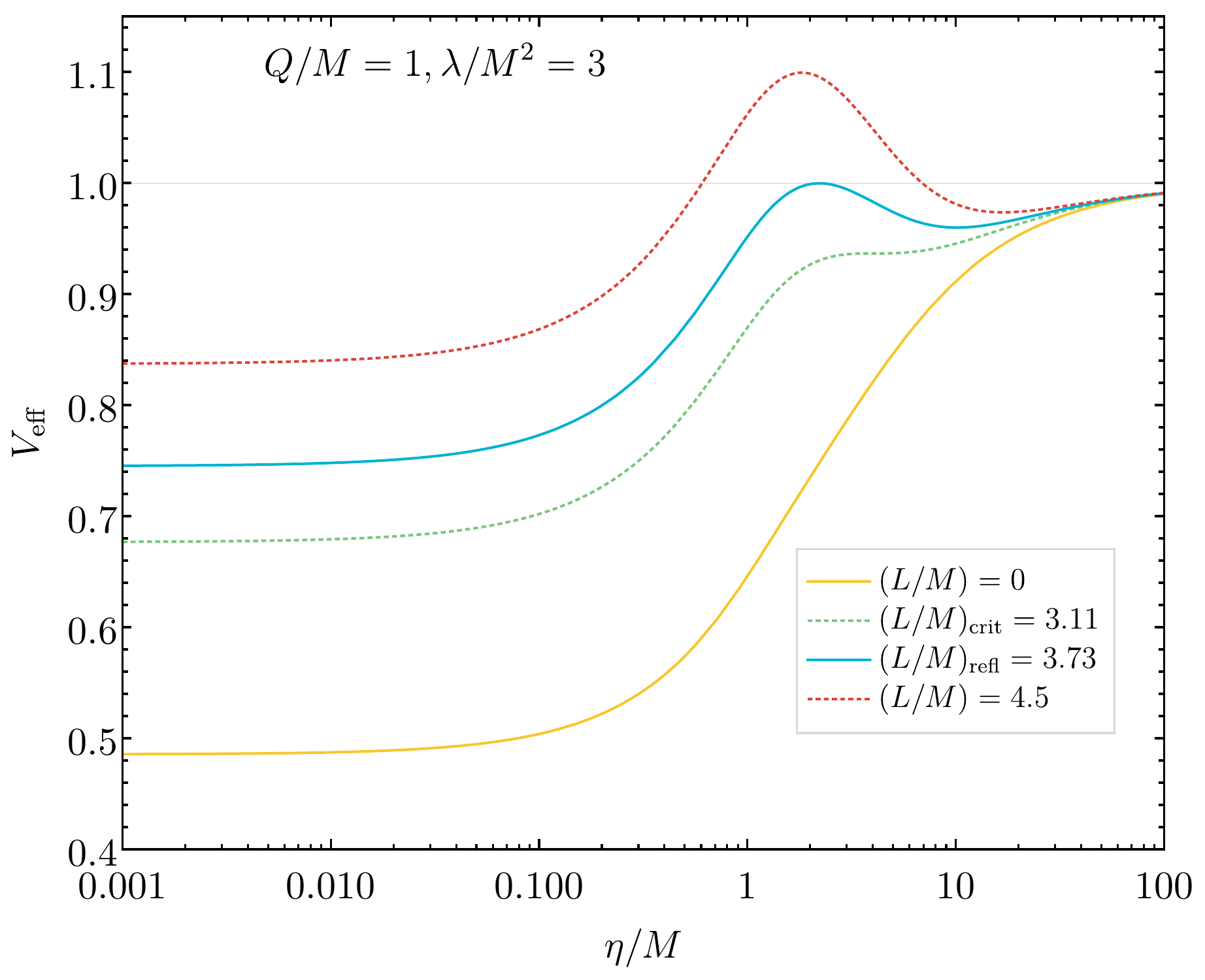}
		\caption{}
	\end{subfigure}%
	\begin{subfigure}{0.45269\textwidth}
		\includegraphics[width=\textwidth]{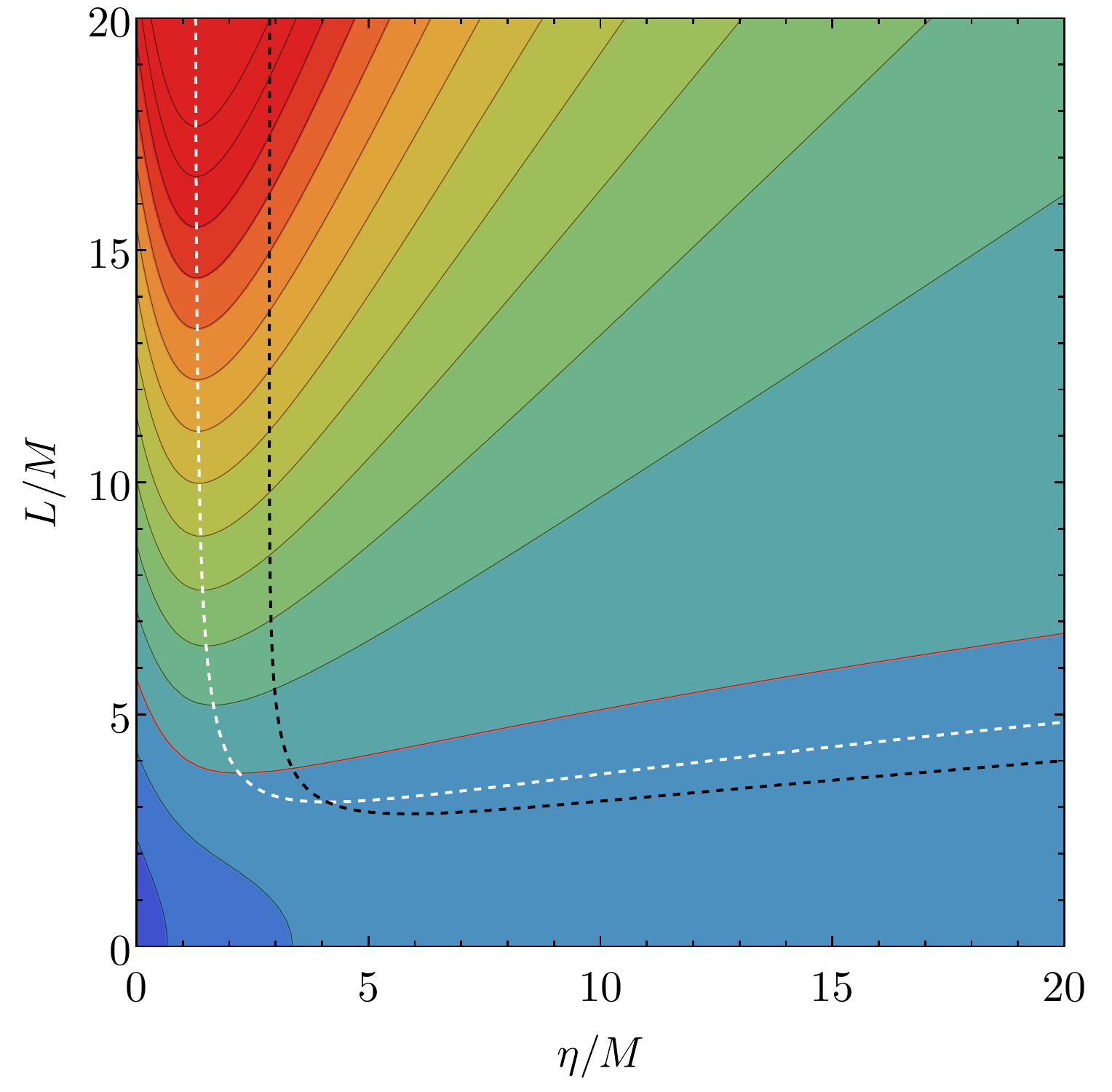}
		\caption{}
		\label{fig:effpot_con_n=1_i}
	\end{subfigure}\\
	\begin{subfigure}{.54731\textwidth}
		\includegraphics[width=\textwidth]{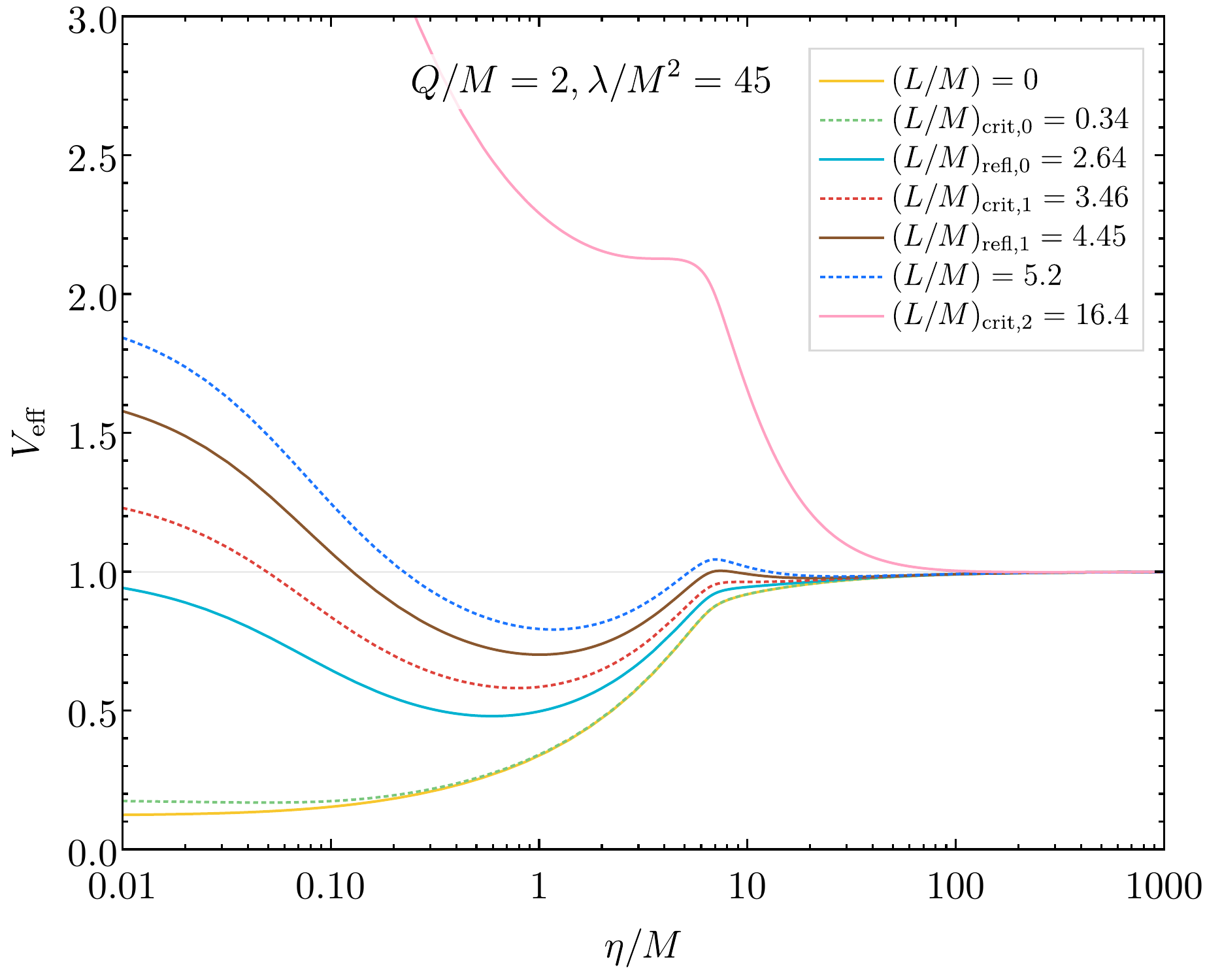}
		\caption{}
	\end{subfigure}%
	\begin{subfigure}{.45269\textwidth}
		\includegraphics[width=\textwidth]{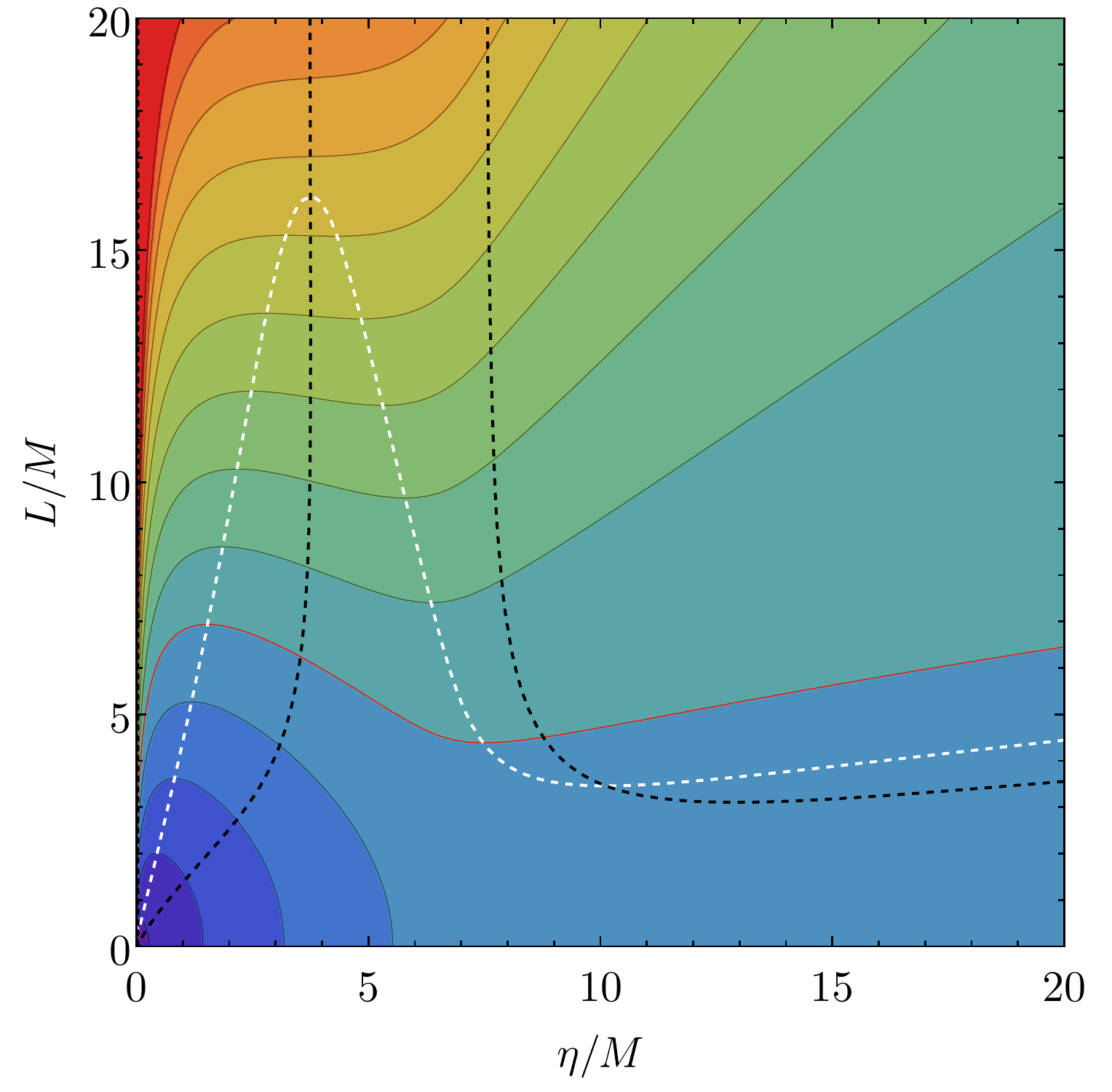}
		\caption{}
		\label{fig:effpot_con_n=3_i}
	\end{subfigure}\\
	\caption{Effective potential $V_{\text{eff}}$ for massive particles ($\kappa=-1$): 
	The upper figures (a) and (b) show the first case (parameters $Q/M=1$, $\lambda/M^2=3$),
	the lower figures (c) and (d) the second case  (parameters $Q/M=2$, $\lambda/M^2=45$).
	    Left: $V_{\text{eff}}$ vs $\eta/M$ for a selection of angular momenta $L$.
    	Right: Contours of $V_{\text{eff}}$ in the $L$--$\eta$ plane.
    	The solid red contour shows $V_{\text{eff}}=1$, the dashed white and black lines show $\partial_\eta V_{\text{eff}}=0$ and $\partial_\eta^2 V_{\text{eff}}=0$, respectively. 
    	The crossings of the white and black lines indicate the critical angular momenta $L_{\text{crit}}$.
    	The crossings of the red line and the maxima indicate the reflective angular momenta $L_{\text{refl}}$.
     The contour differences are $\Delta V_{\text{eff}}=0.2$.}
	\label{fig:veff}
\end{figure}
\begin{figure}[t]
     \begin{subfigure}{.54731\textwidth}
		\includegraphics[width=\textwidth]{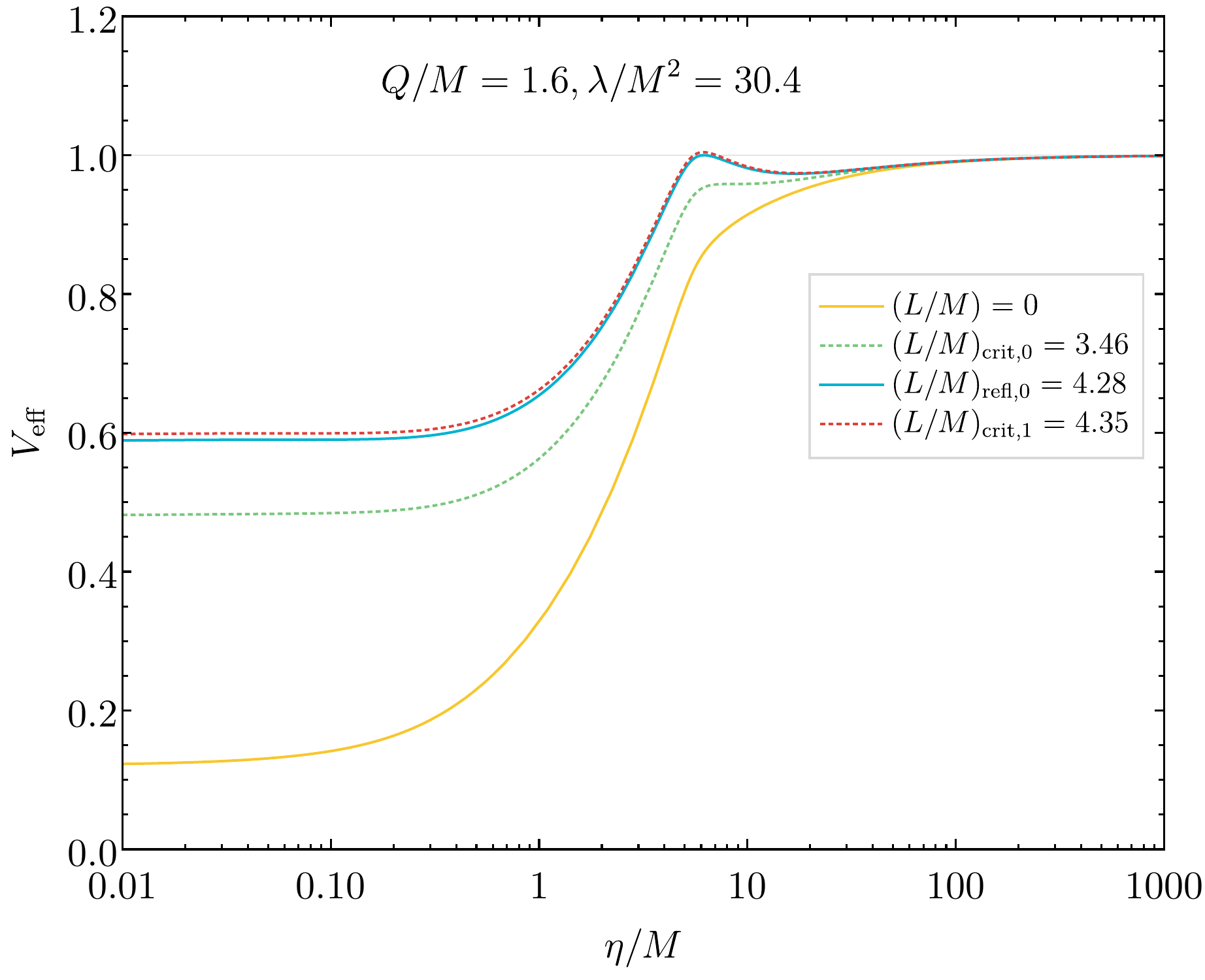}
		\caption{}
	\end{subfigure}%
	\begin{subfigure}{.45269\textwidth}
    	\includegraphics[width=\textwidth]{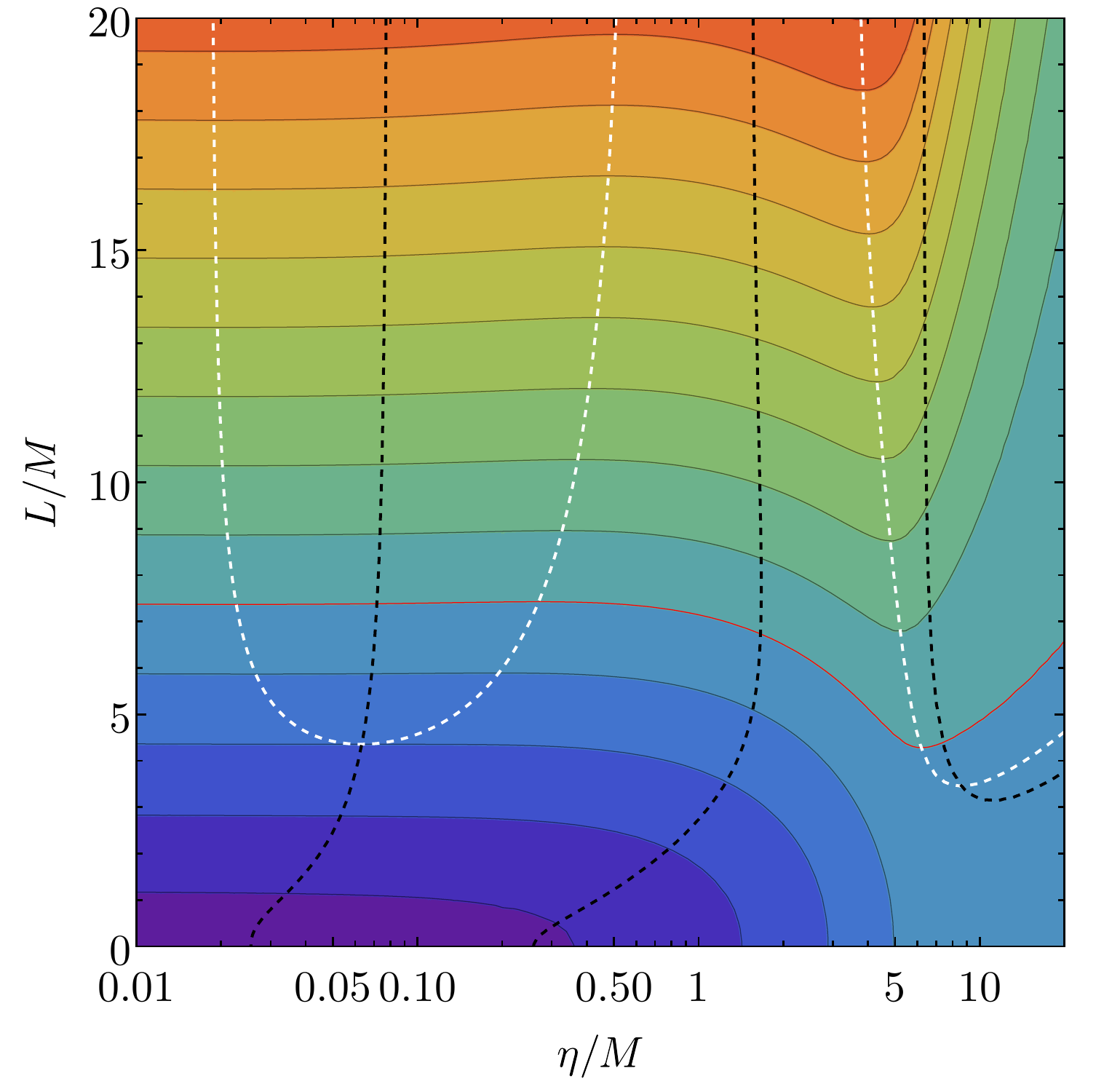}
		\caption{}
	\end{subfigure}
    \begin{subfigure}{0.54731\textwidth}
		\includegraphics[width=\textwidth]{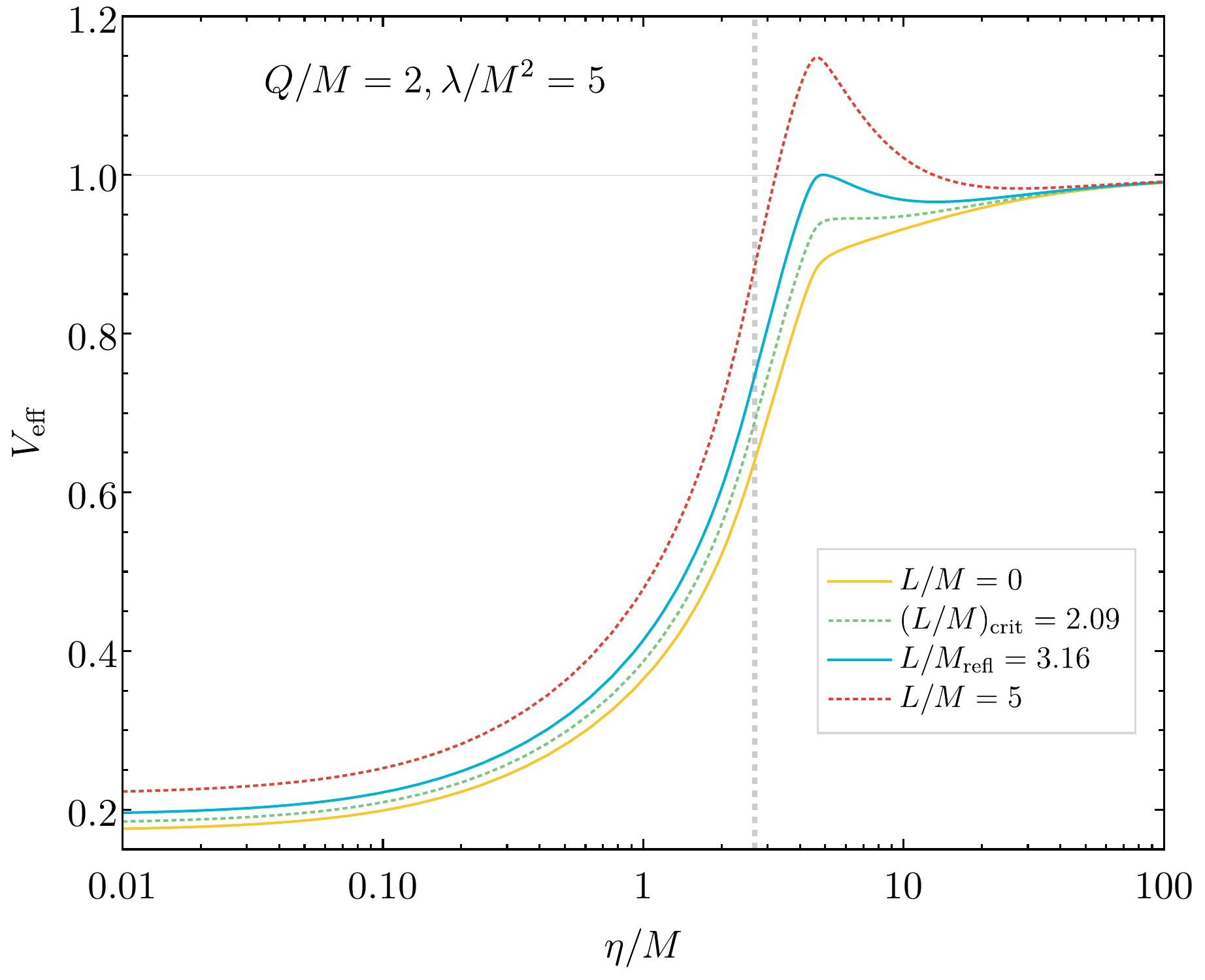}
		\caption{}
	\end{subfigure}%
	\begin{subfigure}{.45269\textwidth}
		\includegraphics[width=\textwidth]{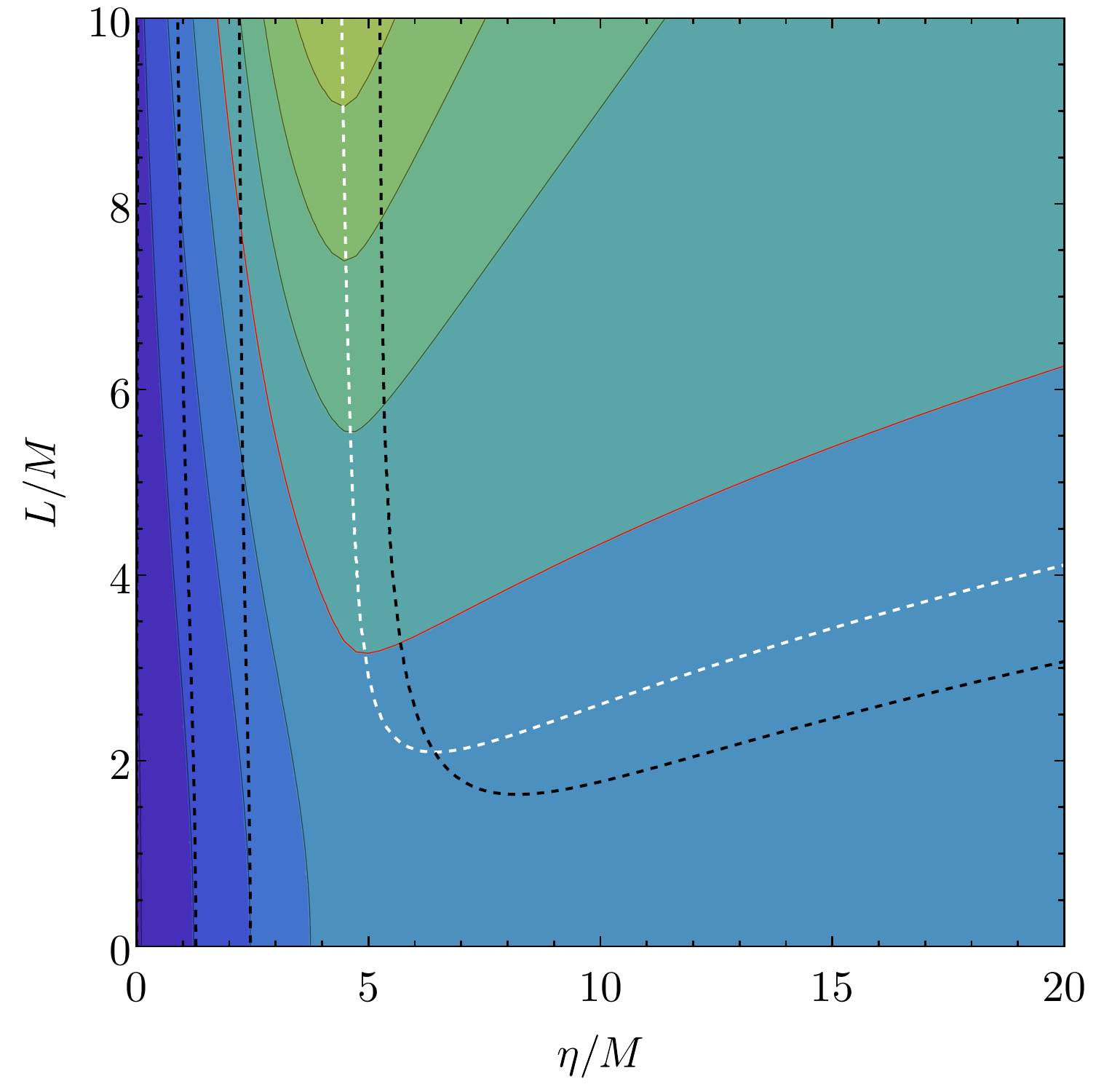}
		\caption{}
		\label{fig:effpot_con_n=1_iii}
	\end{subfigure}
	\caption{Analogous to Fig.~\ref{fig:veff}.
	The upper figures (a) and (b) show the third case (parameters $Q/M=1.6$, $\lambda/M^2=30.4$).
	The lower figures (c) and (d) show a wormhole with equator for coupling function (iii), where the dotted vertical line is the throat.}
	\label{fig:veff_e}
\end{figure}

We finally consider the different types of geodesics that arise in these wormhole spacetimes, considering first the motion of massive particles.
As discussed in subsection \ref{geodesics}, the motion is characterized by an effective potential (\ref{veff}), where $\kappa=-1$ for massive particles.
Here we identify three qualitatively different cases for the effective potential that are visualized in Figs.~\ref{fig:veff} and \ref{fig:veff_e}.
On the left hand side of the figures the effective potential is shown versus the wormhole coordinate for a selection of values of the particle angular momentum $L$, while on the right hand side contours of the effective potential are shown in the $L$--$\eta$ plane.

The first case corresponds to the example with parameters $Q/M=1$, $\lambda/M^2=3$, and is demonstrated in Figs.~\ref{fig:veff}(a)-(b). 
For $L=0$ the effective potential is monotonic and tends to one for $\eta\to\infty$.
At the throat $V_{\text{eff}}$ is finite, and its derivative $\partial_\eta V_{\text{eff}} > 0$.
A particle could sit at rest at the throat or oscillate  radially across the throat.
For small angular momenta bound rosetta orbits across the throat arise.
These orbits are retained also for large angular momenta.
At some critical angular momentum $L_{\text{crit}}$ the effective potential develops a saddle point.
For $L>L_{\text{crit}}$ the saddle point splits into an inner local maximum and an outer local minimum, thus in addition bound orbits exist, that do not cross the throat.
With further increasing $L$ the maximum reaches a value of $V_{\text{eff}}=1$ at some $L_{\text{refl}}$.
For $L>L_{\text{refl}}$ the effective potential acts as a reflective barrier, such that there are unbound orbits that do not cross the throat.

The second case corresponds to the example with parameters $Q/M=2$, $\lambda/M^2=45$, and is exhibited in Figs.~\ref{fig:veff}(c)-(d).
As in the first case, for $L=0$ the effective potential is monotonic and tends to one for $\eta\to\infty$, it is finite at the throat and $\partial_\eta V_{\text{eff}} > 0$.
Nonstatic bound orbits cross the throat.
For some small $L_{\text{crit,0}}$ the derivative at the throat then vanishes.
When $L>L_{\text{crit,0}}$ the potential has a local maximum at the throat and bound orbits do no longer cross the throat.
At some $L_{\text{refl,0}}$ the potential assumes the value $V_{\text{eff}}=1$ at the throat.
With further increasing $L$ a second critical $L$, $L_{\text{crit,1}}$ is reached, where the effective potential develops a saddle point.
Above $L_{\text{crit,1}}$ this saddle has split into a maximum and a minimum, and thus a second region with bound orbits is present.
Next, at $L_{\text{refl,1}}$ the local maximum of the effective potential reaches the value $V_{\text{eff}}=1$.
Subsequently, at $L_{\text{crit,2}}$ the inner minimum and the outer local maximum merge to a saddle point.
Beyond this saddle point only the outer bound orbits remain.

The third case corresponds to the example with parameters $Q/M=1.6$, $\lambda/M^2=30.4$, and  is shown in Figs.~\ref{fig:veff_e}(a)-(b). 
Here the throat is always a minimum, allowing for bound throat crossing orbits.
At a first critical angular momentum $L_{\text{crit,1}}$ a first saddle point arises, allowing for additional bound orbits for $L>L_{\text{crit,1}}$.
Soon a second saddle point arises, allowing for the third set of bound orbits for $L>L_{\text{crit,2}}$.

For the coupling function (iii) solutions with  an equator and a double throat appear for certain values of the parameters.
As an example we illustrate the effective potential in Figs.~\ref{fig:veff_e}(c)-(d) for parameters $Q/M=2$, $\lambda/M^2=5$.
In wormhole spacetimes with an equator and a double throat the equator is now at the center of the motion and the two throats will be passed when crossing from one asymptotic region to the other.
However, the presence of the equator surrounded by two throats now also allows for motion between the two throats, crossing the equator.


\begin{figure}[t]
	\begin{subfigure}{.5\textwidth}
		\includegraphics[width=\textwidth]{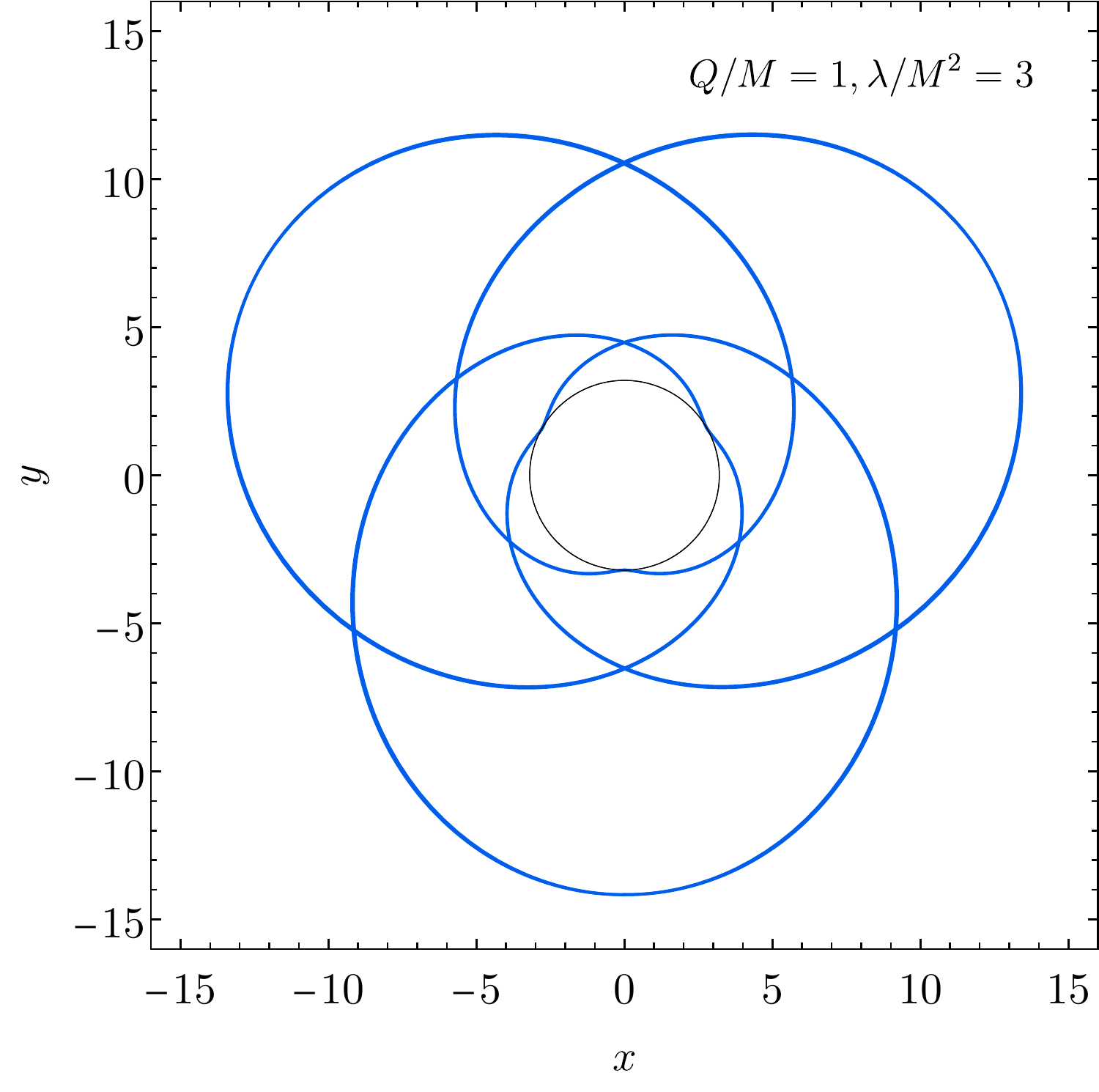}
		\caption{}
	\end{subfigure}%
	\begin{subfigure}{.5\textwidth}
		\includegraphics[width=\textwidth]{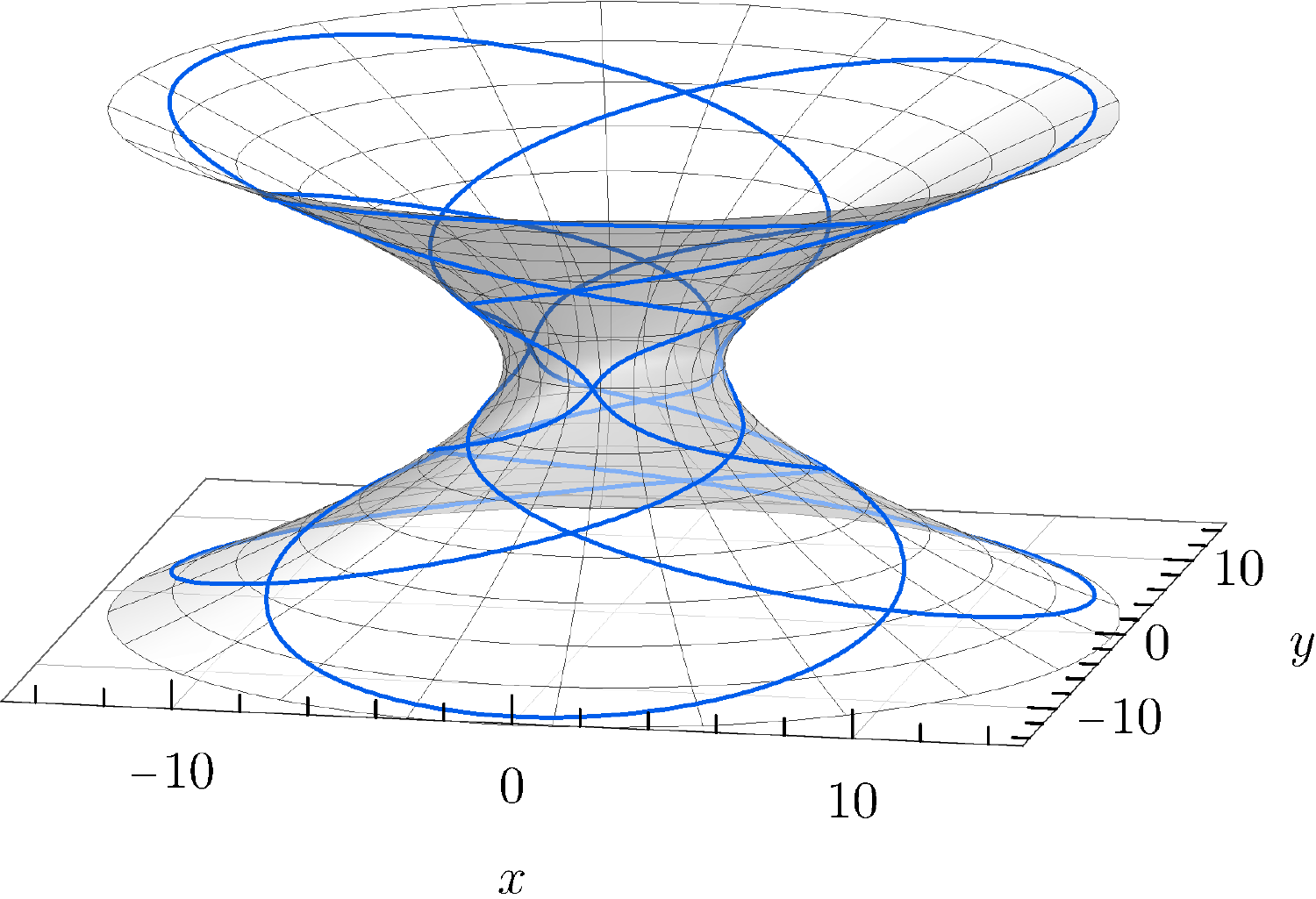}
		\caption{}
	\end{subfigure}\\
	\begin{subfigure}{.5\textwidth}
		\includegraphics[width=\textwidth]{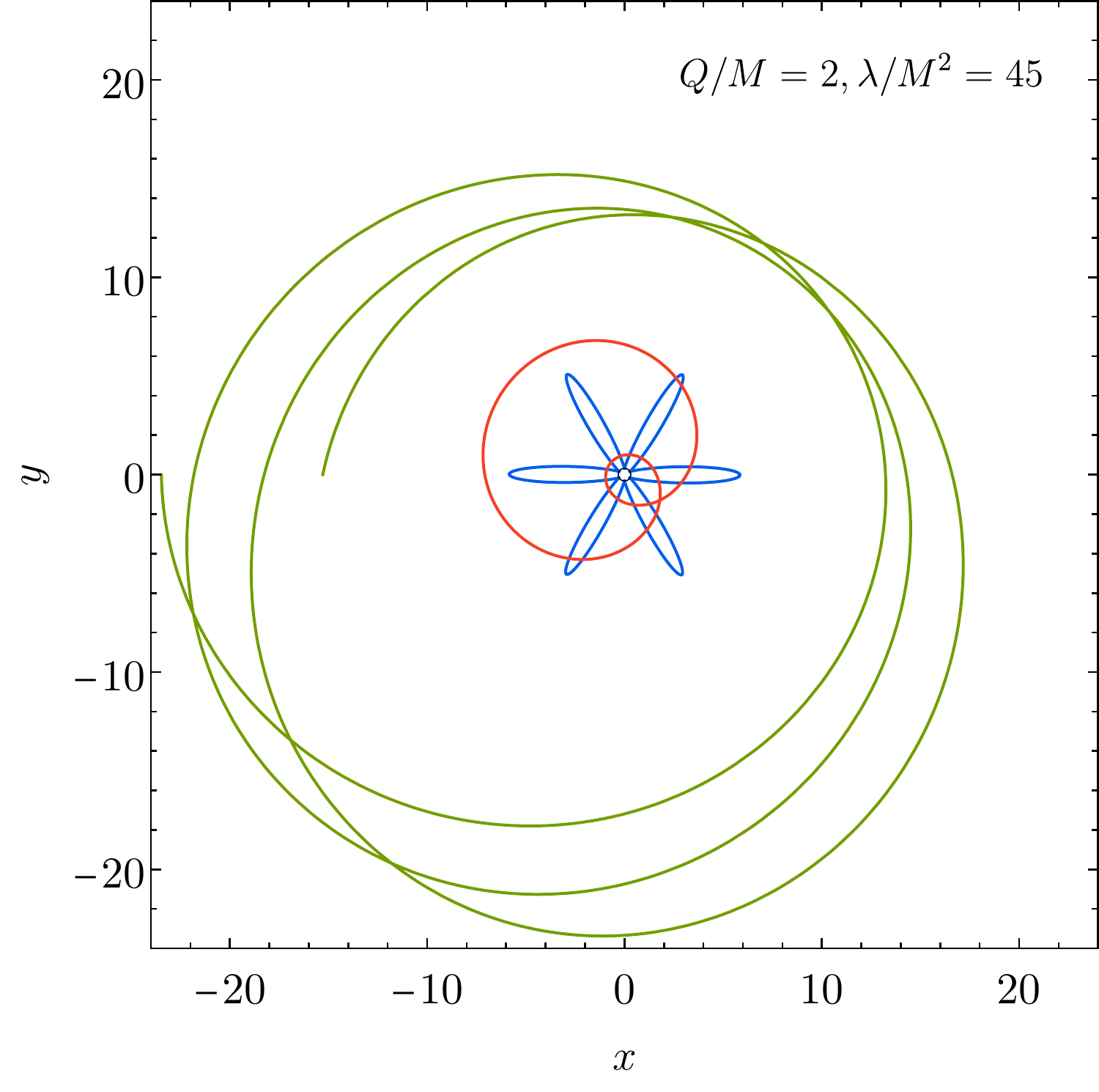}
		\caption{}
	\end{subfigure}%
	\begin{subfigure}{.5\textwidth}
		\includegraphics[width=\textwidth]{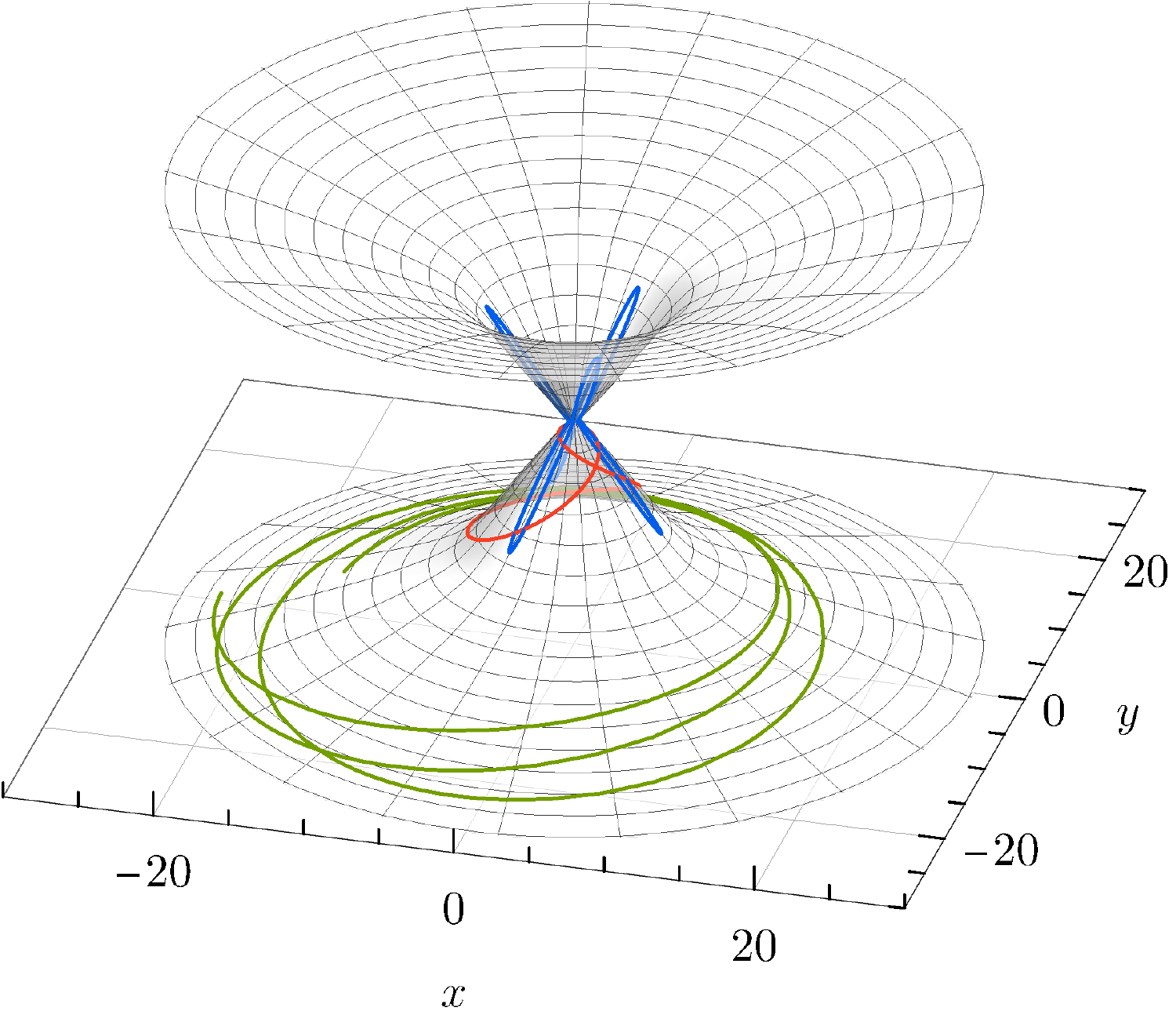}
		\caption{}
	\end{subfigure}
	\caption{Orbits of massive particles in wormhole spacetimes for coupling function (i).
	Left: projections of the orbits. 
	Right: embedding diagrams.
	(a) and (b) show a closed orbit for $L=3,E=0.95$ with 3-fold symmetry, passing the throat 6 times during 8 revolutions.
	(c) and (d) show bound orbits for the second example
    	in blue for $L=0.34,E=0.714$ with 3-fold symmetry but 6 apastra, passing the throat 6 times during one revolution;
    	in green for $L=4,E=0.974$ a precessing `Keplerian' orbit;
    	in red for $L=5.2,E=1.039$ a zoom-whirl orbit. 
    }
    \label{fig:orbits}
\end{figure}
\begin{figure}[t]
	\begin{subfigure}{.5\textwidth}
		\includegraphics[width=\textwidth]{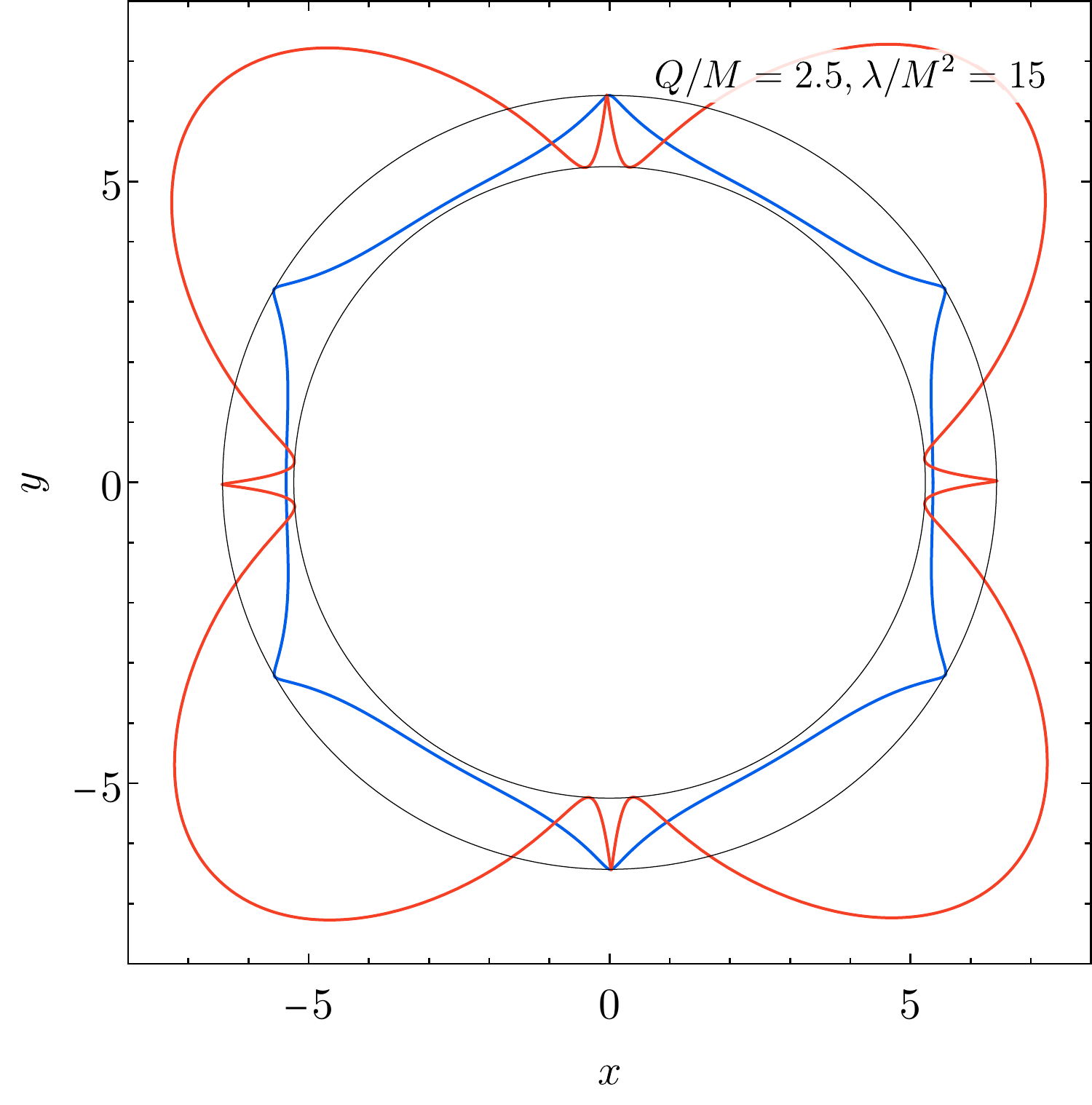}
		\caption{}
	\end{subfigure}%
	\begin{subfigure}{.5\textwidth}
		\includegraphics[width=\textwidth]{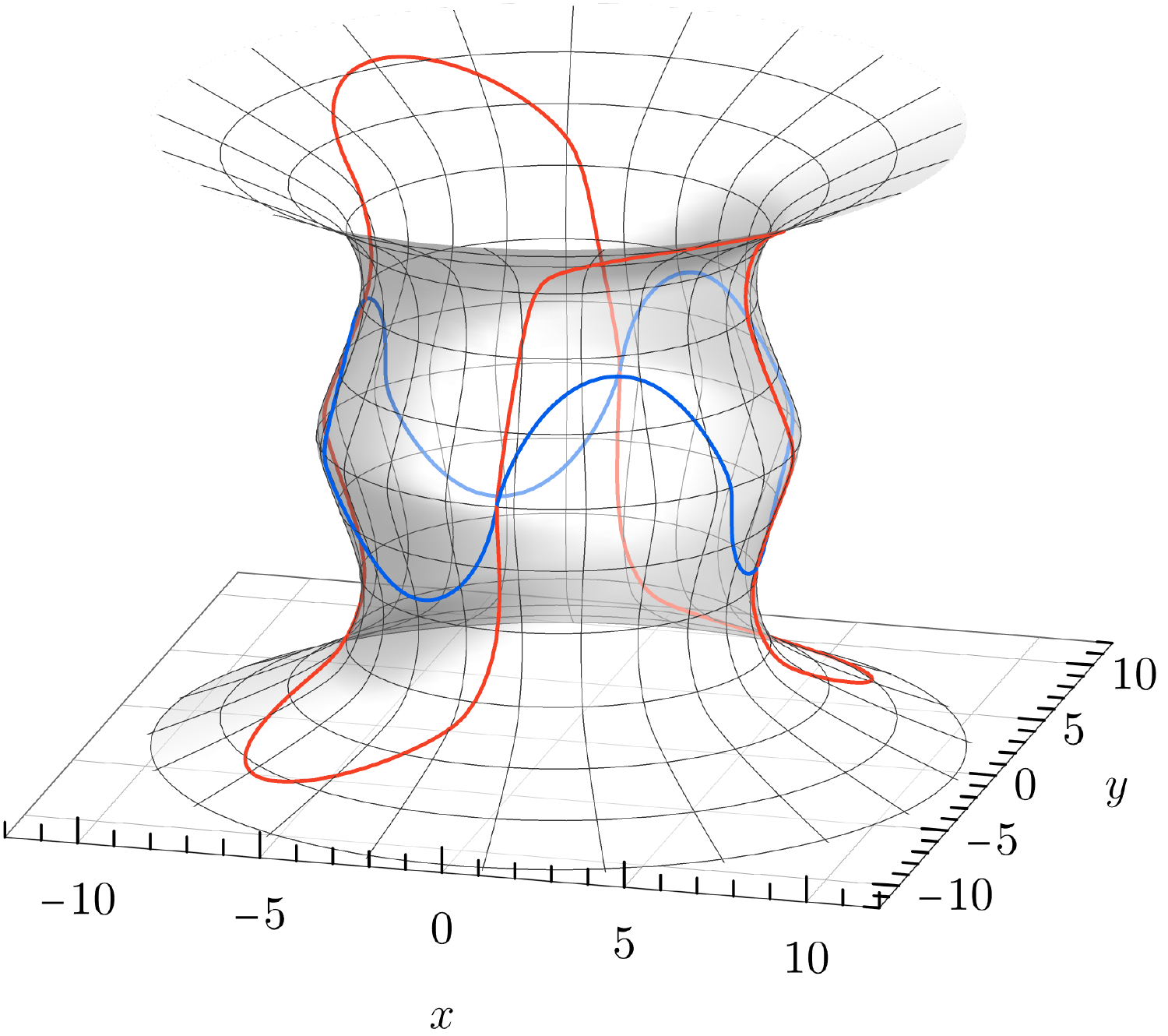}
		\caption{}
	\end{subfigure}
	\caption{Orbits of massive particles in a wormhole spacetime with equator and double throat for coupling function (iii).
	Closed orbits crossing the equator: 
	    in blue for $L=5, E=0.71$ passing only the equator;
	    in red for $L=1, E=0.92789$ passing the equator and the throat.
	    The outer and inner black circles indicate the equator and the throat, respectively.
	    }
	    \label{fig:orbits_e}
\end{figure}

Some examples of bound orbits in single throat wormhole spacetimes are shown in Figs.~\ref{fig:orbits}.
The figures on the left hand side show projections of the orbits, where
\begin{equation}
    x = \frac{1}{M} R_c \cos \varphi, \quad y = \frac{1}{M} R_c \sin \varphi \ ,
\end{equation}
while the figures on the right hand side illustrate the orbits with the help of embedding diagrams. 
Orbits in a wormhole spacetimes with an equator are illustrated in Fig.~\ref{fig:orbits_e}.

\begin{figure}[t]
	\begin{subfigure}{.5\textwidth}
		\includegraphics[width=\textwidth]{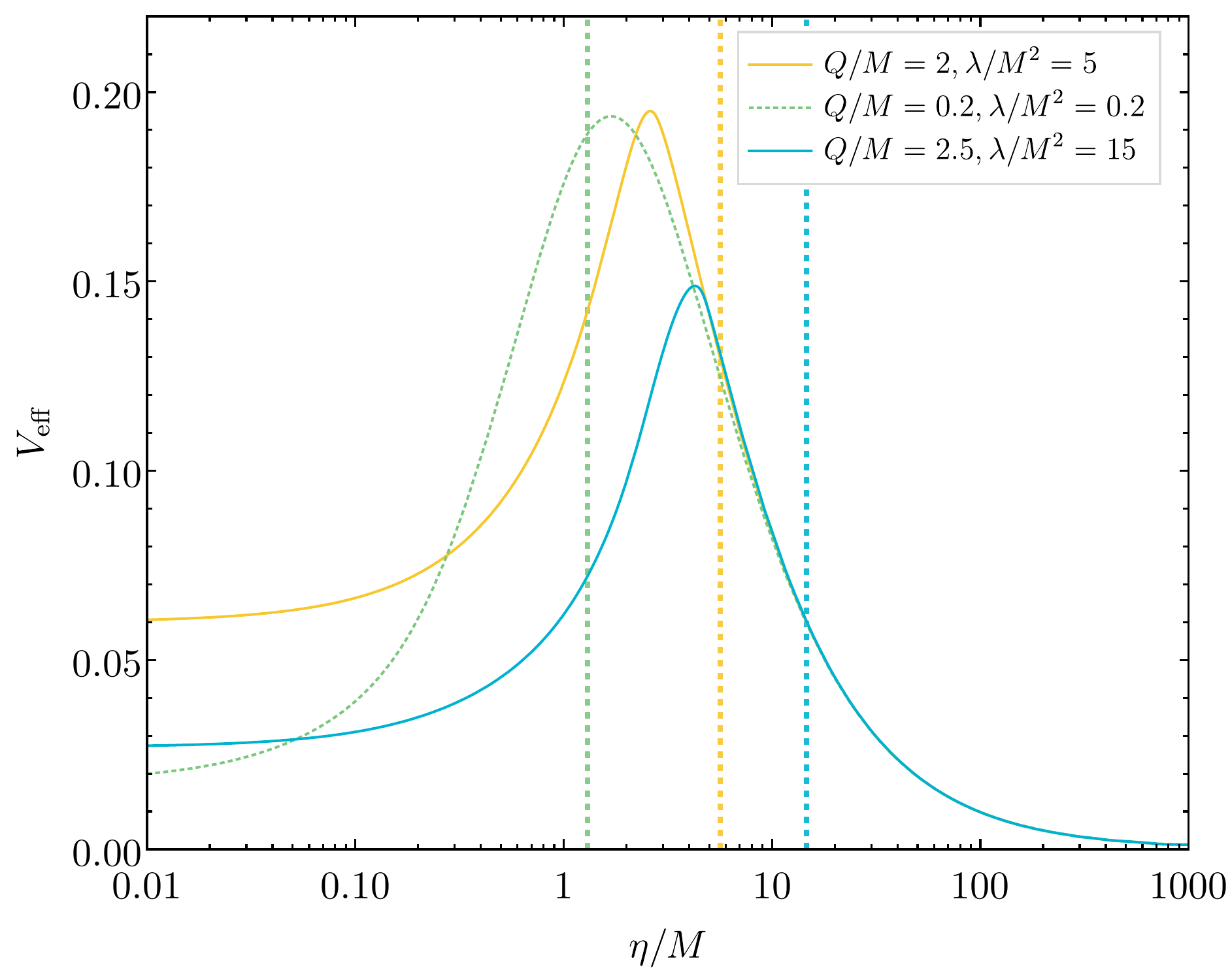}
		\caption{}
	\end{subfigure}%
	\begin{subfigure}{.5\textwidth}
		\includegraphics[width=\textwidth]{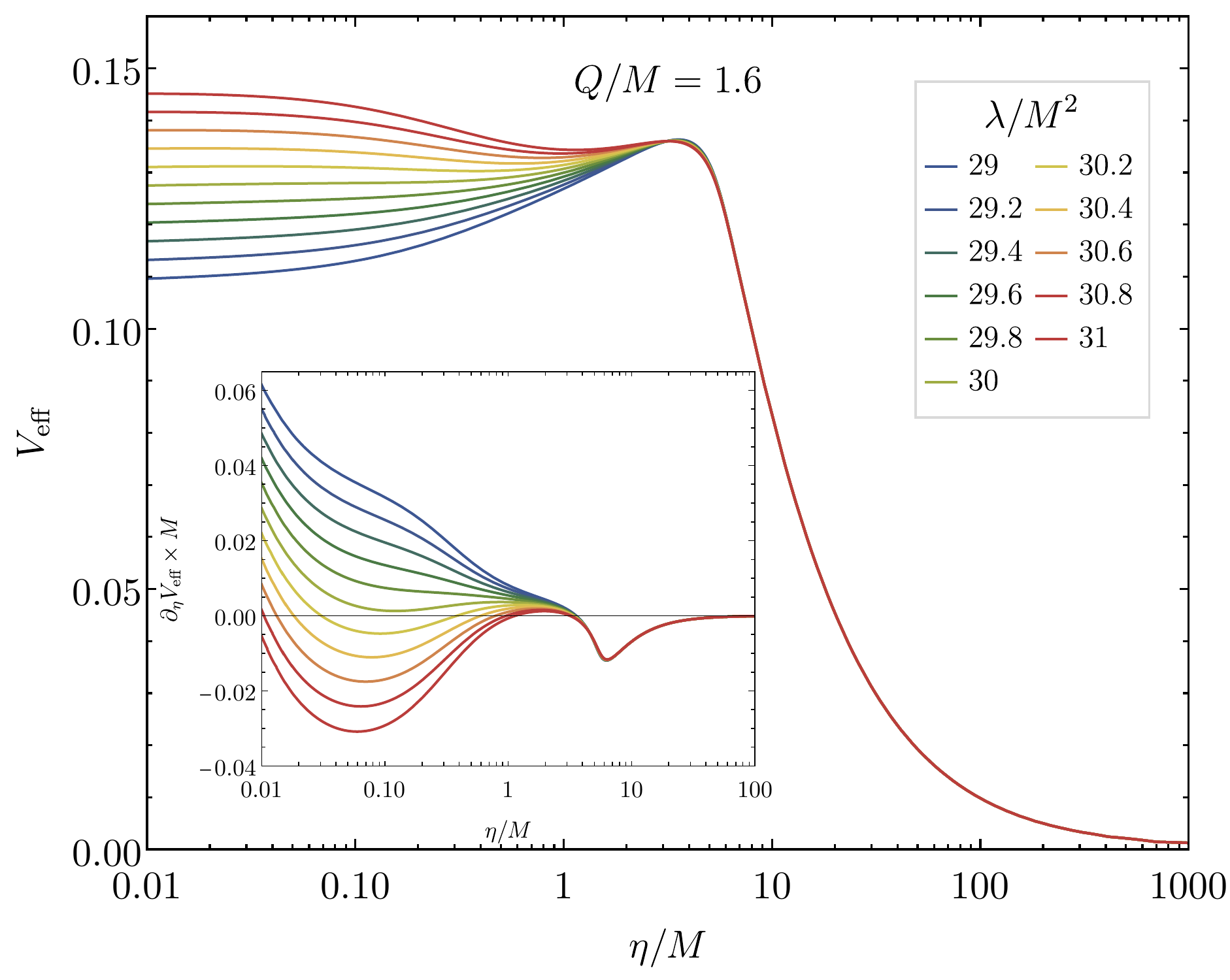}
		\caption{}
	\end{subfigure}
	\caption{Effective potential $V_{\text{eff}}$ for massless particles and light ($\kappa=0$):
	(a) coupling (iii) with equator at $\eta=0$ and throat at the vertical dashed line. 
    (b) coupling (i), subdomain IV for several values of the coupling constant $\lambda$. 
    The extrema of $V_{\text{eff}}$, and thus the zeros of $\partial_\eta V_{\text{eff}}$ (inset), correspond to light rings. 
    Depending on the coupling constant $\lambda$ there are 3,5 or 7 light rings for symmetric wormholes. 
    }
  \label{fig:lightring}  
\end{figure}

At last we turn to the motion of light in these spacetimes, i.e., $\kappa=0$ in \cref{veff}.
In this case we are mostly interested in the circular orbits, since wormholes are often considered black hole mimickers.
Spherical black holes possess an unstable circular orbit, their light ring, which is associated with their shadow size.
Wormholes would have a light ring at their center, but they may possess more circular orbits.
Because of the symmetry of the spacetime, these would always arise in pairs, so there could be 3 or 5 or more light rings present. The location of the light rings is exhibited in Fig.~\ref{fig:lightring} for several typical wormhole solutions (coupling (i)) and a wormhole solution with equator (coupling (iii)).

\section{Conclusions}

We have considered wormhole solutions in a family of Einstein-vector-Gauss-Bonnet theories.
The coupling functions of these theories vanish quadratically for vanishing massless vector field.
Consequently, the Schwarzschild solution remains a solution of the field equations.
In contrast, the Reissner-Nordst\"om solution is only a solution in the limit of vanishing coupling constant.

While General Relativity alone does not lead to classical traversable wormhole solutions, the presence of the Gauss-Bonnet term coupled to the vector field provides the needed violation of the energy conditions to obtain wormholes.
In fact, wormhole solutions arise naturally in these theories when the equations of motion are solved.
In these solutions the circumferential radius develops a minimum, which physically represents a throat, allowing passage to another part of the Universe.
However, typically this other part is not an infinite asymptotically flat region for the solutions of these theories.

Instead, in the construction of the wormhole solutions typically singularities arise.
As long as these singularities reside in the region beyond the throat, they can be avoided by reflecting the regular asymptotically flat side across the throat.
Of course, in this case a thin shell of matter is needed to satisfy the junction conditions (see Appendix).
The set of the resulting wormholes is then referred to as symmetric wormholes.
Also maximal surfaces may arise, across which the solutions can be reflected.
These solutions then represent wormholes with an equator and a double throat.

The domain of existence of these wormholes is bounded by the black holes of General Relativity and by singular solutions, where the singularity either arises on the throat or in the prime part of the Universe.
The vectorized black holes also feature prominently in the domain of existence, forming part of a critical line of solutions, where the vector field and the time component of the metric vanish at the regular horizon or singular throat.
Like the vectorized black holes also the wormholes may possess nodes for sufficiently large coupling.
Moreover, as the coupling constant increases a critical value of the charge is approached.

The wormholes feature a variety of interesting orbits for particles and light.
Unbound and bound motion of massive particles is present in each part of the Universe, but also across the throat and equator.
Light rings are always present at the center of the wormhole, be it a throat or an equator, but in addition pairs of light rings arise or disappear again, when the parameters are varied.
In particular, the presence of unstable light rings, i.e., maxima of the effective potential, then signals that the respective wormholes may be viewed as ultracompact objects (UCOs) \cite{Cardoso:2017cqb}.

\section{Appendix: Junction Conditions}

    When constructing symmetric wormholes by reflecting the solution at the throat, we need to restore differentiability of the metric functions and the vector field by introducing a thin shell of matter at the throat.
    The action then acquires a new source term at the throat ($\eta=0$)
    \begin{equation}
    S=\frac{1}{16\pi}\int d^4x\sqrt{-g}\left[ R-F_{\mu\nu}F^{\mu\nu} + F(A_\mu A^\mu) \rgb + \delta(\eta)( -4 A_\mu j^\mu + 16\pi\mathcal{L}_M) \right],
    \end{equation}
    with the matter Lagrangian $\mathcal{L}_M$ and the current density $j^\mu$.
    Note, that for simplicity of notation, we present this discussion in ordinary wormhole coordinates
       \begin{align}\label{eq_whcoord2}
	ds^2 = -f_0(\eta) dt^2 + f_1(\eta)  d\eta^2 + (\eta^2+\eta_0^2) (d\theta^2 + \sin^2\theta d\varphi^2) \ .
\end{align} 
 (Of course, the radial coordinate and the functions differ from those used previously, although the notation is the same.)
    Varying with respect to $A_\mu$ and $g^{\mu\nu}$ yields the equations of motion,
    \begin{align}
    \nabla_\nu F^{\nu\mu} =& -\frac{1}{2} \lambda \rgb \coupl'(A^2) A^\mu + \delta(\eta) j^\mu, \label{eq:jceom1}\\
    G^\mu_{~\nu} =& \frac{1}{2}T^{(A)\mu}_{~~~~~\nu}-T^{(GB)\mu}_{~~~~~~~\nu}+\delta(\eta) 8\pi T^\mu_{~\nu}. \label{eq:jceom2}
    \end{align}
    where $T^{(A)\mu}_{~~~~~\nu}$ and $T^{(GB)\mu}_{~~~~~~~\nu}$ are as in \cref{tphi,teffi},
    and variation of the matter Lagrangian yields the stress energy tensor
    \begin{equation}
    T_{\mu\nu}:=\frac{-2}{\sqrt{-g}}\frac{\delta(\sqrt{-g}\mathcal{L}_M)}{\delta g^{\mu\nu}}.
    \end{equation}
    
    In these coordinates the throat is located at $\eta=0$, and $\eta_0$ is its circumferential radius.
    $\eta>0$ describes the radial domain on the main side of the throat, that is reflected at the throat to the $\eta<0$ side.
    In practise, we continue all three functions $f\in \{f_0,f_1,A_0\}$ symmetrically in $\eta$ to the other side,
    \begin{equation}
    f(\eta)=f(-\eta), \quad f'=-f'(-\eta), \quad f''=f''(-\eta) \quad \text{with} \quad \eta<0 \ .
    \end{equation}
    Therefore, if some $f$ is part of a solution of the equations of motion, then the replacement
    \begin{align}
    f(\eta) &\mapsto {f} + (2 \Theta (\eta)-1)  \eta {f}' + \mathcal{O}(\eta^2), \label{eq:repl1}\\
    f'(\eta) &\mapsto (2 \Theta(\eta)-1) {f}' + \mathcal{O}(\eta), \label{eq:repl2}\\
    f''(\eta) &\mapsto 2 \delta(\eta) {f}' + \text{const}, \label{eq:repl3}\\
    \eta &\mapsto (2\Theta(\eta)-1) \eta \label{eq:repl4}
    \end{align}
    with $f$ and $f'$ being evaluated at the throat will yield a fully symmetric first order solution that, by construction, satisfies the equations on both sides near the throat.
    Here $\Theta$ is the Heavyside step function and $\delta$ the Dirac delta distribution.
    In order to be a solution over the entire radial regime the jumps of the field equations, introduced by second derivatives as in \cref{eq:repl3}, are required to vanish,
    \begin{align}
    \lim_{L\rightarrow0} \int_{-L}^{L} &(\nabla_\nu F^{\nu\mu} + \frac{1}{2} \lambda \rgb  A^\mu - \delta(l) j^\mu) ds = 0, \label{eq:lim1}\\
    \lim_{L\rightarrow0} \int_{-L}^{L} &(G^\mu_{~\nu}-\frac{1}{2}T^{(A)\mu}_{~~~~~\nu}+T^{(GB)\mu}_{~~~~~~~\nu}-\delta(\eta) 8\pi T^\mu_{~\nu}) ds = 0, \label{eq:lim2}
    \end{align}
    with line element $ds=\sqrt{f_1(\eta)}d\eta$.
    
    The source stress energy tensor and current density 
    that is required to compensate the jumps are attributed to a combination of matter and fields, $j=j^{(M)}+T^{(\Sigma)}$ and  $T=T^{(M)}+T^{(\Sigma)}$.
    For the former we assume the form of a perfect fluid at rest 
    \begin{equation}
    T^{(M)}_{\mu\nu}=(\epsilon+p) u_\mu u_\nu + p g_{\mu\nu}, \quad T^{(M)\mu}_{~~~~~~\nu}=\diag(-\epsilon,p,p,p), \qquad j^{(M)\mu} = (j^0,0,0,0)
    \end{equation}
    with 4-velocity $u^\mu=(f_0^{-1/2},0,0,0)$. 
    Since in wormhole solutions of EsGB theory an additional contribution to the action proved advantageous \cite{Kanti:2011jz,Kanti:2011yv,Antoniou:2019awm,Ibadov:2020btp,Ibadov:2020ajr},
    we suggest the presence of an analogous term on the throat surface described by the action 
    \begin{equation}
    S_\Sigma=\frac{1}{16\pi}\int \sqrt{-\hat{g}} \left[\lambda_1+2\lambda_0 F(A^2) \hat{R} \right] d^3x \ ,
    \end{equation}
    where $\hat{g} = \det(\hat{g}_{\mu\nu})$ denotes the determinant of the induced (2+1)-dimensional metric
    and $\hat{R}$ denotes the associated Ricci scalar,
    \begin{equation}
    	\hat{R}^\mu_{~\nu}= \text{diag}\left(0,\eta_0^{-2},\eta_0^{-2}\right), \quad \hat{R}=2 \eta_0^{-2}.
    \end{equation}
    Variation with respect to the vector field and the metric yields
    \begin{align}
    \frac{\delta S_\Sigma}{\delta A_\mu} &\approx \frac{1}{4\pi} \lambda_0 A^\mu F'(A^2) \hat{R} \sqrt{-\hat{g}},\\
    \frac{\delta S_\Sigma}{\delta \hat{g}^{\mu\nu}} &\approx \frac{1}{16\pi}\left(-\frac{1}{2} \lambda_1 \hat{g}_{\mu\nu} + 2\lambda_0 \left( \coupl(A^2) (\hat{R}_{\mu\nu}-\frac{1}{2}\hat{R} g_{\mu\nu}) + A_\mu A_\nu \coupl'(A^2) \hat{R} \right)  \right) \sqrt{-\hat{g}},
    \end{align}
    respectively.
    
    Expressing the system of \cref{eq:jceom1,eq:jceom2} in wormhole coordinates as specified above, doing the replacements given in \cref{eq:repl1,eq:repl2,eq:repl3,eq:repl4}, and performing the integration \cref{eq:lim1,eq:lim2} finally leads to the junction conditions
   \begin{align}
    	\frac{2 {A_0} \left({A_0} {f_0}'-{f_0} {A_0}'\right) \lambda \coupl'}{\pi {\eta_0}^2 {f_0}^2 \sqrt{{f_1}}} &= -\epsilon + \left(\frac{4 f_0  \lambda \coupl \lambda_0 + 8 {A_0}^2 \lambda_0 \lambda \coupl' }{16 \pi f_0 \eta_0^2} + \frac{\lambda_1}{16\pi}\right),\\
    	\frac{{f_0}'}{8\pi {f_0} \sqrt{{f_1}}} &=  p + \frac{\lambda_1}{16\pi},\\
    	\left(\frac{4{A_0} {f_0}' \lambda \coupl' }{{f_0}^2 {\eta_0}^2 \sqrt{{f_1}}}-
    	\frac{ 2{A_0}'}{{f_0} \sqrt{{f_1}}}\right) &= {j^0}
    	+\frac{{A_0} \lambda _0 \lambda \coupl'}{8\pi {f_0} {\eta_0}^2},
    \end{align}
    where all quantities are evaluated at the throat and $'$ denotes the derivative with respect to $\eta$.
     
Since ordinary wormhole coordinates cannot describe equators, we perform the analogous steps also for isotropic wormhole coordinates.
In that case the jumps are given by
\begin{align}
	\frac{2 {A_t} \left({A_t} {F_{0}}'-{F_{0}} {A_t}'\right) \left(4 {F_{1}}^2-3 \eta _0^2 \left({F_{1}}'\right)^2\right) F'+{F_{0}}^2 {F_{1}}^2 \eta _0^2 {F_{1}}'}{4 \pi  {F_{0}}^2 {F_{1}}^{7/2} \eta _0^2},\\
	\frac{{F_{0}}^2 {F_{1}} \left({F_{1}} {F_{0}}'+{F_{0}} {F_{1}}'\right)-6 {A_t} {F_{0}}' {F_{1}}' \left({A_t} {F_{0}}'-2 {F_{0}} {A_t}'\right) F'}{8 \pi  {F_{0}}^3 {F_{1}}^{5/2}},\\
	\frac{{A_t} {F_{0}}' \left(4 {F_{1}}^2-3 \eta _0^2 \left({F_{1}}'\right)^2\right) F'-2 {F_{0}} {F_{1}}^3 \eta _0^2 {A_t}'}{{F_{0}}^2 {F_{1}}^{7/2} \eta_0^2}.
\end{align}

\section*{Acknowledgement}
	
BK and JK gratefully acknowledge support by the
DFG Research Training Group 1620 {\sl Models of Gravity}
and the COST Actions CA15117 and CA16104.

\end{document}